\documentclass[a4paper,11pt]{article}

\usepackage{jheppub} 
\usepackage[utf8]{inputenc}
\usepackage{hyperref}
\usepackage{hyperref}
\usepackage{xcolor,graphicx}
\definecolor{teal}{rgb}{0, 0.5,0.5}
\hypersetup{
  colorlinks=true,
  linkcolor=teal,
  filecolor=blue,
  citecolor = teal,
  urlcolor=teal,
}
\usepackage{booktabs, dashrule}
\usepackage[T1]{fontenc} 
\usepackage{siunitx} 
\usepackage{physics} 
\usepackage{dcolumn}
\usepackage{bm}
\usepackage[normalem]{ulem}
\usepackage{braket} 
\usepackage{amsmath}
\usepackage{cancel}
\usepackage{slashed}
\usepackage{multicol}
\usepackage{multirow}
\usepackage{cleveref}
\usepackage{arydshln}
\usepackage{lscape}
\usepackage{pdflscape}
\usepackage{makecell}
\usepackage{graphicx}
\usepackage{subcaption}
\usepackage[export]{adjustbox}
\usepackage{tikz}
\usepackage{ytableau}
\usepackage[shortlabels]{enumitem}
\usepackage{dsfont}
\usepackage{comment}
\usepackage{mathrsfs}
\usepackage{csquotes}
\usepackage{placeins}

\newcommand{\lzm}{\left(}
\newcommand{\dzm}{\right)}
\newcommand{\lzs}{\left[}
\newcommand{\dzs}{\right]}
\newcommand{\lzv}{\left\{}
\newcommand{\dzv}{\right\}}
\newcommand{\lzu}{\left|}
\newcommand{\dzu}{\right|}
\newcommand{\cL}{\mathcal{L}}
\newcommand{\cO}{\mathcal{O}}
\newcommand{\cR}{\mathcal{R}}

\newcommand{\cI}{{\mathcal I}}
\newcommand{\cC}{{\mathcal C}}

\newcommand{\cX}{{\mathcal X}}

\newcommand{\cH}{{\mathcal H}}

\newcommand{\mev}{\mathrm{MeV}}
\newcommand{\gev}{\mathrm{GeV}}
\newcommand{\tev}{\mathrm{TeV}}
\newcommand{\re}{{\mathrm{Re}} \,}
\newcommand{\im}{{\mathrm{Im}} \,}
\newcommand{\hermc}{\text{h.c.}}

\newcommand{\U}{\mathrm{U}}
\newcommand{\SU}{\mathrm{SU}}

\newcommand{\eminus}{\vcenter{\hbox{\scalebox{0.6}[1]{$ - $}}}}  
\newcommand{\rep}[1]{\mathbf{#1}}

\newcommand{\sscript}[1]{{\scriptscriptstyle \mathrm{#1}}}

\definecolor{deepskyblue}{rgb}{0.0, 0.75, 1.0}
\definecolor{aqua}{rgb}{0.0, 1.0, 1.0}
\definecolor{bronze}{rgb}{0.8, 0.5, 0.2}
\definecolor{electricyellow}{rgb}{1.0, 1.0, 0.0}
\definecolor{goldenyellow}{rgb}{1.0, 0.87, 0.0}
\definecolor{glaucous}{rgb}{0.38, 0.51, 0.71}

\usepackage{soul}
\colorlet{blueRef}{blue!80!black}
\colorlet{CodeColor}{red!70!black} 

\newcommand{\X}{\mathcal{X}}

\title{
  {\scalebox{0.958}{The KSVZ Atlas: A Unified SMEFT--ALP Framework}}
}

\author[a]{Ajdin Palavri\'c,}
\author[b]{Xavier Ponce D\'iaz,}
\author[b]{Hector Tiblom}

\affiliation[a]{Institut de F\'isica Corpuscular (IFIC), Consejo Superior de Investigaciones Cient\'ificas (CSIC) and Universitat de Val\`encia (UV), 46980 Val\`encia, Spain}
\affiliation[b]{Department of Physics, University of Basel, Klingelbergstrasse 82,  CH-4056 Basel, Switzerland}

\emailAdd{ajdin.palavric@ific.uv.es}
\emailAdd{xavier.poncediaz@unibas.ch}
\emailAdd{hector.tiblom@unibas.ch}

\begin{document}

\abstract
{
  We develop a general framework for matching KSVZ-like ultraviolet completions featuring vector-like fermions and a spontaneously broken $\U(1)_{\sscript{PQ}}$ symmetry onto the Standard Model Effective Field Theory and the low-energy axion-like particle effective theory. The framework applies to arbitrary vector-like fermion representations and PQ-charge assignments, and systematically captures the effective interactions generated in both sectors. We then perform a comprehensive phenomenological analysis of the resulting SMEFT operators, based on global fits to electroweak precision, Higgs, and flavor observables, obtaining robust bounds on the corresponding Wilson coefficients that are largely independent of the details of the ultraviolet realization. These constraints can subsequently be translated into the QCD axion and ALP parameter space, providing indirect probes of ALP couplings. We further investigate several representative examples of the interplay between the SMEFT and ALP sectors, illustrating how direct ALP searches and indirect precision and flavor observables provide complementary information on the same underlying dynamics. We find that, over large regions of parameter space, indirect constraints derived from the SMEFT analysis dominate over direct ALP probes, except in scenarios where the PQ-charge assignment permits mass mixing with Standard Model fermions. Overall, our results establish a unified framework for connecting ultraviolet completions, SMEFT analyses, and ALP searches, enabling both the interpretation of existing constraints and the exploration of future signals within a common theoretical setting.
}

\maketitle
\clearpage
\section{Introduction}
\label{sec:intro}
Kim--Shifman--Vainshtein--Zakharov (KSVZ)~\cite{Kim:1979if,Shifman:1979if} constructions provide a well-motivated and versatile class of ultraviolet (UV) completions in which axions~\cite{Weinberg:1977ma,Wilczek:1977pj}, or more generally axion-like particles (ALPs), arise from the spontaneous breaking of a global $\U(1){\sscript{PQ}}$ symmetry. These scenarios extend the Standard Model by introducing heavy vector-like fermions (VLFs) and a complex scalar field charged under the Peccei--Quinn (PQ) symmetry~\cite{Peccei:1977hh,Peccei:1977ur}. Once the scalar acquires a vacuum expectation value, the $\U(1){\sscript{PQ}}$ symmetry is spontaneously broken, generating masses for the VLFs and giving rise to the pseudo-Nambu--Goldstone Boson (pNGB) associated with the broken symmetry. Depending on the gauge representations and PQ-charge assignments of the Vector-Like Fermions (VLFs)~\cite{DiLuzio:2024xnt}, these models generate a rich variety of effective interactions involving gauge bosons, fermions, and the Higgs sector. As a consequence, KSVZ-like constructions provide a broad framework in which flavor, electroweak, Higgs, and ALP phenomenology are intrinsically connected through a common UV origin.

In the minimal QCD axion scenario, the axion mass is fixed by the QCD topological susceptibility and is therefore tied to the decay constant through the characteristic relation $m_a f_a \sim m_\pi f_\pi$, up to quark-mass factors. This relation, however, need not hold in more general axion or ALP constructions. Additional contributions to the axion potential coming from multiple axion sectors~\cite{Gavela:2023tzu,deGiorgi:2024elx,FernandezNavarro:2026cyu}, additional confining sectors~\cite{Rubakov:1997vp,Gavela:2018paw,Croon:2019iuh,Reig:2019vqh}, and theories with non-trivial embeddings of QCD into a UV gauge group~\cite{Gherghetta:2016fhp,Agrawal:2017evu,Fuentes-Martin:2019bue,DiLuzio:2021gos,DiLuzio:2021pxd} can modify the axion mass while preserving the characteristic derivative and anomalous interactions of a pseudo-Nambu--Goldstone boson. More generally, ALP scenarios allow $m_a$ and $f_a$ to be treated as independent phenomenological parameters, motivating searches beyond the canonical QCD-axion line~\cite{Jaeckel:2012yz,Mimasu:2014nea,Brivio:2017ije,Bauer:2017nlg,Bauer:2018uxu,Bauer:2021mvw,Alda:2025uwo}. This provides a natural setting for the analysis performed in this work, where we consider both the standard QCD axion scenario and a more general ALP setup in which the mass is treated as a free parameter, while the interactions remain correlated with the heavy VLF sector.

From the effective field theory perspective, the structure of these models implies that the low-energy ALP interactions cannot generically be treated independently from the effective operators generated in the Standard Model Effective Field Theory. This connection between axion or axion-like particle phenomenology and UV dynamics has been explored in several scenarios, including Dine--Fischler--Srednicki--Zhitnitsky (DFSZ) models~\cite{Dine:1981rt,Zhitnitsky:1980tq} with flavor violation~\cite{DiLuzio:2023ndz}, Froggatt--Nielsen (FN) ALP constructions~\cite{Froggatt:1978nt,Greljo:2024evt}, and KSVZ-like models~\cite{Alonso-Alvarez:2023wig}. In this work, we extend these studies by integrating out the heavy vector-like fermions while keeping the ALP as a dynamical low-energy degree of freedom, thereby matching simultaneously onto the SMEFT and the ALP effective theory. This procedure generates correlated interactions in both sectors, leading to non-trivial relations between precision, flavor, and direct ALP observables. Motivated by this interplay, we develop a unified SMEFT--ALP framework for KSVZ-like models, applicable to a broad class of VLF representations and PQ-charge assignments. The framework systematically captures the effective interactions generated across the different sectors and provides a common language in which indirect SMEFT constraints and direct ALP probes can be consistently interpreted within the same UV setting.

Building on this framework, we perform a comprehensive phenomenological analysis of the SMEFT operators generated after integrating out the heavy VLFs at tree level. For each representation and flavor structure, the induced Wilson coefficients are correlated by the underlying VLF couplings, leading to predictive rank-one flavor patterns, see also Refs.~\cite{Gherardi:2019zil,Marzocca:2024hua,Biondini:2026ryb}. We constrain these directions using global fits to electroweak precision, Higgs, and flavor observables, deriving bounds on the corresponding SMEFT coefficients for the different KSVZ-like scenarios. Exploiting the common UV origin of the SMEFT and ALP interactions, these bounds can then be mapped onto the ALP parameter space and compared with direct ALP probes. We find that indirect SMEFT constraints provide the leading sensitivity over large regions of parameter space, while scenarios in which the PQ-charge assignment allows for SM--VLF mass mixing constitute an important exception, since flavor-changing meson decays can become the dominant probes of the ALP sector.

The paper is organized as follows. In Section~\ref{sec:UV_IR_Framework}, we classify the KSVZ-like constructions considered in this work and introduce the unified SMEFT--ALP effective description. In Section~\ref{sec:SMEFT_pheno}, we derive constraints on the SMEFT Wilson coefficients generated by the different VLF representations using electroweak precision, Higgs, and flavor observables. Since this part of the analysis depends only on the SMEFT operator structure, the resulting bounds are independent of the ALP mass and decay phenomenology, and provide a comprehensive set of constraints on the VLF-induced SMEFT directions. In Section~\ref{sec:Axion_Pheno}, we discuss the corresponding QCD axion constraints and translate the SMEFT bounds into the axion parameter space. In Section~\ref{sec:Interplay}, we investigate the correlated SMEFT--ALP phenomenology in selected benchmark scenarios beyond the QCD axion setup, highlighting the complementarity between indirect precision probes and direct ALP searches. We summarize our results and conclude in Section~\ref{sec:Conclusions}. Additional technical material is collected in the appendices: Appendix~\ref{app:gen_match_framework} presents the general matching framework, Appendix~\ref{app:det_mat_proc_app} contains details of the matching procedure, and Appendices~\ref{app:SMEFT_obs_overview} and \ref{app:Q7_SMEFT_ALP} collect additional phenomenological details relevant to the analyses presented in the main text.

\section{UV Completions and Effective Descriptions}
\label{sec:UV_IR_Framework}
\subsection{Overview of the UV Models}

We consider a class of UV completions of the SM containing VLFs and a complex scalar field $\Phi$ charged under a global $\U(1)_\sscript{PQ}$ symmetry. The scalar acquires a vacuum expectation value, spontaneously breaking $\U(1)_\sscript{PQ}$ and giving rise to the pseudo-Nambu--Goldstone boson associated with the broken symmetry. We go beyond the original KSVZ construction by allowing VLFs in general SM gauge representations and admitting a broad class of consistent $\U(1)_\sscript{PQ}$ charge assignments, which enables a systematic study of the interplay between ALP physics and SMEFT effects. In particular, we focus on KSVZ-like models in which the VLFs can decay through renormalizable interactions with SM fields, thereby avoiding stable colored relics in the early universe. A complete catalogue of such KSVZ models can be found in Ref.~\cite{DiLuzio:2024xnt}.

We begin by classifying the viable models emerging from the setup described above. To this end, we consider seven vector-like fermion representations under the SM gauge group, each charged under $\SU(3)_\sscript{C}$ and possibly carrying non-trivial electroweak quantum numbers~\cite{deBlas:2017xtg}.\footnote{While we restrict our analysis to colored VLFs, the formalism can be straightforwardly extended to leptonic counterparts.} For each gauge representation, the allowed $\U(1)_{\sscript{PQ}}$ charge assignments are fixed by requiring the existence of a $\Phi$-induced VLF mass term of the form
\begin{equation}\label{eq:UV_mass_term}
  -\cL_{\sscript{UV}}^{(4)}\supset y_\Psi\,\Phi\,\bar\Psi_L\Psi_R+\hermc\,,
\end{equation}
where $\Psi_{L,R}$ denote the left- and right-handed components of the VLF. Once the $\U(1)_{\sscript{PQ}}$ charges are specified, the complete set of renormalizable VLF-SM interactions can be constructed, including mixing terms between the two sectors. These interactions ultimately determine the pattern of effective ALP and SMEFT couplings that arise after integrating out the heavy degrees of freedom.

For the purpose of classification, we fix the PQ charge normalization by assigning $\cX_\Phi=+1$ and taking all SM fields to be neutral. Invariance of the mass term in Eq.~\eqref{eq:UV_mass_term} under $\U(1)_{\sscript{PQ}}$ then requires the VLF charges to satisfy
\begin{equation}\label{eq:PQ_charge_constraint}
  \cX_{\Psi_L} - \cX_{\Psi_R} = \cX_\Phi = 1\,.
\end{equation}
This condition fixes only the relative PQ charge between the left- and right-handed VLF components, leaving the absolute PQ charges free up to a common shift. Fixing this residual freedom by choosing a representative assignment for the unfixed $\U(1)_{\sscript{PQ}}$ charges, together with gauge invariance uniquely determines the allowed set of renormalizable operators involving SM fields and VLFs.\footnote{One can extend this setup by including dimension-five operators as done in Refs.~\cite{Alonso-Alvarez:2023wig,DiLuzio:2024xnt}.} Applying this procedure to each of the seven representations yields a systematic classification of viable interaction structures and their corresponding model realizations.

\begin{table}[t]
  \centering
  \renewcommand{\arraystretch}{1.3}
\scalebox{0.795}{
\begin{tabular}{cccc}
\toprule
\textbf{\textbf{VLF}}&$\bm{\cX_{L}}$ &$\bm{\cX_R}$& $\bm{-\cL^{(4)}_{\sscript{UV}}\supset}$
\\
\midrule
\multirow{3}{*}{\vspace{-0.7cm}$D\sim(\bm 3,\bm 1)_{-1/3}$}
&1
&0
&$y_D\Phi\,\bar D_LD_R+H\,\bar q_Ly_{1,d}D_R+\,\Phi\bar D_L\,y_{2,d}d_R+H\,\bar q_LY_dd_R$
\vspace{+0.3cm}\\
&0
&-1
&$y_D\Phi\,\bar D_LD_R+\bar D_LM_dd_R+H\,\bar q_LY_dd_R$
\vspace{+0.3cm}\\
&-1
&-2
&$y_D\Phi\,\bar D_LD_R+\Phi^*\bar D_Ly_{1,d}d_R+H\,\bar q_LY_dd_R$
\\
\midrule
\multirow{3}{*}{\vspace{-0.7cm}$U\sim(\bm 3,\bm 1)_{2/3}$}
&1
&0
&$y_U\Phi\,\bar U_LU_R+\widetilde H\,\bar q_Ly_{1,u}U_R+\Phi\,\bar U_Ly_{2,u}u_R+\widetilde H\,\bar q_LY_uu_R$
\vspace{+0.3cm}\\
&0
&-1
&$y_U\,\Phi\,\bar U_LU_R+\bar U_L M_u u_R+\widetilde H\,\bar q_LY_u u_R$
\vspace{+0.3cm}\\
&-1
&-2
&$y_U\,\Phi\,\bar U_LU_R+\Phi^*\bar U_L y_{1,u}u_R+\widetilde H\,\bar q_L Y_u u_R$
\\
\midrule
\multirow{3}{*}{\vspace{-0.7cm}$Q\sim(\bm 3,\bm 2)_{1/6}$}
&1
&0
&$y_Q\,\Phi^*\bar Q_RQ_L+\bar Q_R M_qq_L+\widetilde H\,\bar q_LY_uu_R+H\,\bar q_LY_dd_R$
\vspace{+0.3cm}\\
&\multirow{1}{*}{0}
&\multirow{1}{*}{-1}
&$y_Q\,\Phi^*\bar Q_RQ_L+\Phi^*\bar Q_Ry_q q_L+H\,\bar Q_Ly_Q^dd_R+\widetilde H\,\bar Q_Ly_Q^uu_R+\widetilde H\,\bar q_LY_uu_R+H\,\bar q_LY_dd_R$
\vspace{+0.3cm}\\
&2
&1
&$y_Q\,\Phi^*\bar Q_RQ_L+\Phi \,\bar Q_Ry_q q_L+\widetilde H\,\bar q_LY_uu_R+H\,\bar q_LY_dd_R$
\\
\midrule
$Q_5\sim(\bm 3,\bm 2)_{-5/6}$
&0&-1
&$y_{Q_5}\Phi\,\bar Q_{5L}Q_{5R}+\widetilde H\,\bar Q_{5L} \lambda_{Q_5} d_R+H\,\bar q_LY_dd_R$
\\
\midrule
$Q_7\sim(\bm 3,\bm 2)_{7/6}$
&0&-1
&$y_{Q_7}\Phi\,\bar Q_{7L}Q_{7R}+H\,\bar Q_{7L} \lambda_{Q_7} u_R+\widetilde H\,\bar q_LY_uu_R$
\\
\midrule
$T_1\sim(\bm 3,\bm 3)_{-1/3}$
&1&0
&$y_{T_1}\Phi\, \bar T_{1L}T_{1R}+\frac{1}{2}H^\dagger\, \bar T_{1R}^a \sigma^a \lambda_{T_1}q_L+H\,\bar q_LY_dd_R+\widetilde H\,\bar q_LY_uu_R$
\\
\midrule
$T_2\sim(\bm 3,\bm 3)_{2/3}$
&1&0
&$y_{T_2}\Phi\, \bar T_{2L}T_{2R}+\frac{1}{2}\widetilde H^\dagger \bar T_{2R}^a \sigma^a \lambda_{T_2}q_L+H\,\bar q_LY_dd_R+\widetilde H\,\bar q_LY_uu_R$
\\
\bottomrule
\end{tabular}

}
  \caption{Overview of the model classification based on different VLF gauge representations and their $\U(1)_{\sscript{PQ}}$ charge assignments. For each representation, the table displays the allowed PQ charges satisfying $\cX_L - \cX_R = 1$, along with the corresponding dimension-four interaction terms permitted by gauge invariance and PQ symmetry. Flavor indices for the SM fields have been suppressed.}
  \label{tab:models_VLFs}
\end{table}
The results of this classification are summarized in Table~\ref{tab:models_VLFs}. For each of the seven VLF representations, we list the representative $\U(1)_{\sscript{PQ}}$ charge assignments consistent with Eq.~\eqref{eq:PQ_charge_constraint} and admitting renormalizable interactions with SM fermions. We restrict attention to charge assignments that allow mixing terms with the Higgs doublet and SM quarks, as these are the cases relevant for generating effective ALP and SMEFT operators. Configurations that forbid such interactions are not considered.

Moreover, any alternative PQ charge assignment satisfying Eq.~\eqref{eq:PQ_charge_constraint} and leading to the same set of gauge- and PQ-invariant interactions can be obtained from those listed in Table~\ref{tab:models_VLFs} by a common shift of the VLF charges. Consequently, the configurations listed form a minimal and complete basis of physically inequivalent interaction structures, which fully determine the pattern of low-energy effective operators discussed below.

Finally, throughout this analysis the VLFs are treated as flavor singlets.\footnote{The general case with flavored VLFs is treated in the general matching procedure described in Appendix~\ref{app:gen_match_framework}.} As a consequence, the flavor structure of the coupling tensors, and hence of the renormalizable interactions listed in Table~\ref{tab:models_VLFs}, is entirely determined by the SM fields appearing in each operator. This type of construction leads to what is commonly referred to as rank-1 flavor violation~\cite{Gherardi:2019zil, Marzocca:2024hua,Biondini:2026ryb}, since the Wilson coefficients are built from a single vector structure and are therefore rank-1.\footnote{As an illustrative example, consider the $D$-type VLF with PQ charge assignment $(\cX_L,\cX_R)=(1,0)$. In this case, the coupling $y_{1,d}$ transforms as a $3\times1$ column vector in flavor space, while $y_{2,d}$ is a $1\times3$ row vector, reflecting that the flavor dependence originates solely from the SM fields.}

\subsection{Spontaneous \(\U(1)_{\sscript{PQ}}\) Breaking and Field Redefinitions}
\label{subsec:spont_U1br_FRs}
We now turn to the consequences of spontaneous $\U(1)_{\sscript{PQ}}$ breaking in the class of models introduced above. The complex scalar field $\Phi$ acquires a vacuum expectation value (VEV), which can be parametrized as
\begin{equation}\label{eq:phi_vev}
  \langle\Phi\rangle=\frac{v_\Phi+\rho_\Phi}{\sqrt2}e^{i\frac{a}{v_\Phi}}\,,
\end{equation}
where $v_\Phi$ denotes the PQ breaking scale, $\rho_\Phi$ is the radial excitation, and $a$ is the pseudo-Nambu-Goldstone boson associated with the broken symmetry. This breaking generates VLF masses through the dimension-four interaction in Eq.~\eqref{eq:UV_mass_term} and gives rise to a dynamical pseudoscalar degree of freedom identified as an ALP. The radial mode $\rho_\Phi$ typically acquires a mass of order the PQ scale and decouples from the low-energy theory. The scale $v_\Phi$ therefore determines both the VLF mass scale and the effective ALP couplings to SM fields.

For certain gauge representations and PQ charge assignments in Table~\ref{tab:models_VLFs}, the renormalizable Lagrangian contains interaction terms of the form
\begin{equation}\label{eq:generic_mass_mix}
  -\cL_{\sscript{UV}}^{(4)}\supset \Phi \,\bar \Psi_L\, y_{\Psi} \psi_R+ M_\psi\bar\Psi_L\psi_R+\hermc\,,
\end{equation}
which induce mass mixing between SM fermions and VLFs after $\Phi$ acquires a VEV. Such mixings can be removed by field redefinitions that diagonalize the fermion mass matrix. Although this procedure eliminates explicit mass-mixing terms, it generally induces additional interactions in the low-energy theory, including modified Higgs and gauge couplings as well as contributions to higher-dimensional operators. To illustrate this procedure, we focus on a $D$-type VLF with PQ charges $(\cX_L,\cX_R)=(0,-1)$. For this choice, the renormalizable Lagrangian contains the following interaction terms
\begin{equation}\label{eq:D_VLF_ex_FRs_intro}
  -\cL_{\sscript{UV}}^{(4)}\supset y_D\Phi\,\bar D_LD_R+\bar D_LM_dd_R+H\,\bar q_LY_dd_R+\hermc\,.
\end{equation}
Upon symmetry breaking, the interactions in Eq.~\eqref{eq:D_VLF_ex_FRs_intro} become
\begin{equation}\label{eq:D_redef_example}
  -\cL_{\sscript{UV}}^{(4)}\supset \frac{ y_D v_\Phi}{\sqrt2}\,\bar D_LD_R+\bar D_LM_dd_R+H\,\bar q_LY_dd_R+\hermc\,.
\end{equation}
As indicated by Eq.~\eqref{eq:generic_mass_mix}, the $\bar D_LM_dd_R$ term induces a mass mixing between the VLF and the SM down quark, which can be removed by performing the field redefinitions of the form
\begin{equation}
  \label{eq:field_redefinition}
  {\scalebox{0.97}{
      $
      \begin{alignedat}{2}
        D_R\to \lzs1+\frac{2 M_dM_d^\dag}{y_D^2v_\Phi^2}\dzs^{\eminus\frac{1}{2}} D_R-\frac{\sqrt2 M_d}{y_Dv_\Phi}\cR_d d_R\,,
        \quad
        d_R\to \cR_d d_R+\frac{\sqrt2 M_d^\dag}{y_D v_\Phi} \lzs1+\frac{2 M_dM_d^\dag}{y_D^2v_\Phi^2}\dzs^{\eminus\frac{1}{2}} D_R\, ,
      \end{alignedat}
      $
  }}
\end{equation}
where $\mathcal{R}_d$ denotes the matrix introduced to canonically normalize the kinetic terms after the field redefinitions and remove the non-trivial mixing induced in the kinetic sector, which can be defined as
\begin{equation}
  \mathcal{R}_d \equiv \left[1+\frac{2M_d^\dagger M_d}{y_D^2 v_\Phi^2}\right]^{\eminus 1/2}\,.
\end{equation}
Upon substituting these relations into the Lagrangian given by Eq.~\eqref{eq:D_redef_example} and performing the expansion, the interaction terms can be rewritten as
\begin{equation}
  \begin{alignedat}{2}
    -\cL_{\sscript{UV}}^{(4)}&\supset
    \frac{y_D v_\Phi}{\sqrt2} \lzs 1+\frac{2 M_dM_d^\dag}{y_D^2 v_\Phi^2} \dzs^{\frac{1}{2}} \bar D_L D_R
    +\frac{\sqrt2}{y_D v_\Phi} \lzs 1+\frac{2 M_dM_d^\dag}{y_D^2 v_\Phi^2} \dzs^{\eminus\frac{1}{2}} \bar q_L Y_dM_d^\dag D_R H
    \\[3pt]
    &+\bar q_L Y_d \cR_d d_RH+\hermc\,.
  \end{alignedat}
\end{equation}
As a result of the field redefinitions, the mass mixing between $D_L$ and $d_R$ is removed, while the mass term for the VLF becomes rescaled. At the same time, a new dimension-four interaction involving $q_L$, $H$, and $D_R$ is induced, which plays an important role at low energies, as integrating out the heavy $D$ fermion generates effective operators in the SMEFT via tree-level matching.

The relevant field redefinitions for all configurations that exhibit SM-VLF mass mixing are collected in Appendix~\ref{app:det_mat_proc_app}. From the interaction terms presented in Table~\ref{tab:models_VLFs}, we observe that no such redefinitions are required for the $Q_5$, $Q_7$, $T_1$, and $T_2$ fields, as their PQ charge assignments forbid renormalizable mass-mixing terms with SM fields. For all other representations, the mass mixing is removed using field redefinitions analogous to the example outlined above. Depending on the structure of the interactions, these redefinitions may either induce new terms or modify existing ones in the interaction Lagrangian.

\subsection{Unified SMEFT and ALP Effective Descriptions}
\label{sec:unified_effdctive_descriptions}
Having outlined the UV setup and the effects of $\U(1)_{\sscript{PQ}}$ breaking, we turn to the effective descriptions in the SMEFT and ALP sectors. While a fully general matching framework is presented in Appendix~\ref{app:gen_match_framework}, here we adopt a unified perspective tailored to a representative scenario, making the correspondence between the two descriptions explicit. This approach allows for a direct mapping between UV parameters and low-energy observables, enabling correlated effects across different probes to be systematically traced. Before presenting the combined analysis, we briefly summarize the aspects of the ALP sector relevant for what follows.

Breaking $\U(1)_{\sscript{PQ}}$ spontaneously yields a light ALP that remains in the low-energy spectrum. The structure of its interactions with SM fields is fixed by the UV completion. In particular, inserting the VEV parametrization of Eq.~\eqref{eq:phi_vev} into the interaction in Eq.~\eqref{eq:UV_mass_term} shows that the VLF mass term acquires a phase, which encodes the ALP couplings. This phase can be absorbed by a chiral rotation of the form
\begin{equation}
  \Psi_{L,R}\to e^{i\cX_{L,R}\frac{a}{v_\Phi}}\Psi_{L,R}\,,
\end{equation}
which in turn induces derivative couplings between the ALP and fermion currents through the kinetic terms. The resulting interaction reads
\begin{equation}\label{eq:ALP_VLF_der_UV}
  \mathcal{L}_{\sscript{ALP}} \supset \frac{\partial_\mu a}{ v_\Phi}\bar \Psi_{R,\, L}  \X_{R,\, L} \gamma^\mu \Psi_{R,\, L}\,.
\end{equation}
In addition to the derivative couplings to fermions, the ALP inherits anomaly-induced couplings to gauge bosons from the same field redefinitions. These contributions can be captured by the effective Lagrangian
\begin{equation}
  \cL_{\sscript{an.}}\supset \frac{a}{v_\Phi}\lzm \frac{\alpha_s}{4\pi} c_G\, G_{\mu\nu} \widetilde{G}^{\mu\nu} + \frac{\alpha_1}{4\pi} c_B\, B_{\mu\nu} \widetilde{B}^{\mu\nu} + \frac{\alpha_2}{4\pi} c_W \,W_{\mu\nu} \widetilde{W}^{\mu\nu} \dzm\,,
\end{equation}
where the anomaly coefficients $c_{G,W,B}$ are determined by the PQ charges and gauge quantum numbers of the heavy fermions~\cite{DiLuzio:2020wdo}:
\begin{equation}
  c_G = -\sum_i \mathcal{X}_i\, d(\mathcal{I}_i)\, T(\mathcal{I}_i)\,, \quad
  c_W = -\sum_i \mathcal{X}_i\, d(\mathcal{I}_i)\, T(\mathcal{I}_i)\,, \quad
  c_B = -\sum_i \mathcal{X}_i\, d(\mathcal{I}_i)\, Y_i^2\,.
\end{equation}
Here $d(\mathcal{I}_i)$ and $T(\mathcal{I}_i)$ denote the dimension and Dynkin index of the corresponding gauge representation, and $Y_i$ the hypercharge of the fermion.

With the ALP interaction structure in place, we can now specify the complete set of interactions defining our class of UV completions. These comprise the couplings between VLFs and SM fields, together with the derivative ALP interactions induced by spontaneous $\U(1)_{\sscript{PQ}}$ breaking. The resulting interaction structures, collected in Table~\ref{tab:interactions_after_redef}, provide the starting point for the EFT analysis that follows in both the SMEFT and ALP sectors, and underpin the phenomenological results presented in the remainder of this work.

\begin{table}[t]
  \centering
  \centering
\renewcommand{\arraystretch}{1.5}
\scalebox{0.71}{
\begin{tabular}{cccc}
\toprule
\textbf{\textbf{VLF}}&$\bm{\cX_{L}}$ &$\bm{\cX_R}$& $\bm{\cL_{\sscript{UV+ALP}}\supset}$
\\
\midrule
\multirow{3}{*}{\vspace{-2.7cm}$D\sim(\bm 3,\bm 1)_{-1/3}$}
&\multirow{1}{*}{\vspace{-0.1cm}1}
&\multirow{1}{*}{\vspace{-0.1cm}0}
&$\lzs -\widetilde m_D\bar D_L D_R-H\bar q_L \widehat Y_d d_R-H\bar q_L \lambda_D D_R+\hermc \dzs+\frac{\partial_\mu a}{v_\Phi}C_{D,L}^{(1)}\bar D_L\gamma_\mu D_L$
\vspace{+0.1cm}\\
\noalign{\vskip 2pt}
\cdashline{2-4}[.4pt/2pt]
\noalign{\vskip 2pt}
\vspace{+0.0cm}
&\multirow{2}{*}{\vspace{-0.6cm}0}
&\multirow{2}{*}{\vspace{-0.6cm}-1}
&$\lzs -\widetilde m_D\bar D_L D_R-H\bar q_L \widehat Y_d d_R-H\bar q_L \lambda_D D_R+\hermc \dzs+\frac{\partial_\mu a}{v_\Phi}C_{D,R}^{(1)}\bar D_R\gamma_\mu D_R$
\vspace{+0.1cm}\\
&&&$+\frac{\partial_\mu a}{v_\Phi}\bar d_R C^{(2)}_{D,R}\gamma_\mu d_R+\frac{\partial_\mu a}{v_\Phi}\lzs \bar D_R C_{D,R}^{(3)}\gamma_\mu d_R+\hermc \dzs$
\vspace{+0.1cm}\\
\noalign{\vskip 2pt}
\cdashline{2-4}[.4pt/2pt]
\noalign{\vskip 2pt}
\vspace{+0.0cm}
&\multirow{2}{*}{\vspace{-0.4cm}-1}
&\multirow{2}{*}{\vspace{-0.4cm}-2}
&$\lzs -\widetilde m_D\bar D_L D_R-H\bar q_L \widehat Y_d d_R-H\bar q_L \lambda_D D_R+\hermc \dzs+\frac{\partial_\mu a}{v_\Phi}C_{D,L}^{(1)}\bar D_L \gamma_\mu D_L+\frac{\partial_\mu a}{v_\Phi}C_{D,R}^{(1)}\bar D_R\gamma_\mu D_R$
\\[6pt]
&&&$+\frac{\partial_\mu a}{v_\Phi}\bar d_R C^{(2)}_{D,R}\gamma_\mu d_R+\frac{\partial_\mu a}{v_\Phi}\lzs \bar D_R C_{D,R}^{(3)}\gamma_\mu d_R+\hermc \dzs$
\vspace{+0.1cm}\\
\midrule
\multirow{3}{*}{\vspace{-2.6cm}$U\sim(\bm 3,\bm 1)_{2/3}$}
&1
&0
&$\lzs -\widetilde m_U\bar U_L U_R-\widetilde H\bar q_L \widehat Y_u u_R-\widetilde H\bar q_L \lambda_U U_R+\hermc \dzs+\frac{\partial_\mu a}{v_\Phi}C_{U,L}^{(1)}\bar U_L\gamma_\mu U_L$

\vspace{+0.1cm}\\
\noalign{\vskip 2pt}
\cdashline{2-4}[.4pt/2pt]
\noalign{\vskip 2pt}
\vspace{+0.0cm}
&\multirow{2}{*}{\vspace{-0.5cm}0}
&\multirow{2}{*}{\vspace{-0.5cm}-1}
&$\lzs -\widetilde m_U\bar U_L U_R-\widetilde H\bar q_L \widehat Y_u u_R-\widetilde H\bar q_L \lambda_U U_R+\hermc \dzs+\frac{\partial_\mu a}{v_\Phi}C_{U,R}^{(1)}\bar U_R\gamma_\mu U_R$
\vspace{+0.1cm}\\
&&&$+\frac{\partial_\mu a}{v_\Phi}\bar u_R C^{(2)}_{U,R}\gamma_\mu u_R+\frac{\partial_\mu a}{v_\Phi}\lzs \bar U_R C_{U,R}^{(3)}\gamma_\mu u_R+\hermc \dzs$
\vspace{+0.1cm}\\
\noalign{\vskip 2pt}
\cdashline{2-4}[.4pt/2pt]
\noalign{\vskip 2pt}
\vspace{+0.0cm}
&\multirow{2}{*}{\vspace{-0.5cm}-1}
&\multirow{2}{*}{\vspace{-0.5cm}-2}
&$\lzs -\widetilde m_U\bar U_L U_R-\widetilde H\bar q_L \widehat Y_u u_R-\widetilde H\bar q_L \lambda_U U_R+\hermc \dzs+\frac{\partial_\mu a}{v_\Phi}C_{U,L}^{(1)}\bar U_L \gamma_\mu U_L+\frac{\partial_\mu a}{v_\Phi}C_{U,R}^{(1)}\bar U_R\gamma_\mu U_R$
\vspace{+0.1cm}\\
&&&$+\frac{\partial_\mu a}{v_\Phi}\bar u_R C^{(2)}_{U,R}\gamma_\mu u_R+\frac{\partial_\mu a}{v_\Phi}\lzs \bar U_R C_{U,R}^{(3)}\gamma_\mu u_R+\hermc \dzs$
\\
\midrule
\multirow{3}{*}{\vspace{-2.8cm}$Q\sim(\bm 3,\bm 2)_{1/6}$}
&\multirow{2}{*}{\vspace{-0.5cm}1}
&\multirow{2}{*}{\vspace{-0.5cm}0}
&$\lzs -\widetilde m_Q \bar Q_R Q_L-\widetilde H\bar q_L \widehat Y_u u_R-H\bar q_L \widehat Y_d d_R -\widetilde H\bar Q_L \lambda_Q^u u_R-H\bar Q_L \lambda_Q^d d_R+\hermc \dzs$
\vspace{+0.1cm}\\
&&&$+\frac{\partial_\mu a}{v_\Phi}C_{Q,L}^{(1)}\bar Q_L\gamma_\mu Q_L+\frac{\partial_\mu a}{v_\Phi}\bar q_L C_{Q,L}^{(2)}\gamma_\mu q_L +\frac{\partial_\mu a}{v_\Phi}\lzs \bar Q_L C_{Q,L}^{(3)}\gamma_\mu  q_L +\hermc\dzs$
\\[5pt]
\noalign{\vskip 2pt}
\cdashline{2-4}[.4pt/2pt]
\noalign{\vskip 2pt}
\vspace{+0.0cm}
&\multirow{1}{*}{\vspace{-0.1cm}0}
&\multirow{1}{*}{\vspace{-0.1cm}-1}
&$\lzs -\widetilde m_Q \bar Q_R Q_L-\widetilde H\bar q_L \widehat Y_u u_R-H\bar q_L \widehat Y_d d_R -\widetilde H\bar Q_L \lambda_Q^u u_R-H\bar Q_L \lambda_Q^d d_R+\hermc \dzs+\frac{\partial_\mu a}{v_\Phi}C_{Q,R}^{(1)}\bar Q_R \gamma_\mu Q_R$
\\[5pt]
\noalign{\vskip 2pt}
\cdashline{2-4}[.4pt/2pt]
\noalign{\vskip 2pt}
\vspace{+0.0cm}
&\multirow{2}{*}{\vspace{-0.4cm}2}
&\multirow{2}{*}{\vspace{-0.4cm}1}
&$\lzs -\widetilde m_Q \bar Q_R Q_L-\widetilde H\bar q_L \widehat Y_u u_R-H\bar q_L \widehat Y_d d_R -\widetilde H\bar Q_L \lambda_Q^u u_R-H\bar Q_L \lambda_Q^d d_R+\hermc \dzs+\frac{\partial_\mu a}{v_\Phi}C_{Q,R}^{(1)}\bar Q_R \gamma_\mu Q_R$
\vspace{+0.1cm}\\
&&&$+\frac{\partial_\mu a}{v_\Phi}C_{Q,L}^{(1)}\bar Q_L\gamma_\mu Q_L+\frac{\partial_\mu a}{v_\Phi}\bar q_L C_{Q,L}^{(2)}\gamma_\mu q_L +\frac{\partial_\mu a}{v_\Phi}\lzs \bar Q_L C_{Q,L}^{(3)}\gamma_\mu  q_L +\hermc\dzs$
\\
\midrule
$Q_5\sim(\bm 3,\bm 2)_{-5/6}$
&0&-1
&$\lzs -\widetilde m_{Q_5}\bar Q_{5L}Q_{5R}-\widetilde H\bar Q_{5L} \lambda_{Q_5} d_R-H\bar q_L\widehat Y_dd_R +\hermc\dzs+\frac{\partial_\mu a}{v_\Phi}C_{Q_5,R}^{(1)}\bar Q_{5R}\gamma_\mu Q_{5R}$
\\
\midrule
$Q_7\sim(\bm 3,\bm 2)_{7/6}$
&0&-1
&$\lzs -\widetilde m_{Q_7}\bar Q_{7L}Q_{7R}-H\bar Q_{7L} \lambda_{Q_7} u_R-\widetilde H\bar q_L \widehat Y_uu_R +\hermc\dzs+\frac{\partial_\mu a}{v_\Phi}C_{Q_7,R}^{(1)}\bar Q_{7R}\gamma_\mu Q_{7R}$
\\
\midrule
$T_1\sim(\bm 3,\bm 3)_{-1/3}$
&1&0
&$\lzs -\widetilde m_{T_1} \bar T_{1L}T_{1R}-\frac{1}{2}H^\dagger \bar T_{1R}^a \sigma^a \lambda_{T_1}q_L-H\bar q_L\widehat Y_dd_R-\widetilde H\bar q_L \widehat Y_uu_R +\hermc\dzs+\frac{\partial_\mu a}{v_\Phi}C_{T_1,L}^{(1)}\bar T_{1L}\gamma_\mu T_{1L}$
\\
\midrule
$T_2\sim(\bm 3,\bm 3)_{2/3}$
&1&0
&$\lzs -\widetilde m_{T_2} \bar T_{2L}T_{2R}-\frac{1}{2}\widetilde H^\dagger \bar T_{2R}^a \sigma^a \lambda_{T_2}q_L-H\bar q_L\widehat Y_dd_R+\widetilde H\bar q_L\widehat Y_uu_R +\hermc\dzs+\frac{\partial_\mu a}{v_\Phi}C_{T_2,L}^{(1)}\bar T_{2L}\gamma_\mu T_{2L}$
\\
\bottomrule
\end{tabular}
}

  \caption{Unified interaction Lagrangians for all VLF representations considered in this work. Each row lists the $\U(1)_{\sscript{PQ}}$ charge assignments $(\cX_L, \cX_R)$ and the resulting interaction terms involving SM fields and axion-like particles (ALPs), after performing the relevant field redefinitions. These Lagrangians encode the leading contributions to both the SMEFT and ALP effective sectors. The mapping between the parameters appearing in this table and the underlying UV couplings is provided in Appendix~\ref{app:det_mat_proc_app}.}
  \label{tab:interactions_after_redef}
\end{table}

Starting from the UV interaction structures summarized in Table~\ref{tab:interactions_after_redef}, we derive the effective low-energy theory obtained by integrating out the heavy VLFs at tree level. The resulting description is formulated in terms of SM fields and the ALP and organized as an expansion in inverse powers of the VLF mass. We perform this matching systematically for all representations introduced above, retaining the leading operators in the SMEFT and ALP sectors relevant for phenomenology.

In the SMEFT sector, integrating out the VLFs at the scale $v_\Phi$ generates dimension-six operators at tree level. The leading contributions originate from the renormalizable VLF-SM couplings listed in Table~\ref{tab:interactions_after_redef} and populate two classes of operators in the Warsaw basis, namely $\psi^2 H^2 D$ and $\psi^2 H^3$~\cite{Grzadkowski:2010es,deBlas:2017xtg}. The set of operators generated is fixed by the gauge quantum numbers of the VLF representation, while the structure of the Wilson coefficients depends on the specific $\U(1)_{\sscript{PQ}}$ charge assignments through their impact on the underlying couplings.

Conversely, in the ALP sector, effective couplings to SM fermions arise from two distinct mechanisms, both stemming from the derivative interaction between the ALP and the heavy VLFs. The first source of ALP-SM interactions emerges from the field redefinitions required to remove SM-VLF mixing terms following PQ breaking. In representations that permit such mixing, part of the ALP coupling is transferred to the SM fermions, leading to tree-level derivative interactions. A second class of contributions arises in representations without mass mixing, where the ALP couples to SM fermions only after integrating out the VLFs. These effects generate operators suppressed by $\widetilde{m}_\Psi^2$, with additional dependence on the underlying UV couplings, as indicated in Appendix~\ref{app:det_mat_proc_app}.

To illustrate these two mechanisms in a concrete setting, we examine the $D \sim (\rep 3,\rep 1)_{-1/3}$ representation with the generic PQ charges $(\cX_{D_L},\cX_{D_R})$. In this case, the ALP-VLF derivative couplings take the form
\begin{equation}
  \cL_{\sscript{ALP}}\supset \cX_{D_L}\frac{\partial_\mu a}{v_\Phi}\bar D_L\gamma^\mu D_L+\cX_{D_R}\frac{\partial_\mu a}{v_\Phi}\bar D_R\gamma^\mu D_R\,.
\end{equation}
Performing the field redefinitions in order to remove the mass-mixing terms, the generic interaction Lagrangian reads
\begin{equation}\label{eq:D-KSVZ_lag}
  {\scalebox{0.975}{
      $
      \begin{alignedat}{2}
        \cL_{D}&\supset
        \lzs -\widetilde m_D\bar D_L D_R-H\bar q_L \widehat Y_d d_R-H\bar q_L \lambda_D D_R+\hermc \dzs
        +\frac{\partial_\mu a}{v_\Phi}\Big[ C_{D,L}^{(1)}\bar D_L\gamma_\mu D_L
          +C_{D,R}^{(1)}\bar D_R\gamma_\mu D_R
          \\[2pt]
          &~~~~~+\bar d_R C^{(2)}_{D,R}\gamma_\mu d_R+\big( \bar D_R C_{D,R}^{(3)}\gamma_\mu d_R+\hermc \big)
        \Big]\,,
      \end{alignedat}
      $
  }}
\end{equation}
where the specific ALP interactions are determined by the $\U(1)_{\sscript{PQ}}$ charge assignments. As seen from this Lagrangian, the only direct coupling to SM quarks at this stage is generated by the $C_{D,R}^{(2)}$ coupling, which arises in scenarios where $D_R$ carries a non-vanishing $\U(1)_{\sscript{PQ}}$ charge. Beyond these direct terms, further interactions arise at low energies once the heavy fields are integrated out. These effects can be systematically captured by performing the tree-level matching to both the ALP EFT and the SMEFT, which can be carried out simultaneously through the use of the equations of motion, ensuring that all contributions are consistently included within a non-redundant operator basis.

For concreteness, we exemplify the discussion with the generic $D$-VLF case, taking the interactions in Eq.~\eqref{eq:D-KSVZ_lag} as the starting point. The tree-level integration of the heavy VLF relies on the equation of motion given by
\begin{equation}
  \label{eq:EoM_D_VLF}
  \left[i\slashed{D}+\frac{\slashed{\partial} a}{v_\Phi}\left(C^{(1)}_{D,R}P_R+C^{(1)}_{D,L}P_L\right)- \widetilde{m}_D\right]D = \frac{\slashed{\partial} a}{v_\Phi}C^{(3)}_{D,R} d_R + \lambda_D^\dagger H^\dagger q_L \, .
\end{equation}
The structure of the equations of motion reveals that the ALP contribution effectively enters as an extra term in the covariant derivative. Consequently, this ensures a one-to-one correspondence between the leading SMEFT operators and their ALP counterparts after matching. This correspondence becomes manifest upon solving the equation of motion perturbatively in $1/\widetilde m_D$, applying gamma-matrix identities and the SM fermion equations of motion to simplify derivative terms. Neglecting higher-order contributions and terms proportional to $\square a$, the resulting expression can be substituted back into the starting Lagrangian, yielding the set of effective operators given by
\begin{equation}\label{eq:D_ex_SMEFT_Lagr}
  \begin{alignedat}{2}
    \cL_{\sscript{SMEFT}}&\supset [\cC_{Hq}^{(1)}]_{ij}[\cO_{Hq}^{(1)}]_{ij}+ [\cC_{Hq}^{(3)}]_{ij}[\cO_{Hq}^{(3)}]_{ij} +  \Big\{ [\cC_{dH}]_{ij}[\cO_{dH}]_{ij}+\hermc \Big\} \,,
  \end{alignedat}
\end{equation}
where the Wilson coefficients read
\begin{equation}\label{eq:D_WC}
  [\cC_{Hq}^{(1)}]_{ij}=[\cC_{Hq}^{(3)}]_{ij}=-\frac{1}{4\widetilde m_D^2}[\lambda_D]^*_i[\lambda_D]_j\,,
  \qquad
  [\cC_{dH}]_{ij}=\frac{1}{2\widetilde m_D^2}[\widehat Y_d]^*_{jk}[\lambda_D]^*_i[\lambda_D]_k\,.
\end{equation}
Similarly, the interactions in the ALP sector can be written as
\begin{equation}\label{eq:D_ex_ALP_Lagr}
  \small
  \begin{alignedat}{2}
    \cL_{\sscript{ALP}}&\supset
    [\cC_{aHq}^{(1)}]_{ij}\frac{\partial_\mu a}{v_\Phi}(\bar q_L^i \gamma^\mu q_L^j)(H^\dag H)
    +[\cC_{aHq}^{(3)}]_{ij}\frac{\partial_\mu a}{v_\Phi}(\bar q_L^i\sigma^I \gamma^\mu q_L^j)(H^\dag\sigma^I H)
    +[C^{(2)}_{D,R}]_{ij}\frac{\partial_\mu a}{v_\Phi}\bar d_R^i \gamma^\mu d_R^j
    \\[3pt]&
    +[\cC_{aHd}]_{ij}\frac{\partial_\mu a}{v_\Phi}(H^\dag H)(\bar d_R^i \gamma^\mu d_R^j)
    -\lzs \frac{\partial_\mu a}{v_\Phi}[\cC_{qDd}]_{ij}(\bar q_L^i D_\nu H)(\gamma^\mu \gamma^\nu d_R^j)+\hermc \dzs\,,
  \end{alignedat}
\end{equation}
where
\begin{equation}\small
  \label{eq:WC_alp_EFT}
  \begin{alignedat}{2}
    [\cC_{aHq}^{(1,3)}]_{ij}&=\frac{C_{D,L}^{(1)}}{2\widetilde m_D^2} [\lambda_D]_i [\lambda_D]^*_j\,,
    \quad
    [\cC_{aHd}]_{ij}=\frac{1}{\widetilde m_D^2}[\widehat Y_d^\dag \lambda_D C_{D,R}^{(3)}+\textrm{h.c.} ]_{ij}\,,
    \quad
    [\cC_{qDd}]_{ij}=\frac{i}{\widetilde m_D^2}[\lambda_D]_i[C_{D,R}^{(3)}]_j\,.
  \end{alignedat}
\end{equation}
As seen from Eqs.~\eqref{eq:D_ex_SMEFT_Lagr} and \eqref{eq:D_ex_ALP_Lagr}, the ALP interaction proportional to $C_{D,L}^{(1)}$ exhibits a structure identical to that of the SMEFT coefficients $\cC_{Hq}^{(1,3)}$. The exact relation between the coefficients reads
\begin{equation}
  [\cC_{aHq}^{(1,3)}]_{ij}=-2C_{D,L}^{(1)} \, [\cC_{Hq}^{(1)}]_{ij}\, .
\end{equation}
This correspondence reflects the common origin of the two sets of operators. In the SMEFT, they arise from integrating out heavy fields, and, as shown by Eq.~\eqref{eq:EoM_D_VLF}, the same mechanism governs the ALP case. This contrasts with ALP operators induced purely by field redefinitions.

Analogous mappings can be derived for the remaining terms in the ALP Lagrangian, though for clarity it is more instructive to illustrate the correspondence within specific benchmark setups. As indicated in Table~\ref{tab:interactions_after_redef}, for the $D\sim(\rep3,\rep1)_{-1/3}$ representation, three distinct PQ charge assignments are possible. In the $(\cX_L,\cX_R)=(1,0)$ case, only the term proportional to $C_{D,L}^{(1)}$ remains after the field redefinitions, and the correspondence with the SMEFT operators is immediate, as discussed above. By contrast, when the right-handed component carries a non-trivial PQ charge, additional contributions proportional to $C_{D,R}^{(2)}$ and $C_{D,R}^{(3)}$ arise through mass mixing. In this case, establishing the mapping to the SMEFT requires a more detailed treatment of the induced interactions.

\begin{table}[t]
  \centering
  \renewcommand{\arraystretch}{1.5}
\scalebox{0.678}{
\begin{tabular}{cccc}
\toprule
\textbf{\textbf{VLF}}& $\bm{\cL_{\sscript{SMEFT}}\supset}$
\\
\midrule
\multirow{1}{*}{\vspace{-0.0cm}$D\sim(\bm 3,\bm 1)_{-1/3}$}
&$-\frac{1}{4\widetilde m_D^2}[\lambda_D]^*_i[\lambda_D]_j\lzs[\cO_{Hq}^{(1)}]_{ij}+[\cO_{Hq}^{(3)}]_{ij}\dzs+\lzv\frac{1}{2\widetilde m_D^2}[\widehat Y_d]^*_{jk}[\lambda_D]^*_i[\lambda_D]_k[\cO_{dH}]_{ij}+\hermc\dzv$
\\
\midrule
\multirow{1}{*}{\vspace{-0.0cm}$U\sim(\bm 3,\bm 1)_{2/3}$}
&$\frac{1}{4\widetilde m_U^2}[\lambda_U]^*_i[\lambda_U]_j\lzs[\cO_{Hq}^{(1)}]_{ij}-[\cO_{Hq}^{(3)}]_{ij}\dzs+\lzv\frac{1}{2\widetilde m_U^2}[\widehat Y_u]^*_{jk}[\lambda_U]^*_i[\lambda_U]_k[\cO_{uH}]_{ij}+\hermc\dzv$
\\
\midrule
\multirow{2}{*}{\vspace{-0.3cm}$Q\sim(\bm 3,\bm 2)_{1/6}$}
&$\frac{1}{2\widetilde m_Q^2}[\lambda_Q^d]^*_i[\lambda_Q^d]_j[\cO_{Hd}]_{ij}-\frac{1}{2\widetilde m_Q^2}[\lambda_Q^u]^*_i[\lambda_Q^u]_j[\cO_{Hu}]_{ij}+\Big\{\frac{1}{\widetilde m_Q^2}[\lambda_Q^u]^*_i[\lambda_Q^d]_j[\cO_{Hud}]_{ij}$
\\
&$+\frac{1}{2\widetilde m_Q^2}[\widehat Y_d]^*_{ki}[\lambda_Q^d]_j[\lambda_Q^d]^*_k[\cO_{dH}]_{ij}
+\frac{1}{2\widetilde m_Q^2}[\widehat Y_u]^*_{ki}[\lambda_Q^u]_j[\lambda_Q^u]^*_k[\cO_{uH}]_{ij}+\hermc\Big\}~~~~~~~~~~~~$
\\
\midrule
$Q_5\sim(\bm 3,\bm 2)_{-5/6}$
&$-\frac{1}{2\widetilde m_{Q_5}^2}[\lambda_{Q_5}]^*_i[\lambda_{Q_5}]_j[\cO_{Hd}]_{ij}+\lzv \frac{1}{2\widetilde m_{Q_5}^2}[\widehat Y_d]^*_{ki}[\lambda_{Q_5}]_j[\lambda_{Q_5}]^*_k[\cO_{dH}]_{ij}+\hermc \dzv$
\\
\midrule
$Q_7\sim(\bm 3,\bm 2)_{7/6}$
&$\frac{1}{2\widetilde m_{Q_7}^2}[\lambda_{Q_7}]^*_i[\lambda_{Q_7}]_j[\cO_{Hu}]_{ij}+\lzv \frac{1}{2\widetilde m_{Q_7}^2}[\widehat Y_u]^*_{ki}[\lambda_{Q_7}]_j[\lambda_{Q_7}]^*_k[\cO_{uH}]_{ij}+\hermc \dzv$
\\
\midrule
$T_1\sim(\bm 3,\bm 3)_{-1/3}$
&$\frac{1}{16\widetilde m^2_{T_1}}[\lambda_{T_1}]^*_i[\lambda_{T_1}]_j\lzs [\cO_{Hq}^{(3)}]_{ij}-3[\cO_{Hq}^{(1)}]_{ij} \dzs+\lzv \frac{1}{8\widetilde m_{T_1}^2}[\widehat Y_d]^*_{jk}[\lambda_{T_1}]_k[\lambda_{T_1}]^*_i [\cO_{dH}]_{ij}+\frac{1}{4\widetilde m^2_{T_1}}[\widehat Y_u]^*_{jk}[\lambda_{T_1}]_k [\lambda_{T_1}]_i^*[\cO_{uH}]_{ij}+\hermc \dzv$
\\
\midrule
$T_2\sim(\bm 3,\bm 3)_{2/3}$
&$\frac{1}{16\widetilde m^2_{T_2}}[\lambda_{T_2}]^*_i[\lambda_{T_2}]_j\lzs [\cO_{Hq}^{(3)}]_{ij}+3[\cO_{Hq}^{(1)}]_{ij} \dzs+\lzv \frac{1}{4\widetilde m_{T_2}^2}[\widehat Y_d]^*_{jk}[\lambda_{T_2}]_k[\lambda_{T_2}]^*_i [\cO_{dH}]_{ij}+\frac{1}{8\widetilde m^2_{T_2}}[\widehat Y_u]^*_{jk}[\lambda_{T_2}]_k [\lambda_{T_2}]_i^*[\cO_{uH}]_{ij}+\hermc \dzv$
\\
\bottomrule
\end{tabular}
}
  \caption{Tree-level dimension-six SMEFT operators generated by integrating out heavy VLFs for each gauge representation considered. The operator structures arise from the UV interactions in Table~\ref{tab:interactions_after_redef} and are organized in terms of the SMEFT Warsaw basis. The absence of $\U(1)_{\sscript{PQ}}$ charge assignments in this table reflects the fact that SMEFT operators depend only on the gauge quantum numbers of the VLFs and not on their PQ charges, which do not enter the couplings between VLFs and SM fermions. The explicit mapping to UV parameters is given in Appendix~\ref{app:det_mat_proc_app}.
  }
  \label{tab:VLFs_SMEFT_tree}
\end{table}

Tables~\ref{tab:SMEFTmap} and~\ref{tab:ALPmap} provide the explicit mapping of the UV parameters to effective operators in the SMEFT and ALP sectors, obtained after the field redefinitions. As an example, for the $(\cX_L,\cX_R)=(0,-1)$ assignment, the $C_{D,R}^{(3)}$ coefficient can be written as
\begin{equation}
  C_{D,R}^{(3)}=\frac{1}{\widetilde m_D}M_d\cR_d
  =\lambda_D^\dag (Y_d^\dag)^{\eminus1}\cR_d
  =\lambda_D^\dag (Y_d^\dag)^{\eminus1}\left[1+\frac{2M_d^\dagger M_d}{y_D^2 v_\Phi^2}\right]^{\eminus 1/2}\,.
\end{equation}
Furthermore, this relation allows us to express the $\cC_{aHd}$ as
\begin{equation}
  {\scalebox{0.96}{
      $
      \begin{alignedat}{2}
        [\cC_{aHd}]_{ij}
        =\frac{1}{\widetilde m_D^2}[\cR_d^\dag Y_d^{\dag}\lambda_D\lambda_D^\dag (Y_d^\dag)^{\eminus1}]_{ij}+\cO(\widetilde m_D^{\eminus4})+\text{h.c.}=4 \left[Y_d^{\dag} \cC_{Hq}^{(1)} (Y_d^\dagger)^{-1}\right]_{ij}+\cO( \widetilde m_D^{\eminus4}) +\text{h.c.}\,.
      \end{alignedat}$
  }}
\end{equation}
Analogously, $C_{D,R}^{(2)}$ and $\cC_{qDd}$ can be expressed as
\begin{equation}
  \begin{alignedat}{2}
    {\scalebox{0.99}{
        $  C_{D,R}^{(2)}=-\lzm 1+\frac{2M_dM_d^\dag}{y_D^2v_\Phi^2} \dzm \cR_d^\dag (Y_d)^{\eminus1}\lambda_D \lambda_D^\dag (Y_d^{\eminus1})^\dag \cR_d\,,
        \quad
        [\cC_{qDd}]_{ij}=\frac{i}{\widetilde m_D^2}[\lambda_D]_i [\lambda_D^\dag (Y_d^\dag)^{\eminus1}\cR_d]_j\,.
        $
      }
    }
  \end{alignedat}
\end{equation}
Expanding in the limit $y_D v_\Phi \gg M_d$, the leading contribution matches the SMEFT expressions as
\begin{equation}
  C_{D,R}^{(2)}= 4\, m_D^2 (Y_d)^{-1}\cC_{Hq}^{(1)} (Y_d^{\dagger})^{-1}+\cO( m_D^{\eminus4}),
  \quad
  \cC_{qDd}= 4i\, \cC_{Hq}^{(1)}(Y_{d}^\dagger)^{-1}+\cO( m_D^{\eminus4})\,,
\end{equation}
thereby demonstrating the intrinsic link between SMEFT and ALP operators.

While the preceding discussion focused on the $D$-type VLF, the $U$-type case follows directly upon replacing down-type with up-type indices. For $Q$-type VLFs, the analysis proceeds analogously after exchanging left- and right-handed components, corresponding in the SMEFT to $\cO_{Hq}\to \cO_{Hu},\, \cO_{Hd}$, with the ALP correspondence unchanged apart from this chirality exchange. All other representations contribute solely through direct matching, leading to an exact one-to-one mapping to effective operators, as exemplified by the $D_L$ case with $\cX_L=1$.

\begin{table}[t]
  \centering
  \centering
\renewcommand{\arraystretch}{1.2}
\scalebox{0.72}{
\begin{tabular}{cccc}
\toprule
\textbf{\textbf{VLF}}&$\bm{\cX_{L}}$ &$\bm{\cX_R}$& $\bm{\cL_{\sscript{ALP}}\supset}$
\\
\midrule
\multirow{6}{*}{\vspace{-0.65cm}$D\sim(\bm 3,\bm 1)_{-1/3}$}
&1
&0
&$\frac{C_{D,L}^{(1)}}{2\widetilde m_D^2} [\lambda_D]_i [\lambda_D]^*_j\frac{\partial_\mu a}{v_\Phi}\lzs (\bar q_L^i \gamma_\mu q_L^j)(H^\dag H)+(\bar q_L^i\sigma^I \gamma_\mu q_L^j)(H^\dag\sigma^I H) \dzs$
\vspace{+0.1cm}\\
\noalign{\vskip 2pt}
\cdashline{2-4}[.4pt/2pt]
\noalign{\vskip 2pt}
\vspace{+0.1cm}
&\multirow{2}{*}{\vspace{-0.5cm}0}
&\multirow{2}{*}{\vspace{-0.5cm}-1}
&$[C^{(2)}_{D,R}]_{ij}\frac{\partial_\mu a}{v_\Phi}\bar d_R^i \gamma_\mu d_R^j
+\frac{1}{\widetilde m_D^2}[\widehat Y_d^\dag \lambda_D C_{D,R}^{(3)}]_{ij}\frac{\partial_\mu a}{v_\Phi}(\bar d_R^i \gamma^\mu d_R^j)(H^\dag H)$
\\[2pt]
&&&$-\lzs \frac{\partial_\mu a}{v_\Phi}\frac{i}{\widetilde m_D^2}[\lambda_D]_i[C_{D,R}^{(3)}]_j(\bar q_L^i D_\nu H)(\gamma^\mu\gamma^\nu d_R^j)+\hermc \dzs$
\vspace{+0.1cm}\\
\noalign{\vskip 2pt}
\cdashline{2-4}[.4pt/2pt]
\noalign{\vskip 2pt}
&\multirow{2}{*}{\vspace{-0.5cm}-1}
&\multirow{2}{*}{\vspace{-0.5cm}-2}
&$\frac{C_{D,L}^{(1)}}{2\widetilde m_D^2} [\lambda_D]_i [\lambda_D]^*_j\frac{\partial_\mu a}{v_\Phi}\lzs (\bar q_L^i \gamma_\mu q_L^j)(H^\dag H)+(\bar q_L^i\sigma^I \gamma_\mu q_L^j)(H^\dag\sigma^I H) \dzs+[C^{(2)}_{D,R}]_{ij}\frac{\partial_\mu a}{v_\Phi}\bar d_R^i \gamma_\mu d_R^j$
\\[5pt]
&&&$+\frac{1}{\widetilde m_D^2}[\widehat Y_d^\dag \lambda_D C_{D,R}^{(3)}]_{ij}\frac{\partial_\mu a}{v_\Phi}(\bar d_R^i \gamma^\mu d_R^j)(H^\dag H)
-\lzs \frac{\partial_\mu a}{v_\Phi}\frac{i}{\widetilde m_D^2}[\lambda_D]_i[C_{D,R}^{(3)}]_j(\bar q_L^i D_\nu H)(\gamma^\mu\gamma^\nu d_R^j)+\hermc \dzs$
\\[5pt]
\midrule
\multirow{6}{*}{\vspace{-0.55cm}$U\sim(\bm 3,\bm 1)_{2/3}$}
&1
&0
&$\frac{C_{U,L}^{(1)}}{2\widetilde m_U^2} [\lambda_U]_i [\lambda_U]^*_j\frac{\partial_\mu a}{v_\Phi}\lzs (\bar q_L^i \gamma_\mu q_L^j)(H^\dag H)-(\bar q_L^i\sigma^I \gamma_\mu q_L^j)(H^\dag\sigma^I H) \dzs$
\vspace{+0.1cm}\\
\noalign{\vskip 2pt}
\cdashline{2-4}[.4pt/2pt]
\noalign{\vskip 2pt}
\vspace{+0.1cm}
&\multirow{2}{*}{\vspace{-0.3cm}0}
&\multirow{2}{*}{\vspace{-0.3cm}-1}
&$[C^{(2)}_{U,R}]_{ij}\frac{\partial_\mu a}{v_\Phi}\bar u_R^i \gamma_\mu u_R^j
+\frac{1}{\widetilde m_U^2}[\widehat Y_u^\dag \lambda_U C_{U,R}^{(3)}]_{ij}\frac{\partial_\mu a}{v_\Phi}(\bar u_R^i \gamma^\mu u_R^j)(H^\dag H)$
\\
&&&$-\lzs \frac{\partial_\mu a}{v_\Phi}\frac{i}{\widetilde m_U^2}[\lambda_U]_i[C_{U,R}^{(3)}]_j(\bar q_L^i D_\nu H)(\gamma^\mu\gamma^\nu u_R^j)+\hermc \dzs$
\vspace{+0.1cm}\\
\noalign{\vskip 2pt}
\cdashline{2-4}[.4pt/2pt]
\noalign{\vskip 2pt}
&\multirow{2}{*}{\vspace{-0.3cm}-1}
&\multirow{2}{*}{\vspace{-0.3cm}-2}
&$\frac{C_{U,L}^{(1)}}{2\widetilde m_U^2} [\lambda_U]_i [\lambda_U]^*_j\frac{\partial_\mu a}{v_\Phi}\lzs (\bar q_L^i \gamma_\mu q_L^j)(H^\dag H)-(\bar q_L^i\sigma^I \gamma_\mu q_L^j)(H^\dag\sigma^I H) \dzs+[C^{(2)}_{U,R}]_{ij}\frac{\partial_\mu a}{v_\Phi}\bar u_R^i \gamma_\mu u_R^j$
\\[5pt]
&&&$+\frac{1}{\widetilde m_U^2}[\widehat Y_u^\dag \lambda_U C_{U,R}^{(3)}]_{ij}\frac{\partial_\mu a}{v_\Phi}(\bar u_R^i \gamma^\mu u_R^j)(H^\dag H)
-\lzs \frac{\partial_\mu a}{v_\Phi}\frac{i}{\widetilde m_U^2}[\lambda_U]_i[C_{U,R}^{(3)}]_j(\bar q_L^i D_\nu H)(\gamma^\mu\gamma^\nu u_R^j)+\hermc \dzs$
\\[5pt]
\midrule
\multirow{10}{*}{\vspace{-0.7cm}$Q\sim(\bm 3,\bm 2)_{1/6}$}
&\multirow{3}{*}{\vspace{-0.4cm}1}
&\multirow{3}{*}{\vspace{-0.4cm}0}
&$[C_{Q,L}^{(2)}]_{ij}\frac{\partial_\mu a}{v_\Phi}\bar q_L^i\gamma_\mu q_L^j
-\frac{1}{2\widetilde m_Q^2}[\widehat Y_d \lambda_Q^{d\dag}C_{Q,L}^{(3)}]_{ij}\frac{\partial_\mu a}{v_\Phi}\lzs (\bar q_L^i \gamma_\mu q_L^j)(H^\dag H)+(\bar q_L^i\sigma^I \gamma_\mu q_L^j)(H^\dag\sigma^I H) \dzs$
\\[5pt]
&&&$-\frac{1}{2\widetilde m_Q^2}[\widehat Y_u \lambda_Q^{u\dag}C_{Q,L}^{(3)}]_{ij}\frac{\partial_\mu a}{v_\Phi}\lzs (\bar q_L^i \gamma_\mu q_L^j)( H^\dag  H){\color{black}-}(\bar q_L^i\sigma^I \gamma_\mu q_L^j)( H^\dag\sigma^I  H) \dzs$
\\[5pt]
&&&$-\lzs 
\frac{\partial_\mu a}{v_\Phi}\frac{i}{2\widetilde m_Q^2}[\lambda_Q^u]_i[C_{Q,L}^{(3)}]_j(\bar q_L^i D_\nu \widetilde H)(\gamma^\mu\gamma^\nu u_R^j)
+ 
\frac{\partial_\mu a}{v_\Phi}\frac{i}{2\widetilde m_Q^2}[\lambda_Q^d]_i[C_{Q,L}^{(3)}]_j(\bar q_L^i D_\nu H)(\gamma^\mu\gamma^\nu d_R^j)+\hermc\dzs$
\\[5pt]
\noalign{\vskip 2pt}
\cdashline{2-4}[.4pt/2pt]
\noalign{\vskip 2pt}
&\multirow{1}{*}{\vspace{-0.0cm}0}
&\multirow{1}{*}{\vspace{-0.0cm}-1}
&$\frac{C_{Q,R}^{(1)}}{\widetilde m_Q^2}[\lambda_Q^d]^*_i[\lambda_Q^d]_j\frac{\partial_\mu a}{v_\Phi}(\bar d_R^i\gamma_\mu d_R^j)(H^\dag H)+\frac{C_{Q,R}^{(1)}}{\widetilde m_Q^2}[\lambda_Q^u]^*_i[\lambda_Q^u]_j\frac{\partial_\mu a}{v_\Phi}(\bar u_R^i\gamma_\mu u_R^j)( H^\dag  H)$
\\
\noalign{\vskip 2pt}
\cdashline{2-4}[.4pt/2pt]
\noalign{\vskip 2pt}
&\multirow{4}{*}{\vspace{-0.5cm}2}
&\multirow{4}{*}{\vspace{-0.5cm}1}
&$[C_{Q,L}^{(2)}]_{ij}\frac{\partial_\mu a}{v_\Phi}\bar q_L^i\gamma_\mu q_L^j
-\frac{1}{2\widetilde m_Q^2}[\widehat Y_d \lambda_Q^{d\dag}C_{Q,L}^{(3)}]_{ij}\frac{\partial_\mu a}{v_\Phi}\lzs (\bar q_L^i \gamma_\mu q_L^j)(H^\dag H)+(\bar q_L^i\sigma^I \gamma_\mu q_L^j)(H^\dag\sigma^I H) \dzs$
\\[5pt]
&&&$-\frac{1}{2\widetilde m_Q^2}[\widehat Y_u \lambda_Q^{u\dag}C_{Q,L}^{(3)}]_{ij}\frac{\partial_\mu a}{v_\Phi}\lzs (\bar q_L^i \gamma_\mu q_L^j)( H^\dag  H)-(\bar q_L^i\sigma^I \gamma_\mu q_L^j)( H^\dag\sigma^I  H) \dzs$
\\[5pt]
&&&$-\lzs 
\frac{\partial_\mu a}{v_\Phi}\frac{i}{2\widetilde m_Q^2}[\lambda_Q^u]_i[C_{Q,L}^{(3)}]_j(\bar q_L^i D_\nu \widetilde H)(\gamma^\mu\gamma^\nu u_R^j)
+ 
\frac{\partial_\mu a}{v_\Phi}\frac{i}{2\widetilde m_Q^2}[\lambda_Q^d]_i[C_{Q,L}^{(3)}]_j(\bar q_L^i D_\nu H)(\gamma^\mu\gamma^\nu d_R^j)+\hermc\dzs$
\\
&&&$+\frac{C_{Q,R}^{(1)}}{\widetilde m_Q^2}[\lambda_Q^d]^*_i[\lambda_Q^d]_j\frac{\partial_\mu a}{v_\Phi}(\bar d_R^i\gamma_\mu d_R^j)(H^\dag H)+\frac{C_{Q,R}^{(1)}}{\widetilde m_Q^2}[\lambda_Q^u]^*_i[\lambda_Q^u]_j\frac{\partial_\mu a}{v_\Phi}(\bar u_R^i\gamma_\mu u_R^j)( H^\dag  H)$
\\
\midrule
$Q_5\sim(\bm 3,\bm 2)_{-5/6}$
&0&-1
&$\frac{C_{Q_5,R}^{(1)}}{\widetilde m_{Q_5}^2}[\lambda_{Q_5}]^*_i[\lambda_{Q_5}]_j\frac{\partial_\mu a}{v_\Phi}(\bar d_R^i \gamma_\mu d_R^j)(H^\dag H)$
\\
\midrule
$Q_7\sim(\bm 3,\bm 2)_{7/6}$
&0&-1
&$\frac{C_{Q_7,R}^{(1)}}{\widetilde m_{Q_7}^2}[\lambda_{Q_7}]^*_i[\lambda_{Q_7}]_j\frac{\partial_\mu a}{v_\Phi}(\bar u_R^i \gamma_\mu u_R^j)(H^\dag H)$
\\
\midrule
$T_1\sim(\bm 3,\bm 3)_{-1/3}$
&1&0
&$\frac{C_{T_1,L}^{(1)}}{8\widetilde m_{T_1}^2}[\lambda_{T_1}]^*_i[\lambda_{T_1}]_j\frac{\partial_\mu a}{v_\Phi}\lzs3\,(\bar q_L^i\gamma_\mu q_L^j)(H^\dag H)-(\bar q_L^i\sigma^I \gamma_\mu q_L^j)(H^\dag\sigma^I H)\dzs$
\\
\midrule
$T_2\sim(\bm 3,\bm 3)_{2/3}$
&1&0
&$\frac{C_{T_2,L}^{(1)}}{8\widetilde m_{T_2}^2}[\lambda_{T_2}]^*_i[\lambda_{T_2}]_j\frac{\partial_\mu a}{v_\Phi}\lzs3\,(\bar q_L^i\gamma_\mu q_L^j)(H^\dag H)+(\bar q_L^i\sigma^I \gamma_\mu q_L^j)(H^\dag\sigma^I H)\dzs$
\\
\bottomrule
\end{tabular}
}
  \caption{Effective ALP-SM interactions obtained after integrating out different VLF representations at tree level. The table collects the leading contributions for different VLF representations, specified together with their possible PQ charge assignments $(\cX_L,\cX_R)$. The explicit mapping to UV parameters is given in Appendix~\ref{app:det_mat_proc_app}.}
  \label{tab:ALP_EFT_Lagr}
\end{table}

For clarity, the EFT matching results are presented in two separate tables. Table~\ref{tab:VLFs_SMEFT_tree} collects the leading tree-level contributions to SMEFT operators, while Table~\ref{tab:ALP_EFT_Lagr} summarizes the corresponding ALP-SM interactions. Each entry shows the dominant matching contribution of a single VLF representation, under the assumption that only a single species is active at a time. The Wilson coefficients are expressed in terms of the effective parameters introduced in Table~\ref{tab:interactions_after_redef}, ensuring straightforward use in the subsequent phenomenological analysis. Crucially, both SMEFT and ALP interactions depend on a common set of UV couplings and mass parameters, which leads to correlated predictions and allows for a unified treatment in the combined SMEFT--ALP analysis. The explicit mapping of these coefficients to the underlying UV parameters is provided in Appendix~\ref{app:det_mat_proc_app}.

The framework developed in this section, namely the systematic matching of KSVZ-like VLF completions onto both the SMEFT and the ALP EFT, constitutes one of the main results of this work. Although the fact that KSVZ models lead to correlated low-energy ALP couplings has been discussed previously~\cite{Alonso-Alvarez:2023wig}, the gauge-invariant matching performed here provides a more complete starting point for phenomenological analyses. In particular, the leading fermionic ALP interactions generated by direct VLF exchange arise, above the electroweak scale, from dimension-seven ALP-SMEFT operators. This differs from the commonly used dimension-five ALP EFT treatment~\cite{Bauer:2020jbp,Chala:2020wvs,DasBakshi:2023lca}, and affects the running from the VLF scale $m_\Psi$ down to the electroweak scale. Moreover, the simultaneous SMEFT and ALP matching provides a consistent framework for organizing the leading phenomenological effects. As we will see later, these may arise either from modified SM currents, from direct ALP interactions, or from one-loop RGE effects, and the present formalism makes their common UV origin explicit, which may be missed by not using this framework.

\section{Overview of SMEFT Constraints}
\label{sec:SMEFT_pheno}
With the EFT framework established in the previous section, we now turn to its phenomenological implications within the SMEFT. Integrating out the heavy VLFs induces a definite set of dimension-six operators, summarized in Table~\ref{tab:VLFs_SMEFT_tree}. Their phenomenological impact can be quantified through global SMEFT analyses of precision observables. Importantly, once expressed in terms of SMEFT Wilson coefficients, the resulting constraints are independent of the details of the underlying UV realization and depend only on the operator structure associated with each VLF representation. As such, they can be directly applied to a broad class of models featuring VLFs~\cite{Smolkovic:2019jow,Cornella:2023zme,Greljo:2024evt,Loisa:2024xuk,Cornella:2024jaw,Greljo:2025mwj,Arkani-Hamed:2026wwy,Calibbi:2025rxn,Greljo:2024zrj,Asadi:2023ucx,Fedele:2020fvh,Calibbi:2020jvd,Bonnefoy:2019lsn,Giarnetti:2025idu,Greljo:2023bix,Mohanta:2023soi,Mohanta:2022seo}.\footnote{In order to achieve this, the relevant couplings entering the SMEFT matching must be correctly identified. In the present work, VLFs are assumed to be flavor singlets, such that the couplings are described by vectors $\lambda \sim \mathbb{C}^{3\times 1}$. In more general scenarios where VLFs carry a non-trivial flavor multiplicity $N$, these couplings are promoted to matrices $\lambda \sim \mathbb{C}^{3\times N}$, with the mapping onto SMEFT Wilson coefficients proceeding analogously~\cite{Greljo:2023adz,Greljo:2023bdy,Palavric:2024gvu,Moreno-Sanchez:2025bzz,Kosnik:2025srw,Beneito:2025fzf}.}

In this section, we derive bounds on the corresponding SMEFT Wilson coefficients across the different VLF scenarios, performing both a global fit and dedicated analyses of selected subsets of observables in order to disentangle the phenomenological impact of each case. We first outline the methodology of our analysis before presenting and discussing the results.

\subsection{Methodology}
\label{sec:SMEFT_meth}

The tree-level SMEFT Wilson coefficients generated by each VLF representation are constrained using global fits to electroweak precision, Higgs, and flavor observables. For this purpose, the analysis is carried out using the \texttt{smelli}~\cite{Aebischer:2018iyb} and \texttt{flavio}~\cite{Straub:2018kue} frameworks, which provide a consistent implementation of the SMEFT likelihood and allow for the extraction of confidence intervals from the corresponding $\chi^2$ function.

For a given VLF representation, the tree-level matching relations correlate all induced SMEFT operators through an underlying parameter $\mathscr{C}^F_{ij}$ defined as
\begin{equation}\label{eq:cfij}
  \mathscr{C}^F_{ij}\equiv \frac{[\lambda_F]_i^* [\lambda_F]_j}{\widetilde m_F^2}\,,
\end{equation}
where $[\lambda_F]_i$ denotes the coupling of the $F$-type VLF to SM fermions (see Table~\ref{tab:interactions_after_redef}) and $\widetilde m_F$ is the VLF mass. As a result, all generated Wilson coefficients can be expressed in terms of $\mathscr{C}^F_{ij}$, up to representation-dependent numerical prefactors. For example, as highlighted in Table~\ref{tab:VLFs_SMEFT_tree}, for the $D$-type VLF, one finds
\begin{equation}\label{eq:D_case_CHQ_CDH_coeff}
  [\cC_{Hq}^{(1)}]_{ij}=[\cC_{Hq}^{(3)}]_{ij}=-\frac{\mathscr{C}^D_{ij}}{4}\,,
  \qquad
  [\cC_{dH}]_{ij}=\frac{1}{2}[\widehat Y_d]^*_{jk}\mathscr{C}^D_{ik}
  =\frac{1}{2}[\mathscr{C}^D \widehat Y_d^\dag]_{ij}\,.
\end{equation}
This structure reduces the fit to a one-dimensional problem for each flavor entry and each VLF scenario. We note that in order to extend our results to scenarios with multiple flavors of the same VLF, the single mass parameter $\widetilde m_F$ is promoted to a mass matrix $\widetilde{\mathcal M}_F$, and Eq.~\eqref{eq:cfij} generalizes to
\begin{equation}
  \mathscr{C}^F_{ij}\equiv [\lambda_F^\dag \widetilde{\mathcal M}_F^{\eminus2} \lambda_F]_{ij}\,.
\end{equation}
Therefore our bounds on $\mathscr{C}^F_{ij}$ can then be translated to bounds on the UV parameters even in the flavored VLF case.

The global $\chi^2$ is therefore evaluated as a function of $\mathscr{C}^F_{ij}$, retaining only the Wilson coefficients generated at tree level for the representation under consideration. For each VLF scenario, the analysis is performed separately for every flavor combination $(ij)$ by activating a single parameter $\mathscr{C}^F_{ij}$ at a time, while setting all other flavor entries to zero. In this way, each fit probes one definite direction in the SMEFT parameter space associated with a specific flavor structure. The best-fit point is obtained by minimizing the $\chi^2$ along this direction, and confidence intervals at $1\sigma$, $2\sigma$, and $3\sigma$ are derived from $\Delta \chi^2 \equiv \chi^2 - \chi^2_{\mathrm{min}}$ using the standard one-parameter thresholds. Some observables exhibit a quadratic dependence on the Wilson coefficient, which means the resulting $\chi^2$ function gets a quartic dependence, causing it to develop multiple local minima. In such cases, the confidence-level threshold may be crossed in two disconnected regions, leading to separate allowed intervals centered around distinct minima.

Before presenting the results, we comment on two aspects of the operator content entering the fits:
\begin{itemize}
  \item For the $Q$-type scenarios, we treat the $Q_u$ and $Q_d$ configurations separately by activating either $\lambda_Q^u$ or $\lambda_Q^d$. As a consequence, the operator $\cO_{Hud}$, which requires the simultaneous presence of both couplings, is not included in the fit. In addition, since $\cO_{Hud}$ predominantly contributes to right-handed charged-current interactions, which are currently subject to comparatively weak experimental constraints, its omission is not expected to qualitatively affect the resulting bounds.
  \item For the flavor structures $(ij)=(13)$ and $(23)$, the operator $\cO_{uH}$ is omitted from the fit. In these cases, the resulting likelihood does not exhibit a sufficiently smooth behavior along the scanned direction, preventing a stable extraction of reliable confidence intervals within the one-parameter analysis adopted in this work. We therefore restrict the fit to the remaining tree-level operators for these flavor configurations. From a phenomenological perspective, the inclusion of $[\cO_{uH}]_{13,23}$ would primarily induce flavor-violating top-Higgs interactions after rotating to the mass basis, leading to processes such as $t\to h (u,c)$ and off-shell Higgs decays $h\to t^* (u,c)$, which do not provide constraints competitive with those already captured by the remaining operators included in the analysis~\cite{Harnik:2012pb}. By contrast, for $(ij)=(12)$ the contribution from $\cO_{uH}$ is suppressed by the small up-type Yukawa couplings, leading to a smooth likelihood profile and allowing for a consistent implementation within the standard one-parameter $\Delta\chi^2$ analysis.
\end{itemize}

\subsection{Results and Discussion}
\label{ref:SMEFT_res_dis}
\begin{table}[t]
  \centering
\makebox[\textwidth][c]{%
\begin{minipage}[t]{0.42\textwidth}
\centering
\scalebox{0.78}{
\begin{tabular}{ll|ccc}
\hline
\textbf{VLF} &\textbf{CL} & \textbf{11} & \textbf{22} & \textbf{33} 
\\
\hline
$D$ & $1\sigma$ & $[17.8, 56.1]$ & $[0, 8.54]$ & $[0, 5.21]$ \\
$(\times10^{\eminus3})$ & $2\sigma$ & $[0, 75.2]$ & $[0, 31.6]$ & $[0, 18.6]$ \\
 & $3\sigma$ & $[0, 94.4]$ & $[0, 64.1]$ & $[0, 36.6]$ \\
\hline
$U$ & $1\sigma$ & $[2.10, 6.01]$ & $[0, 1.10]$ & $[0, 5.39]$ \\
$(\times10^{\eminus2})$ & $2\sigma$ & $[0.14, 7.96]$ & $[0, 4.04]$ & $[0, 9.16]$ \\
 & $3\sigma$ & $[0, 9.91]$ & $[0, 8.13]$ & $[0, 12.9]$ \\
\hline
$Q_u$ & $1\sigma$ & $[0, 3.37]$ & $[0, 3.36]$ & $[0.98, 4.94]$ \\
$(\times10^{\eminus2})$ & $2\sigma$ & $[0, 11.6]$ & $[0, 11.5]$ & $[0, 6.91]$ \\
 & $3\sigma$ & $[0, 22.3]$ & $[0, 22.0]$ & $[0, 8.89]$ \\
\hline
$Q_d$ & $1\sigma$ & $[0, 4.51]$ & $[0, 12.1]$ & $[0, 4.29]$ \\
$(\times10^{\eminus2})$ & $2\sigma$ & $[0, 16.7]$ & $[0, 35.5]$ & $[0, 15.7]$ \\
 & $3\sigma$ & $[0, 34.2]$ & $[0, 61.9]$ & $[0, 31.5]$ \\
\hline
\end{tabular}
}
\end{minipage}
\hfill
\hfill
\begin{minipage}[t]{0.55\textwidth}
\centering
\scalebox{0.78}{
\begin{tabular}{ll|ccc}
\hline
\textbf{VLF} &\textbf{CL} & \textbf{11} & \textbf{22} & \textbf{33} \\
\hline
$Q_5$ & $1\sigma$ & $[4.30, 9.07]$ & $[0, 4.54]$ & $[4.04, 8.02]$ \\
$(\times10^{\eminus1})$ & $2\sigma$ & $[1.83, 11.4]$ & $[0, 6.92]$ & $[1.98, 9.94]$ \\
 & $3\sigma$ & $[0, 13.7]$ & $[0, 9.24]$ & $[0, 11.8]$ \\
\hline
$Q_7$ & $1\sigma$ & $[1.07, 3.66]$ & $[0.99, 3.53]$ & $[0, 0.06]$ \\
$(\times10^{\eminus1})$ & $2\sigma$ & $[0, 4.95]$ & $[0, 4.79]$ & $[0, 0.20]$ \\
 & $3\sigma$ & $[0, 6.23]$ & $[0, 6.04]$ & $[0, 0.36]$ \\
\hline
$T_1$ & $1\sigma$ & $[0, 1.65]$ & $[27.1, 79.9]$ & $[0, 1.51]$ \\
$(\times10^{\eminus2})$ & $2\sigma$ & $[0, 5.93]$ & $[0.43, 106]$ & $[0, 4.89]$ \\
 & $3\sigma$ & $[0, 11.7]$ & $[0, 132]$ & $[0, 8.92]$ \\
\hline
$T_2$ & $1\sigma$ & $[0, 0.20]$ & $[2.99, 6.19]$ & $[0.36, 1.14]$ \\
$(\times10^{\eminus1})$ & $2\sigma$ & $[0, 0.69]$ & $[1.38, 7.78]$ & $[0, 1.52]$ \\
 & $3\sigma$ & $[0, 1.31]$ & $[0, 9.37]$ & $[0, 1.91]$ \\
\hline
\end{tabular}
}
\end{minipage}%
}
  \caption{$1\sigma$, $2\sigma$, and $3\sigma$ confidence intervals for the $\mathscr{C}^F_{ii}$ parameters in units of $\tev^{\eminus2}$ derived from the global SMEFT analysis. For each VLF representation, the bounds are given separately for the indicated flavor combinations, with the overall normalization factor shown in parentheses.}
  \label{tab:SMEFT_intervals_sbs_diag}
\end{table}
\begin{table}[t]
  \centering
\scalebox{0.85}{
\begin{tabular}{ll|cc@{\hspace{1.0cm}}c}
\hline
\textbf{VLF} &\textbf{CL} & \textbf{12} & \textbf{13} & \textbf{23} \\
\hline
$D$ & $1\sigma$ & $[-0.09, 0.02]$ & $[-5.34, -4.18]$ $\cup$ $[-1.58, -0.36]$ & $[1.54, 2.41]$ \\
$(\times10^{\eminus3})$ & $2\sigma$ & $[-0.13, 0.11]$ & $[-5.79, 0.08]$ & $[1.12, 2.86]$ \\
 & $3\sigma$ & $[-0.17, 0.47]$ & $[-6.17, 0.45]$ & $[0.70, 3.32]$ \\
\hline
$U$ & $1\sigma$ & $[-0.11, -0.07]$ & $[-1.52, -1.24]$ & $[1.03, 1.79]$ \\
$(\times10^{\eminus2})$ & $2\sigma$ & $[-0.13, -0.05]$ & $[-1.52, -0.75]$ & $[0.66, 2.19]$ \\
 & $3\sigma$ & $[-0.15, -0.03]$ & $[-1.52, -0.28]$ & $[0.29, 2.58]$ \\
\hline
$Q_u$ & $1\sigma$ & $[-0.03, 0.02]$ & $[-1.67, -0.94]$ $\cup$ $[1.14, 1.47]$ & $[1.39, 1.86]$ \\
$(\times10^{1})$ & $2\sigma$ & $[-0.05, 0.04]$ & $[-1.94, 1.86]$ & $[1.12, 2.07]$ \\
 & $3\sigma$ & $[-0.07, 0.07]$ & $[-2.17, 2.13]$ & $[0.82, 2.26]$ \\
\hline
$Q_d$ & $1\sigma$ & $[0.11, 0.17]$ & $[-2.79, 5.55]$ & $[-6.57, -0.08]$ \\
$(\times10^{\eminus4})$ & $2\sigma$ & $[0.07, 0.20]$ & $[-7.11, 9.52]$ & $[-9.86, 3.23]$ \\
 & $3\sigma$ & $[0.04, 0.23]$ & $[-11.5, 13.3]$ & $[-13.1, 6.52]$ \\
\hline
$Q_5$ & $1\sigma$ & $[-0.17, -0.11]$ & $[-5.74, 2.68]$ & $[0.04, 6.55]$ \\
$(\times10^{\eminus4})$ & $2\sigma$ & $[-0.20, -0.07]$ & $[-9.75, 7.04]$ & $[-3.27, 9.86]$ \\
 & $3\sigma$ & $[-0.23, -0.04]$ & $[-13.6, 11.4]$ & $[-6.57, 13.1]$ \\
\hline
$Q_7$ & $1\sigma$ & $[-0.02, 0.03]$ & $[-1.46, -1.15]$ $\cup$ $[0.94, 1.67]$ & $[-1.86, -1.39]$ \\
$(\times10^{1})$ & $2\sigma$ & $[-0.04, 0.05]$ & $[-1.86, 1.94]$ & $[-2.07, -1.12]$ \\
 & $3\sigma$ & $[-0.06, 0.08]$ & $[-2.13, 2.17]$ & $[-2.26, -0.82]$ \\
\hline
$T_1$ & $1\sigma$ & $[1.47, 1.99]$ & $[-4.24, -0.24]$ & $[7.79, 12.1]$ \\
$(\times10^{\eminus3})$ & $2\sigma$ & $[1.09, 2.18]$ & $[-6.75, 1.45]$ & $[5.70, 14.3]$ \\
 & $3\sigma$ & $[0.48, 2.36]$ & $[-9.88, 3.00]$ & $[3.65, 16.5]$ \\
\hline
$T_2$ & $1\sigma$ & $[-0.92, -0.68]$ & $[0.55, 2.86]$ & $[-5.14, -3.29]$ \\
$(\times10^{\eminus3})$ & $2\sigma$ & $[-1.01, -0.41]$ $\cup$ $[-0.23, 0.16]$ & $[-0.31, 5.10]$ $\cup$ $[8.44, 11.0]$ & $[-6.09, -2.40]$ \\
 & $3\sigma$ & $[-1.08, 0.28]$ & $[-1.07, 12.4]$ & $[-7.06, -1.52]$ \\
\hline
\end{tabular}
}
  \caption{$1\sigma$, $2\sigma$, and $3\sigma$ confidence intervals for the $\mathscr{C}^F_{ij}$ parameters in units of $\tev^{\eminus2}$ derived from the global SMEFT analysis. For each VLF representation, the bounds are given separately for the indicated flavor combinations, with the overall normalization factor shown in parentheses.}
  \label{tab:SMEFT_intervals_sbs_offdiag}
\end{table}
The resulting $1\sigma$, $2\sigma$, and $3\sigma$ intervals for all VLF representations and flavor configurations are collected in Tables~\ref{tab:SMEFT_intervals_sbs_diag} and \ref{tab:SMEFT_intervals_sbs_offdiag}. Each entry is analyzed independently by activating a single flavor structure $\mathscr{C}^F_{ij}$ at a time, corresponding to a single direction in parameter space as discussed in Section~\ref{sec:SMEFT_meth}. We note that the structure of rank-1 flavor violating forces the flavor-diagonal entries to be positive definite, $\mathscr{C}^F_{ii}=[\lambda_F]^*_i[\lambda_F]_i/\widetilde{m}_F^2\geq0$. Consequently, for these entries, the allowed intervals extend from zero towards positive values only. Although we restrict our analysis to real parameters, flavor-violating coefficients remain sensitive to the relative signs of the parameters. The results collected in Tables~\ref{tab:SMEFT_intervals_sbs_diag} and \ref{tab:SMEFT_intervals_sbs_offdiag} exhibit several noteworthy features:
\begin{itemize}
  \item A clear hierarchy is observed between flavor-diagonal $(ij)=(11),(22),(33)$ and off-diagonal $(ij)=(12),(13),(23)$ configurations. The diagonal entries are typically constrained at the level of $\cO(10^{\eminus2}\,\tev^{\eminus2})$, while the off-diagonal ones are generally subject to stronger bounds, often by one to two orders of magnitude. Although this pattern is not uniform across all representations, it constitutes a robust overall trend visible in the table.
  \item The relative strength of diagonal and off-diagonal constraints depends on the quark sector to which the VLF couples (see Table~\ref{tab:VLFs_SMEFT_tree}). For representations coupling to down-type quarks, the off-diagonal entries are systematically more strongly constrained than the diagonal ones. In contrast, for representations coupling to up-type quarks this hierarchy is less pronounced, with diagonal and off-diagonal bounds typically of comparable size.
  \item An additional pattern emerges upon comparing left- and right-handed structures. For VLF representations coupling to down-type quarks, the resulting constraints on left- and right-handed operators are typically of comparable size. In contrast, for representations coupling to up-type quarks, left-handed interactions tend to be more strongly constrained than their right-handed counterparts.
\end{itemize}

We further elucidate the origin of the bounds in Tables~\ref{tab:SMEFT_intervals_sbs_diag} and \ref{tab:SMEFT_intervals_sbs_offdiag} by focusing on the $D$-type VLF as a benchmark and identifying the observables driving the constraints. Rather than relying solely on the global fit, the full likelihood is decomposed into contributions from different observable classes, and the constraints arising from each sector are evaluated separately. The outcome of this decomposition is shown in Figure~\ref{fig:D_VLF_SMEFT_observables_pheno} where the six panels display the bounds obtained from selected classes of observables for flavor-diagonal and off-diagonal configurations, highlighting the relative impact of flavor-conserving and flavor-violating sectors. The main features of the resulting constraints can be summarized as follows:
\begin{figure*}[t]
  \centering
  \begin{tabular}{cc}
    \includegraphics[width=0.475\linewidth]{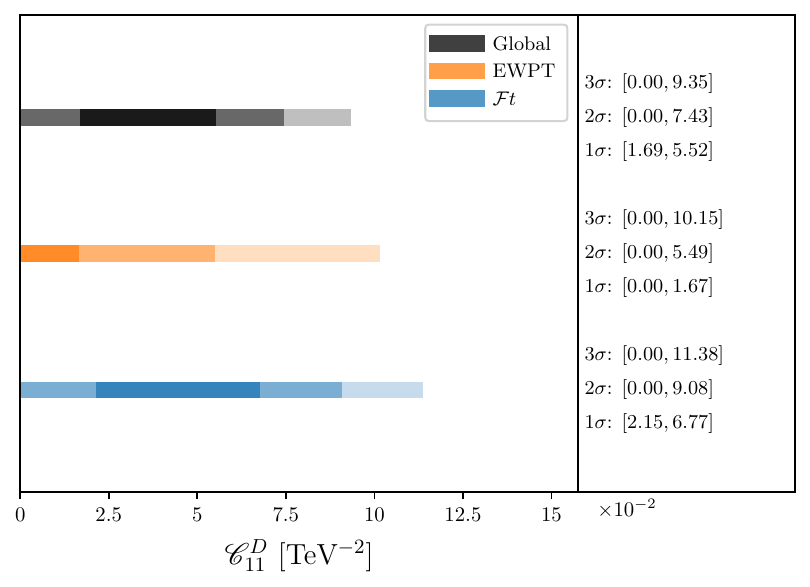}
    &
    \includegraphics[width=0.475\linewidth]{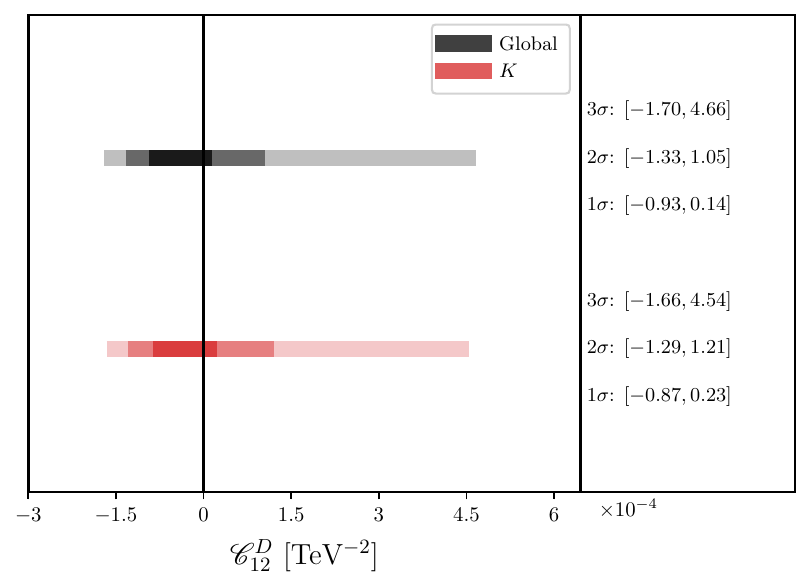}
    \\
    \includegraphics[width=0.475\linewidth]{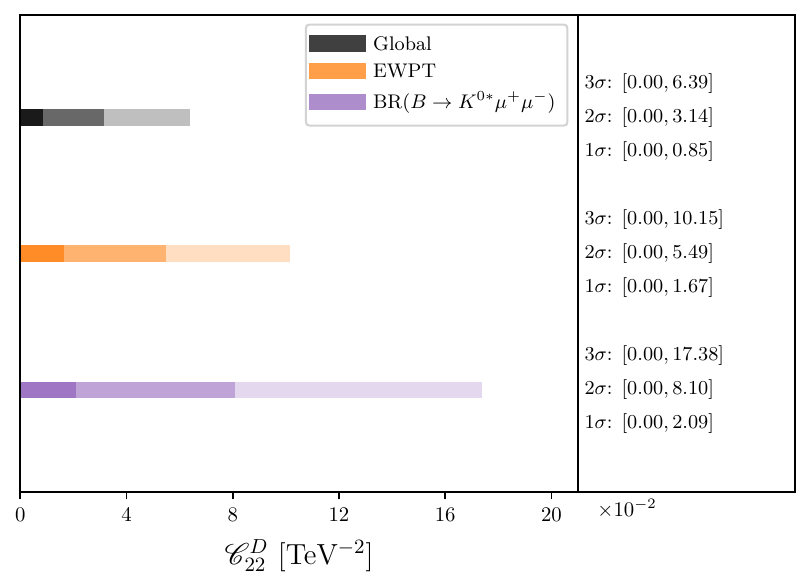}
    &
    \includegraphics[width=0.475\linewidth]{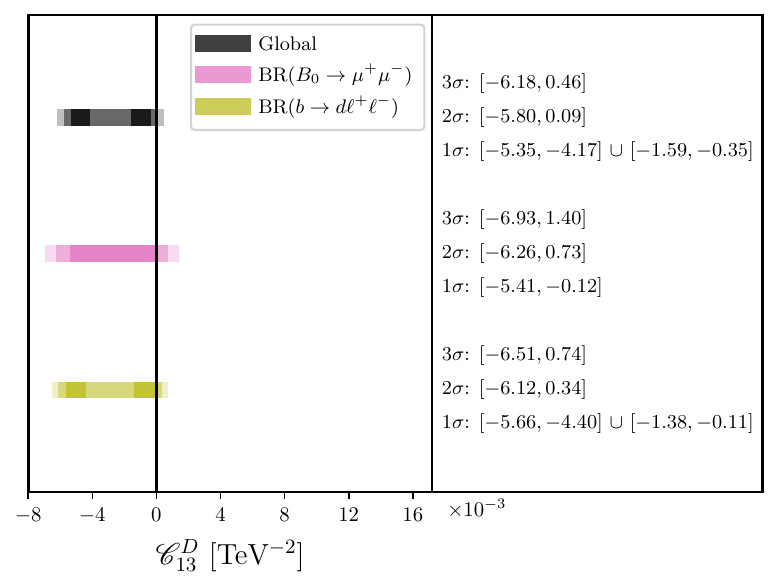}
    \\
    \includegraphics[width=0.475\linewidth]{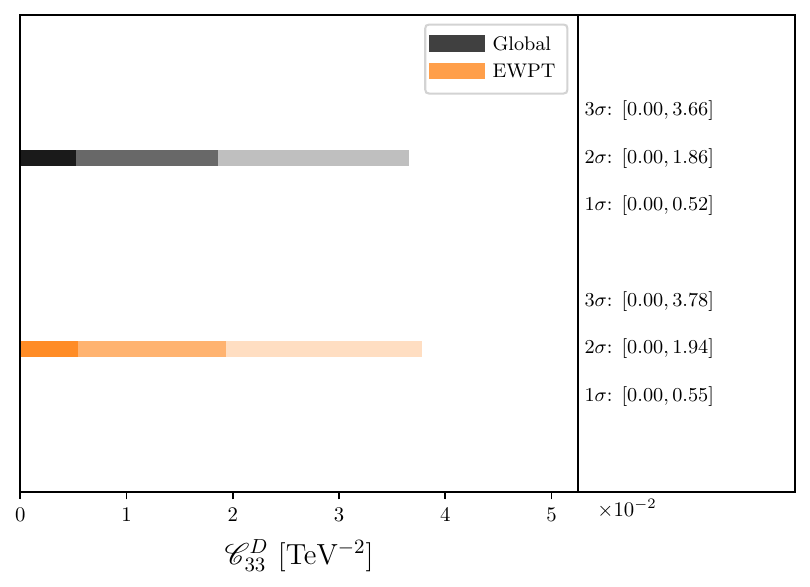}
    &
    \includegraphics[width=0.475\linewidth]{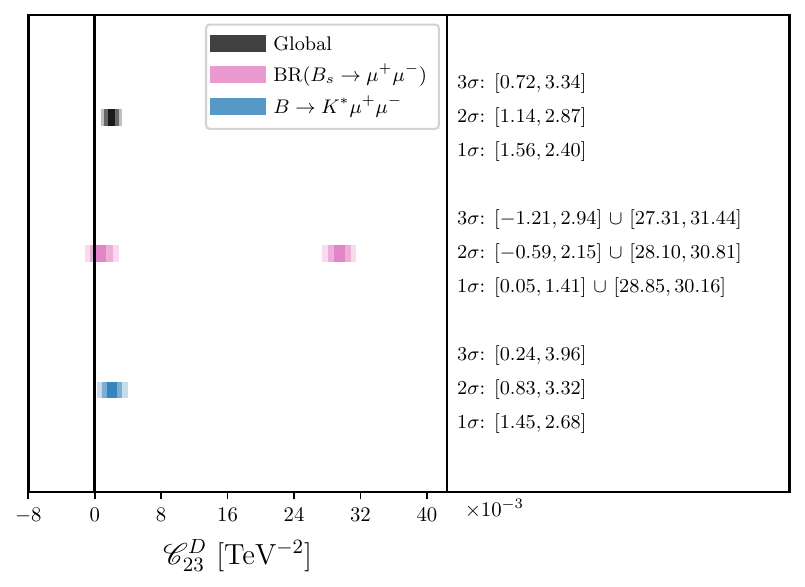}
  \end{tabular}
  \caption{Decomposition of the constraints on the $\mathscr{C}^D_{ij}$ parameter for the $D$-type VLF obtained from different subsets of observables. Each panel corresponds to a specific flavor configuration $(ij)$. The horizontal bars show the $1\sigma$, $2\sigma$, and $3\sigma$ intervals derived from fits restricted to selected classes of observables, compared with the result of the full global fit.}
  \label{fig:D_VLF_SMEFT_observables_pheno}
\end{figure*}
\begin{itemize}
  \item For the flavor-diagonal configurations $(ij)=(11),(22),(33)$, the bounds are consistently of order $\mathcal{O}(10^{\eminus2}\,\tev^{\eminus2})$, with electroweak precision tests (EWPT) providing the dominant sensitivity. In the $(11)$ case, the constraints are further strengthened by first-generation charged-current observables, most notably superallowed $\beta$ decays~\cite{Falkowski:2020pma,Gonzalez-Alonso:2018omy,Alioli:2017ces,Falkowski:2021vdg}. For $(22)$, additional sensitivity arises from flavor observables such as the differential branching ratio of $B^0 \to K^{\ast 0}\mu^+\mu^-$, whose dependence on $\mathscr{C}^D_{22}$ is induced through renormalization group evolution (RGE)~\cite{Jenkins:2013wua,Alonso:2013hga}. By contrast, the $(33)$ configuration remains predominantly constrained by EWPT, reflecting the limited sensitivity of current flavor observables to purely third-generation diagonal couplings.
  \item The flavor off-diagonal configurations $(ij)=(12),(13),(23)$ are subject to more stringent constraints than their diagonal counterparts. For $(12)$, the bounds reach the $\mathcal{O}(10^{-4}\,\tev^{-2})$ level and are predominantly driven by kaon observables, including $\varepsilon'/\varepsilon$ and rare decays such as $K\to\pi\nu\bar{\nu}$ and $K_{L,S}\to\ell^+\ell^-$. By contrast, the $(13)$ and $(23)$ configurations are constrained at the $\mathcal{O}(10^{-3}\,\tev^{-2})$ level, with the dominant sensitivity arising from rare $B$-meson transitions. In particular, the $(13)$ bounds are mainly controlled by processes such as $B_{d,s}\to\mu^+\mu^-$ and inclusive $b\to d\ell^+\ell^-$ decays, while the $(23)$ configuration is primarily constrained by a combination of the branching ratio of $B_s\to \mu^+\mu^-$ and the combination of the branching ratio and angular observables for the process $B\to K^{\ast0}\mu^+\mu^-$, which exhibits a mild tension with the SM prediction.
\end{itemize}
For completeness, Appendix~\ref{app:SMEFT_obs_overview} provides a brief overview of the key observables discussed above, while Appendix~\ref{app:Q7_SMEFT_ALP} includes the analysis of the analogous up-type scenario involving $Q_7$, probing a distinct set of observables. Finally, the collider constraints discussed in Appendix~\ref{sec:coll_constraints} are found to be subleading compared to the indirect probes considered throughout this analysis.

\subsubsection*{Neutral meson mixing}
To conclude our discussion of the relevant observables, let us briefly comment on neutral meson mixing. In the global analysis performed with \texttt{smelli} and \texttt{flavio}, part of the information from meson-mixing observables is already incorporated through the determination of the CKM parameters. More precisely, $\Delta M_d$ enters the determination of the CKM parameters and therefore contributes indirectly to the likelihood, while the remaining $\Delta F=2$ observables appear as outputs. It is nevertheless instructive to estimate the standalone sensitivity of the $\Delta F=2$ transitions (including $\Delta M_d$) induced by the SMEFT operators considered here, in order to directly compare their impact with the bounds obtained in Tables~\ref{tab:SMEFT_intervals_sbs_diag} and \ref{tab:SMEFT_intervals_sbs_offdiag}.

In the scenarios considered here, contributions to $\Delta F=2$ arise from several sources. At tree level, integrating out the VLFs generates Higgs-fermion current operators, which induce modifications of both the $Z$-boson and Higgs couplings to quarks after electroweak symmetry breaking. Upon matching onto the low-energy theory, these modified interactions give rise to $\Delta F=2$ transitions through double insertions, mediated by tree-level $Z$- and Higgs-boson exchange~\cite{Jenkins:2017jig,Harnik:2012pb,Alonso-Alvarez:2023wig,DiLuzio:2019jyq}. Beyond tree level, additional contributions are generated through RGE, which mixes the operators present at the matching scale into four-fermion structures relevant for meson mixing. Furthermore, one-loop matching contributions within the UV model can directly induce $\Delta F=2$ operators. In the following, we estimate these effects for the $D$ VLF as a representative case.

Starting with the tree-level constraints, as indicated in Eq.~\eqref{eq:D_case_CHQ_CDH_coeff}, the relevant coefficient is given by $[\cC_{Hq}^{(1,3)}]_{ij}=-\mathscr{C}^D_{ij}/4$. Using the bounds on the corresponding Wilson coefficients from Ref.~\cite{Silvestrini:2018dos}, we extract the following constraints. From kaon mixing, we obtain $|\mathscr{C}^D_{12}|\lesssim 3.56\times10^{\eminus3}\,\tev^{\eminus2}$, which is weaker than the bound reported in Table~\ref{tab:SMEFT_intervals_sbs_offdiag}. For $B_d$ mixing, the corresponding constraint $|\mathscr{C}^D_{13}|\lesssim 2.96\times10^{\eminus3}\,\tev^{\eminus2}$ is of comparable size to our result, while $B_s$ mixing yields $|\mathscr{C}^D_{23}|\lesssim 1.20\times10^{\eminus2}\,\tev^{\eminus2}$, which remains less constraining.

It is worth emphasizing that the constraints quoted above are obtained from the modified $Z$-boson couplings induced by $\cO_{Hq}^{(1,3)}$, which provide the leading tree-level contribution to meson mixing in the SMEFT description. By contrast, the Higgs-mediated contribution considered in Ref.~\cite{Alonso-Alvarez:2023wig} arises through $\cO_{dH}$ and is proportional to the down-type Yukawa couplings, see Eq.~\eqref{eq:D_case_CHQ_CDH_coeff}. As a result, the corresponding $\Delta F=2$ amplitudes are Yukawa suppressed, whereas the modified-$Z$ contribution is not. This explains why the constraints derived here are generally stronger and why, using the SMEFT, Higgs-exchange effects constitute only a subleading contribution to neutral meson mixing.

\begin{figure}[t]
  \centering
  \includegraphics[width=0.98\textwidth]{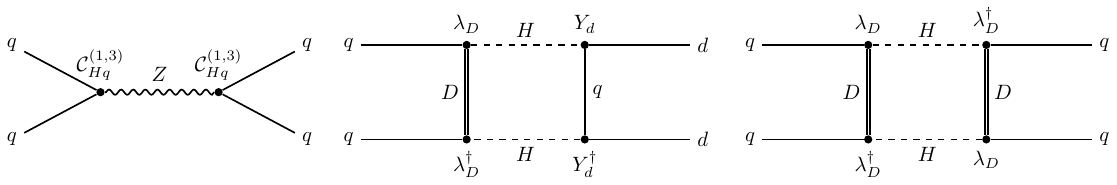}
  \caption{Representative tree-level and one-loop matching diagrams in the $D$ VLF model contributing to $\Delta F=2$ operators relevant for meson mixing. The left diagram corresponds to the tree-level contribution induced by modified $Z$-boson couplings arising from $\cO_{Hq}^{(1,3)}$ insertions, while the middle and right diagrams generate $\cO_{qd}^{(1)}$ and $\cO_{qq}^{(1,3)}$, respectively.}
  \label{fig:meson_mix_diag}
\end{figure}

Furthermore, in order to estimate the impact of the one-loop effects, we consider the direct one-loop matching contributions (see Figure~\ref{fig:meson_mix_diag} for representative diagrams) and focus on the most relevant components entering meson mixing. The corresponding contributions, evaluated at the matching scale $\mu \simeq \widetilde m_D$, are given by~\cite{Gargalionis:2024jaw}
\begin{equation}
  {\scalebox{0.97}{
      $
      \begin{alignedat}{2}
        [\cC_{qd}^{(1)}]_{ijk\ell} &\supset -\frac{1}{16\pi^2}\lzv \frac{3}{8} \mathscr C^D_{ij}[Y_d^\dag Y_d]_{k\ell}+\frac{1}{6}\mathscr C^D_{aa}[Y_d]_{i\ell}[Y_d^\dag]_{kj} \dzv\,,
        \quad
        [\cC_{qq}^{(1,3)}]_{ijk\ell}\supset -\frac{1}{16\pi^2}\frac{\widetilde m_D^2}{16}\mathscr C^D_{ij}\mathscr C^D_{k\ell}\,.
      \end{alignedat}
      $
  }}
\end{equation}
The coefficient $\cC_{qd}^{(1)}$ contains two distinct structures: the first, proportional to $[Y_d^\dag Y_d]_{k\ell}$, does not induce flavor-changing effects since this combination remains diagonal in any flavor basis, while the second, proportional to $[Y_d]_{i\ell}[Y_d^\dag]_{kj}$, can in principle generate flavor violation, although its contribution is controlled by the flavor-diagonal coefficients $\mathscr C^D_{aa}$, with an implicit sum over $a$ understood. A numerical estimate yields, for instance taking kaon mixing,  $[\cC_{qd}^{(1)}]_{1212} \sim 7.7 \times 10^{\eminus13}\,\mathscr{C}^D_{aa}$, leading to constraints that are significantly weaker than those obtained for the diagonal entries in Table~\ref{tab:SMEFT_intervals_sbs_diag}. A similar conclusion applies to the $\cC_{qq}^{(1,3)}$ coefficients. Fixing $\widetilde m_D\sim1\,\tev$, one finds $\mathscr{C}^D_{12}\sim2.80\times10^{\eminus2}\,\tev^{\eminus2}$, which again results in bounds that are weaker than those obtained from the global fit.

Overall, this discussion clarifies the role played by $\Delta F=2$ observables in the present analysis. For the $D$-type scenario considered here, the standalone constraints obtained from meson mixing, including $\Delta M_d$, remain comparable to or weaker than the bounds extracted from the global SMEFT fit, indicating that no additional sensitivity is gained beyond the observables already driving the likelihood. This situation, however, is not generic. In other configurations, such as the $(12)$ flavor structure of the $Q_7$-type scenario, charm-mixing observables provide one of the dominant contributions to the global fit (see Appendix~\ref{app:Q7_SMEFT_ALP}). As mentioned above, such effects are already consistently incorporated within the \texttt{smelli} analysis adopted throughout this work.

\section{Overview of Axion and ALP Constraints}
\label{sec:Axion_Pheno}
\begin{table}[t]
  \centering
\renewcommand{\arraystretch}{1.2}
\scalebox{0.78}{
\begin{tabular}{cccc}
\toprule
\textbf{\textbf{VLF}}&$\bm{\cX_{L}}$ &$\bm{\cX_R}$& $\bm{\cL_{\sscript{ALP}}\supset}$
\\
\midrule
\multirow{3}{*}{\vspace{-0.75cm}$D\sim(\bm 3,\bm 1)_{-1/3}$}
&1
&0
&$\frac{v^2 C_{D,L}^{(1)}}{2\widetilde m_D^2} [\lambda_D]_i [\lambda_D]^*_j\frac{\partial_\mu a}{v_\Phi}(\bar d_L^i \gamma_\mu d_L^j) $
\vspace{+0.1cm}\\
\noalign{\vskip 2pt}
\cdashline{2-4}[.4pt/2pt]
\noalign{\vskip 5pt}
\vspace{+0.1cm}
&\multirow{1}{*}{\vspace{-0.2cm}0}
&\multirow{1}{*}{\vspace{-0.2cm}-1}
&$\lzs C^{(2)}_{D,R}+ \frac{v^2}{2\widetilde m_D^2}\widehat Y_d^\dag \lambda_D C_{D,R}^{(3)}\dzs_{ij}\frac{\partial_\mu a}{v_\Phi}(\bar d_R^i \gamma^\mu d_R^j)$
\\
\noalign{\vskip 1pt}
\cdashline{2-4}[.4pt/2pt]
\noalign{\vskip 4pt}
&\multirow{1}{*}{\vspace{-0.2cm}-1}
&\multirow{1}{*}{\vspace{-0.2cm}-2}
&$\frac{v^2 C_{D,L}^{(1)}}{2\widetilde m_D^2} [\lambda_D]_i [\lambda_D]^*_j\frac{\partial_\mu a}{v_\Phi}(\bar d_L^i\gamma_\mu d_L^j)+\lzs C^{(2)}_{D,R}+\frac{v^2}{2\widetilde m_D^2}\widehat Y_d^\dag \lambda_D C_{D,R}^{(3)} \dzs_{ij}\frac{\partial_\mu a}{v_\Phi}(\bar d_R^i \gamma_\mu d_R^j)$
\\[5pt]
\midrule
\multirow{4}{*}{\vspace{-0.45cm}$U\sim(\bm 3,\bm 1)_{2/3}$}
&1
&0
&$\frac{v^2C_{U,L}^{(1)}}{2\widetilde m_U^2} [\lambda_U]_i [\lambda_U]^*_j\frac{\partial_\mu a}{v_\Phi}(\bar u_L^i\gamma_\mu u_L^j)$
\vspace{+0.1cm}\\
\noalign{\vskip 2pt}
\cdashline{2-4}[.4pt/2pt]
\noalign{\vskip 5pt}
\vspace{+0.1cm}
&\multirow{1}{*}{\vspace{-0.1cm}0}
&\multirow{1}{*}{\vspace{-0.1cm}-1}
&$\lzs C^{(2)}_{U,R}+\frac{v^2}{2\widetilde m_U^2}\widehat Y_u^\dag \lambda_U C_{U,R}^{(3)} \dzs_{ij} \frac{\partial_\mu a}{v_\Phi} (\bar u_R^i \gamma^\mu u_R^j)$
\vspace{+0.0cm}\\
\noalign{\vskip 2pt}
\cdashline{2-4}[.4pt/2pt]
\noalign{\vskip 2pt}
&\multirow{1}{*}{\vspace{-0.0cm}-1}
&\multirow{1}{*}{\vspace{-0.0cm}-2}
&$\frac{v^2C_{U,L}^{(1)}}{2\widetilde m_U^2} [\lambda_U]_i [\lambda_U]^*_j\frac{\partial_\mu a}{v_\Phi}(\bar u_L^i\gamma_\mu u_L^j)+\lzs C^{(2)}_{U,R}+\frac{v^2}{2\widetilde m_U^2}\widehat Y_u^\dag \lambda_U C_{U,R}^{(3)} \dzs_{ij} \frac{\partial_\mu a}{v_\Phi} (\bar u_R^i \gamma^\mu u_R^j)$
\\[5pt]
\midrule
\addlinespace[0.2cm]
\multirow{5}{*}{\vspace{-0.7cm}$Q\sim(\bm 3,\bm 2)_{1/6}$}
&\multirow{1}{*}{\vspace{-0.1cm}1}
&\multirow{1}{*}{\vspace{-0.1cm}0}
&$\lzs C_{Q,L}^{(2)}-\frac{v^2}{2\widetilde m_Q^2}\widehat Y_d \lambda_Q^{d\dag}C_{Q,L}^{(3)} \dzs_{ij}\frac{\partial_\mu a}{v_\Phi}(\bar d_L^i\gamma_\mu d_L^j)
+\lzs C_{Q,L}^{(2)}-\frac{v^2}{2\widetilde m_Q^2}\widehat Y_u \lambda_Q^{u\dag}C_{Q,L}^{(3)} \dzs_{ij}\frac{\partial_\mu a}{v_\Phi}(\bar u_L^i\gamma_\mu u_L^j)$
\\[5pt]
\noalign{\vskip 2pt}
\cdashline{2-4}[.4pt/2pt]
\noalign{\vskip 2pt}
&\multirow{1}{*}{\vspace{-0.0cm}0}
&\multirow{1}{*}{\vspace{-0.0cm}-1}
&$\frac{v^2C_{Q,R}^{(1)}}{2\widetilde m_Q^2}[\lambda_Q^d]^*_i[\lambda_Q^d]_j\frac{\partial_\mu a}{v_\Phi}(\bar d_R^i\gamma_\mu d_R^j)+\frac{v^2 C_{Q,R}^{(1)}}{2\widetilde m_Q^2}[\lambda_Q^u]^*_i[\lambda_Q^u]_j\frac{\partial_\mu a}{v_\Phi}(\bar u_R^i\gamma_\mu u_R^j)$
\\
\noalign{\vskip 2pt}
\cdashline{2-4}[.4pt/2pt]
\noalign{\vskip 5pt}
&\multirow{2}{*}{\vspace{-0.4cm}2}
&\multirow{2}{*}{\vspace{-0.4cm}1}
&$\lzs C_{Q,L}^{(2)}-\frac{v^2}{2\widetilde m_Q^2}\widehat Y_d \lambda_Q^{d\dag}C_{Q,L}^{(3)} \dzs_{ij}\frac{\partial_\mu a}{v_\Phi}(\bar d_L^i\gamma_\mu d_L^j)
+\lzs C_{Q,L}^{(2)}-\frac{v^2}{2\widetilde m_Q^2}\widehat Y_u \lambda_Q^{u\dag}C_{Q,L}^{(3)} \dzs_{ij}\frac{\partial_\mu a}{v_\Phi}(\bar u_L^i\gamma_\mu u_L^j)$
\\[5pt]
&&&$+\frac{v^2C_{Q,R}^{(1)}}{2\widetilde m_Q^2}[\lambda_Q^d]^*_i[\lambda_Q^d]_j\frac{\partial_\mu a}{v_\Phi}(\bar d_R^i\gamma_\mu d_R^j)+\frac{v^2C_{Q,R}^{(1)}}{2\widetilde m_Q^2}[\lambda_Q^u]^*_i[\lambda_Q^u]_j\frac{\partial_\mu a}{v_\Phi}(\bar u_R^i\gamma_\mu u_R^j)$
\\
\midrule
$Q_5\sim(\bm 3,\bm 2)_{-5/6}$
&0&-1
&$\frac{v^2C_{Q_5,R}^{(1)}}{2\widetilde m_{Q_5}^2}[\lambda_{Q_5}]^*_i[\lambda_{Q_5}]_j\frac{\partial_\mu a}{v_\Phi}(\bar d_R^i \gamma_\mu d_R^j)$
\\
\midrule
$Q_7\sim(\bm 3,\bm 2)_{7/6}$
&0&-1
&$\frac{v^2C_{Q_7,R}^{(1)}}{2\widetilde m_{Q_7}^2}[\lambda_{Q_7}]^*_i[\lambda_{Q_7}]_j\frac{\partial_\mu a}{v_\Phi}(\bar u_R^i \gamma_\mu u_R^j)$
\\
\midrule
$T_1\sim(\bm 3,\bm 3)_{-1/3}$
&1&0
&$\frac{v^2 C_{T_1,L}^{(1)}}{4\widetilde m_{T_1}^2}[\lambda_{T_1}]^*_i[\lambda_{T_1}]_j\frac{\partial_\mu a}{v_\Phi}(\bar u_L^i\gamma_\mu u_L^j)+\frac{v^2 C_{T_1,L}^{(1)}}{8\widetilde m_{T_1}^2}[\lambda_{T_1}]^*_i[\lambda_{T_1}]_j\frac{\partial_\mu a}{v_\Phi}(\bar d_L^i\gamma_\mu d_L^j)$
\\
\midrule
$T_2\sim(\bm 3,\bm 3)_{2/3}$
&1&0
&$\frac{v^2 C_{T_2,L}^{(1)}}{8\widetilde m_{T_2}^2}[\lambda_{T_2}]^*_i[\lambda_{T_2}]_j\frac{\partial_\mu a}{v_\Phi}(\bar u_L^i\gamma_\mu u_L^j)+\frac{v^2 C_{T_2,L}^{(1)}}{4\widetilde m_{T_2}^2}[\lambda_{T_2}]^*_i[\lambda_{T_2}]_j\frac{\partial_\mu a}{v_\Phi}(\bar d_L^i\gamma_\mu d_L^j)$
\\
\bottomrule
\end{tabular}
}
  \caption{Effective ALP-fermion interactions after EWSB for the VLF representations considered in this work. The resulting couplings to quark bilinears are expressed in terms of the SMEFT--ALP Wilson coefficients defined in Section~\ref{sec:UV_IR_Framework}.}
  \label{fig:tab_ALP_Lagr_after_ewsb}
\end{table}
The phenomenology of axions and axion-like particles (ALPs) is typically probed at energies well below the electroweak scale. It is therefore necessary to express the effective interactions obtained in the previous section after EWSB. In practice, this corresponds to expanding the ALP-SM effective Lagrangian given in Table~\ref{tab:ALP_EFT_Lagr} and matching it onto the relevant low-energy operators. Since most experimental probes of ALPs involve fermionic couplings of the form $a\bar f f$, we restrict the analysis to operators up to dimension five and neglect higher-dimensional interactions such as $a(\bar f f)^2$. The low-energy Lagrangian corresponds to the derivative interactions
\begin{equation}
  \mathcal{L}_{\sscript{LE-ALP}} =
  \frac{\partial_\mu a}{f_a}[c_{af_R}]_{ij} \bar{f}^i_R \gamma^\mu f^j_R+\frac{\partial_\mu a}{f_a}[c_{af_L}]_{ij} \bar f^i_{L} \gamma^\mu f^j_L\,,
  \qquad
  f_a\equiv\frac{v_\Phi}{2c_{G}}\,,
  \label{eq:LE_alp_lag}
\end{equation}
where the resulting set of effective interactions after EWSB for the different models is summarized in Table~\ref{fig:tab_ALP_Lagr_after_ewsb}.

As discussed in Section~\ref{sec:UV_IR_Framework}, an important feature of the matching is that the Wilson coefficients $\cC_{aHq}^{(1)}$ and $\cC_{aHq}^{(3)}$ originate from a common operator in the UV theory and are therefore related by $\cC_{aHq}^{(1)}=\pm \cC_{aHq}^{(3)}$. This relation implies that the pattern of ALP couplings to quarks depends on the specific VLF representation. For instance, in the case $\Psi=D$ one finds $\cC_{aHq}^{(1)}=\cC_{aHq}^{(3)}$, which removes the coupling to left-handed up-type quarks when matching to the low-energy lagrangian of Eq.~\eqref{eq:LE_alp_lag}, whereas for $\Psi=U$ the relation $\cC_{aHq}^{(1)}=-\cC_{aHq}^{(3)}$ cancels the interaction with left-handed down-type quarks. On the other hand, representations that generate RH operators $\cC_{aHd}$ or $\cC_{aHu}$ match directly to the RH operators of Eq.~\eqref{eq:LE_alp_lag} after EWSB. Consequently, the pattern of low-energy ALP-fermion interactions directly reflects the underlying UV mediator.

\begin{figure}[t]
  \centering
  \includegraphics[width=0.948\textwidth]{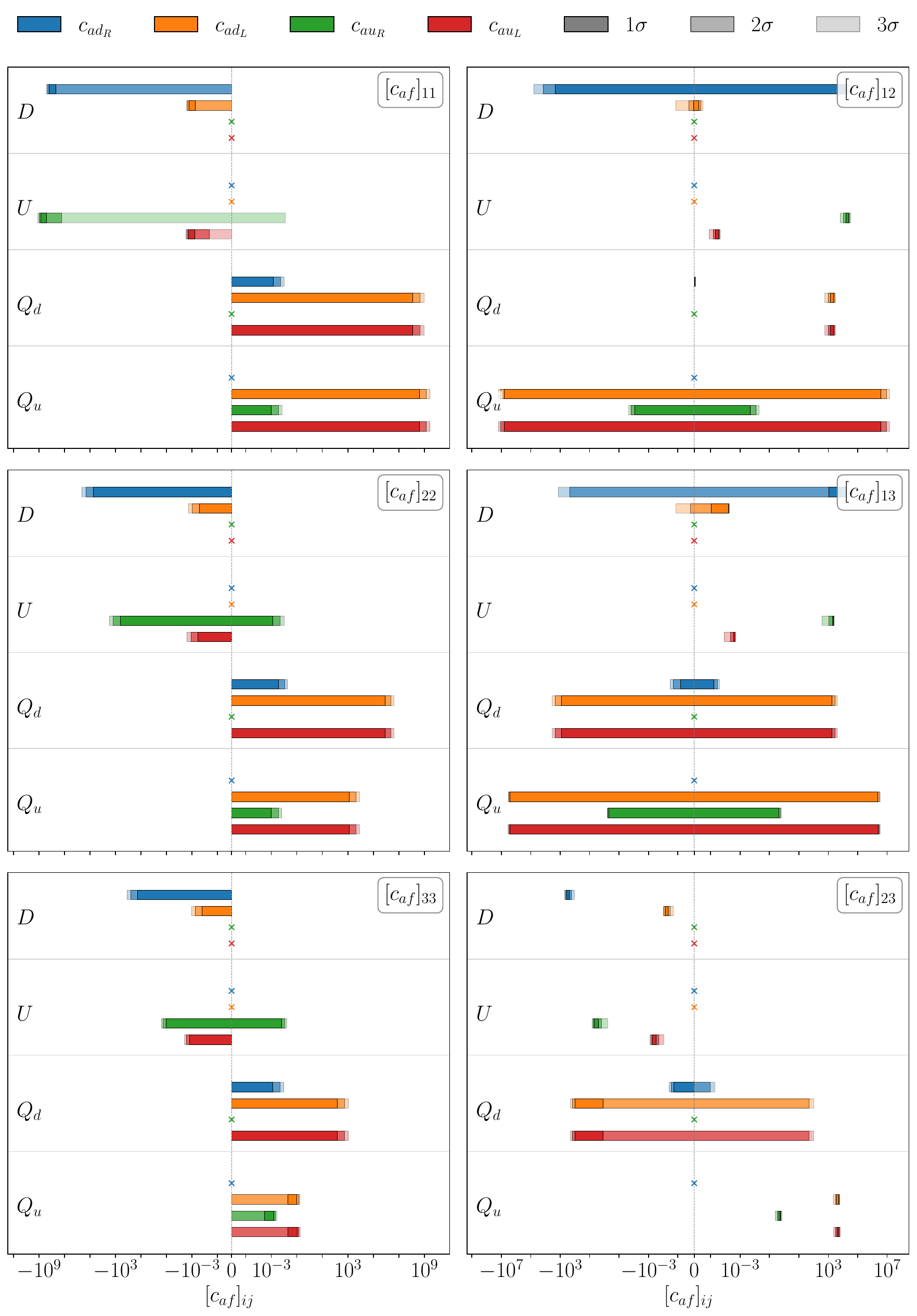}
  \caption{$1\sigma$, $2\sigma$, and $3\sigma$ bounds on the axion-fermion couplings for models containing VLF representations $D$, $U$, $Q_d$, and $Q_u$. The results are shown for the PQ charge assignment in which both chiralities have nonzero PQ charge. Crosses indicate operators that are absent in the corresponding model.}
  \label{fig:bounds_axion1}
\end{figure}

\begin{figure}[t]
  \centering
  \includegraphics[width=0.948\textwidth]{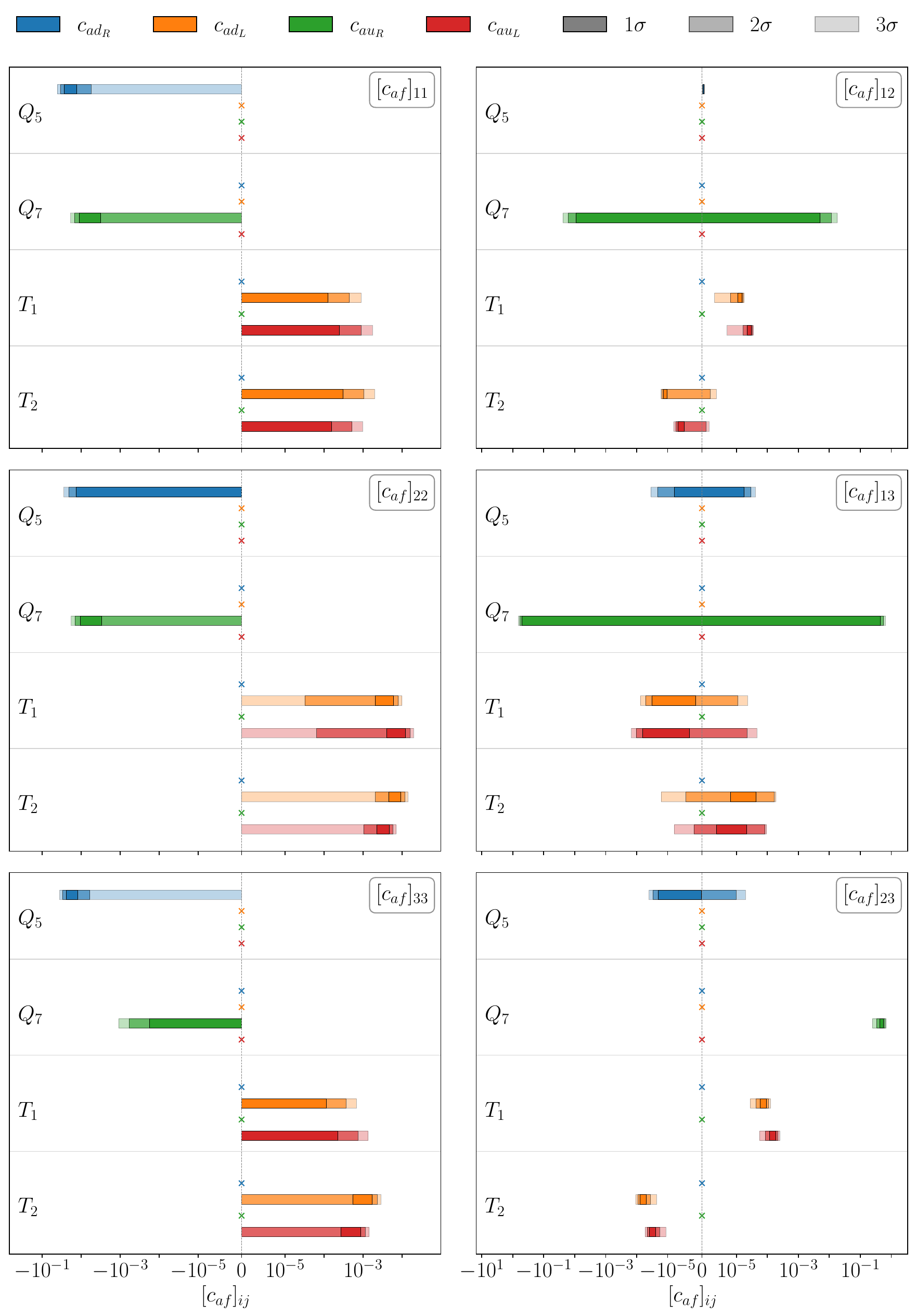}
  \caption{$1\sigma$, $2\sigma$, and $3\sigma$ bounds on the axion-fermion couplings for the models with VLF representations $Q_5$, $Q_7$, $T_1$ and $T_2$. Crosses represent operators which are not generated by the model.}
  \label{fig:bounds_axionQ5_T2}
\end{figure}

Building on the results of Section~\ref{sec:UV_IR_Framework}, we translate the bounds on the SMEFT Wilson coefficients into constraints on the ALP-fermion interactions. Using the fit results of Section~\ref{sec:SMEFT_pheno}, these bounds are mapped onto the corresponding low-energy theory for each model. In Figure~\ref{fig:bounds_axion1} we present the bounds for the models associated with the $D$, $U$, and $Q$ representations.\footnote{As indicated in Section~\ref{sec:SMEFT_pheno}, in the $Q$ model, the representation can couple to both up- and down-type quarks. In our analysis, we treat these two possibilities separately, denoted by $Q_u$ and $Q_d$, corresponding to scenarios in which only the coupling to up- or down-type quarks is switched on, respectively.} We note that, although these models admit three possible PQ-charge configurations, the bounds obtained for the configuration with both left- and right-handed PQ charges are sufficient to infer those for the other two. As an illustrative example, consider the VLF representation $D$ with PQ charges $\left(\chi_R,\,\chi_L\right)=\left(-1,\,-2\right)$, whose results are shown in Figure~\ref{fig:bounds_axion1}. The bounds for the other two assignments, $(1,\,0)$ and $(0,\,-1)$, can be obtained by a simple rescaling: the coefficients of the RH operators are divided by the corresponding right-handed PQ charge, which in this case gives a factor of $2$, while the coefficients of the left-handed operator are rescaled by a factor of $-1$. For the representations that admit only a single PQ-charge assignment, namely $Q_{5,7}$ and $T_{1,2}$, the corresponding results can be read directly from Figure~\ref{fig:bounds_axionQ5_T2}.

For the models shown in Figure~\ref{fig:bounds_axion1}, we observe a marked asymmetry between the bounds on the right- and left-handed coefficients. In particular, for the $D$ and $U$ models the right-handed coefficients are essentially unconstrained, whereas in the $Q$ model the opposite behavior is found. This difference can be traced back to the origin of the corresponding operators. For these models, there exist PQ-charge assignments that allow for a mass mixing between the VLF and its SM counterpart. Whenever such a term is present, the field redefinition of Eq.~\eqref{eq:field_redefinition} used to remove it induces a direct interaction in the ALP Lagrangian, with a coefficient scaling as $\sim M_\Psi^\dagger M_\Psi / m_\Psi^2$. By contrast, the corresponding SMEFT coefficients scale as $\sim Y_f M_\Psi^\dagger M_\Psi Y_f^\dagger / m_\Psi^4$, where $Y_f$ denotes the Yukawa matrix of the SM fermion coupled to the VLF. As a result, when translating the SMEFT bounds into the ALP theory, the corresponding constraints are effectively suppressed by the fermion Yukawa coupling, so that the associated ALP couplings remain essentially unconstrained. This also explains why the bounds become stronger as the flavor index increases: owing to the hierarchical structure of the Yukawa couplings, the strongest constraints are obtained for third-generation interactions, in particular in the $U$ and $Q_u$ models.

The general pattern can thus be summarized as follows: whenever the PQ-charge assignment allows for mixing with the corresponding SM fermion, the SMEFT bounds do not translate into meaningful constraints on the associated ALP coefficients, with the notable exception of interactions involving the top quark. By contrast, the $Q_{5,7}$ and $T_{1,2}$ models do not exhibit this behavior, as their gauge quantum numbers forbid any mixing with SM fermions. On the other hand, in models with couplings to up-type quarks, constraints involving the third generation can remain competitive. This is because the large top Yukawa does not induce the same suppression in the corresponding SMEFT operators.

Although off-diagonal coefficients are generally more tightly constrained, notable exceptions exist. In the $Q_7$ model, for example, the third-generation diagonal coupling is bounded at the level of $\lesssim10^{\eminus3}$, while the off-diagonal entries are constrained only within $[10^{\eminus2},1]$. The stronger diagonal bound originates from top-loop contributions to EWPO, whereas flavor constraints arise only through loop-induced running into the down sector and are therefore significantly suppressed. This pattern is not observed in the $U$ model, where loop-generated flavor-violating operators induce larger effects in flavor observables, leading to correspondingly stronger off-diagonal constraints.

We stress that the bounds obtained in this low-energy basis apply to KSVZ models both in the QCD axion case and in the more general ALP setup with an independent mass. In the next section, we explore the interplay between these bounds and the correlated SMEFT constraints, focusing in particular on the information that precision observables can provide about ALP phenomenology.

\FloatBarrier
\FloatBarrier

\section{Phenomenological Interplay of Axion/ALP and SMEFT Effects}
\label{sec:Interplay}
The analysis presented in Sections~\ref{sec:SMEFT_pheno} and \ref{sec:Axion_Pheno} provides a comprehensive characterization of the SMEFT and axion/ALP parameter spaces associated with the different KSVZ-like scenarios. Beyond the individual constraints derived in each sector, the unified framework developed in this work predicts non-trivial correlations among observables that originate from the same underlying dynamics. These correlations provide an additional layer of phenomenological information, allowing constraints and signals in one sector to be interpreted in the context of the other. In this section, we explore several representative manifestations of this complementary phenomenology and discuss their implications.

Before proceeding, it is useful to recall a key feature that distinguishes ALPs from the QCD axion and underlies much of the phenomenology discussed below. While the QCD axion predicts a characteristic relation between the axion mass and decay constant, $m_a f_a \approx m_\pi f_\pi$, general ALP scenarios allow these quantities to be treated as independent parameters. This enlarged parameter space considerably broadens the range of viable phenomenological scenarios, but at the same time raises the question of the UV origin of the effective ALP interactions. The additional degrees of freedom introduced in these constructions can in turn give rise to a richer phenomenology and reveal complementary signatures beyond those associated with the ALP itself. In this sector, we analyze separately the phenomenological implications for the QCD axion scenario, characterized by smaller masses, and a heavier ALP scenario with a free mass parameter.
\subsection{QCD Axion Phenomenology}
\begin{figure}[t]
  \centering
  \includegraphics[width=1\textwidth]{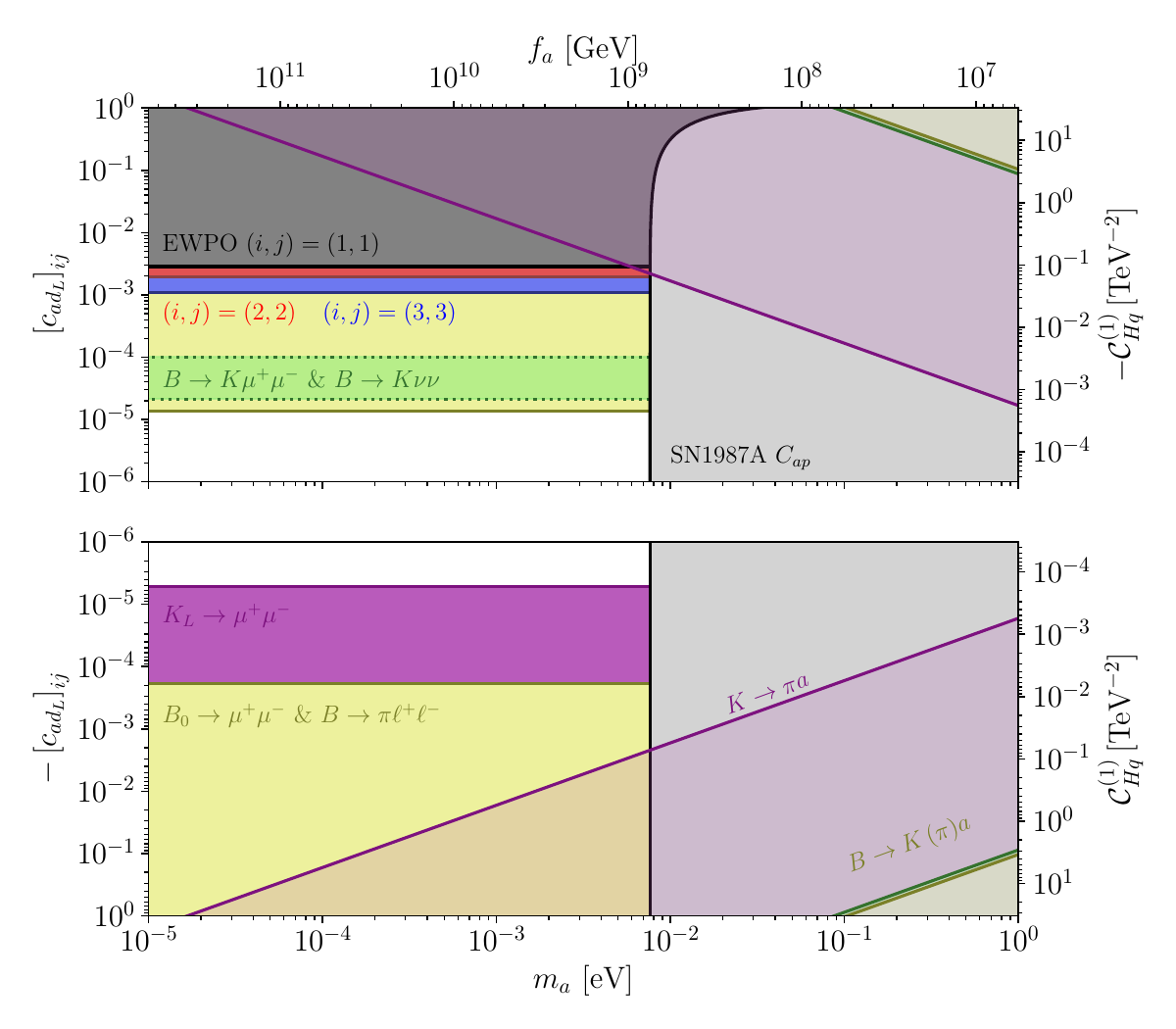}
  \caption{Bounds on the QCD-axion parameter space in the $(m_a,\,c_{ad_L})$ plane for the $D$ model. The black, red, and blue regions denote the constraints on the first-, second-, and third-generation diagonal coefficients, respectively, as derived from EWPO. The purple, yellow, and green regions correspond instead to the flavor-violating transitions $sd$, $db$, and $sb$, respectively, with the latter exhibiting a $3\sigma$ anomaly.}
  \label{fig:axion_D_model}
\end{figure}
As a first illustration of the phenomenological interplay, we consider the QCD axion realization of the $D$ model and focus on the flavor-diagonal couplings. This example provides a particularly transparent setting in which the constraints obtained from the SMEFT analysis can be compared with standard QCD axion bounds. The resulting bounds are shown in Figure~\ref{fig:axion_D_model}. Since the combination entering the diagonal couplings is positive definite, being proportional to $|\lambda_i|^2$, the sign of each diagonal fermionic coupling is determined entirely by the corresponding PQ charge assignment. In the benchmark considered here, we take $(\chi_L,\chi_R)=(1,0)$, such that all diagonal couplings are positive. The gray, red, and blue regions therefore correspond to the first-, second-, and third-generation diagonal couplings, respectively.

From the point of view of axion phenomenology, the most relevant of these is the coupling to the first generation, since it also controls stellar-cooling processes, both in ordinary stars and in supernova cores, through the induced couplings to neutrons and protons. As shown in Figure~\ref{fig:axion_D_model}, however, the strongest constraint, arising from SN1987A~\cite{Lella:2023bfb}, exhibits only a very mild dependence on the first-generation coupling. The reason is that, in this case, there is an additional contribution originating from the gluon anomaly, which effectively constrains the axion scale directly up to order one first-generation couplings.

Furthermore, flavor-violating processes are shown in purple, yellow, and green, corresponding to the $sd$, $db$, and $sb$ transitions, respectively. In this case, the bounds derived from the SMEFT analysis are significantly stronger than those obtained from direct axion searches. Indeed, among axion observables, only processes such as $K\to \pi a$ are capable of constraining the fermionic flavor-violating couplings in part of the parameter space; however, this region is already much more tightly constrained by the corresponding SMEFT-induced flavor observables, for instance through processes such as $K_L\to\mu\mu$.

This behavior is representative of the general picture across all models: searches for axions in meson decays do not yield stronger constraints, nor do they probe regions of parameter space beyond those already excluded by the SMEFT analysis. An important exception arises when the PQ-charge assignment allows for mass mixing with the corresponding SM fermion. In this case, as discussed above, the SMEFT bounds on the associated ALP couplings become strongly suppressed, remaining relevant only for third-generation interactions. Consequently, the decay $K \to \pi a$ emerges as the leading probe, as it directly accesses the flavor-violating couplings that are otherwise only weakly constrained.

\subsection{ALP Phenomenology}
\begin{figure}[t]
  \centering
  \includegraphics[width=0.9\textwidth]{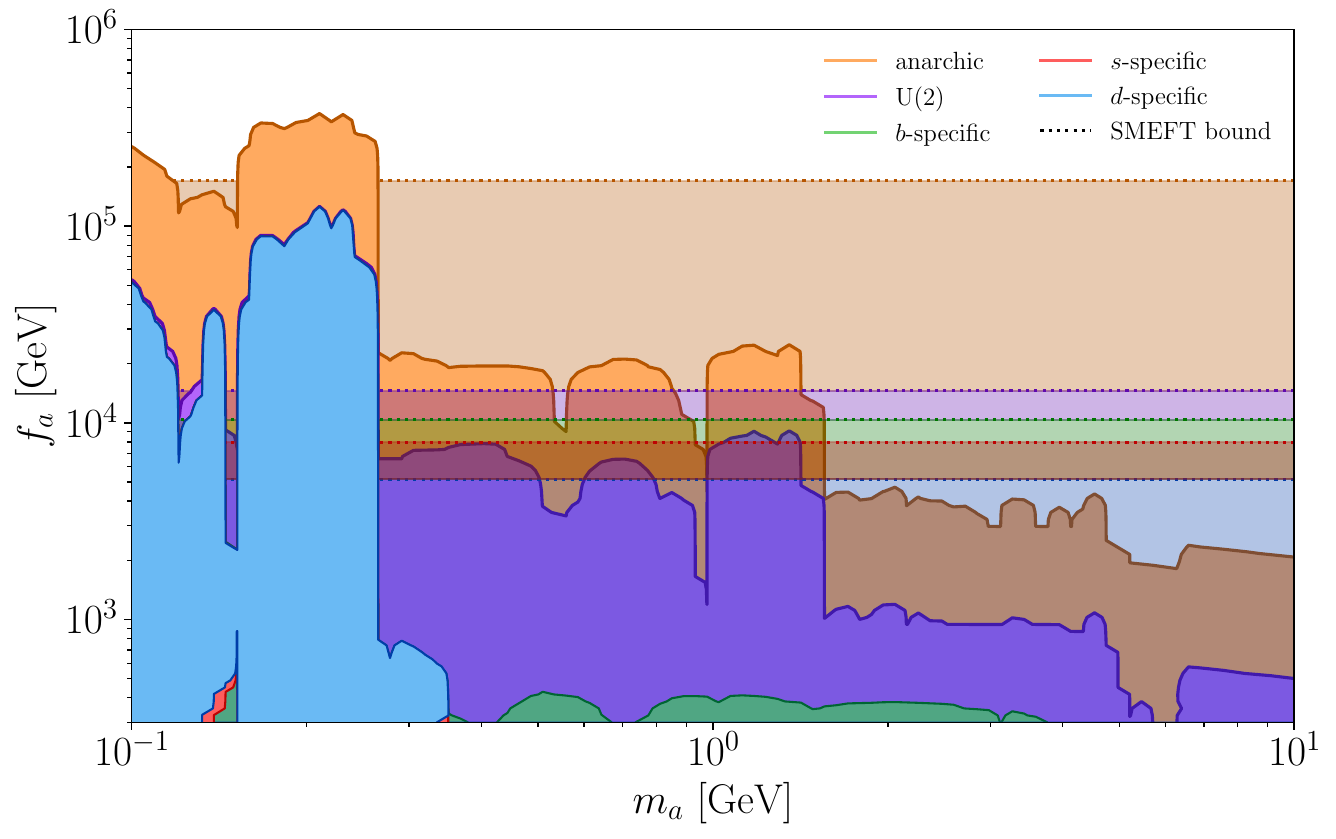}
  \caption{Global overview of the ALP parameter space in the $(m_a,f_a)$ plane for the representative $D$-type VLF scenario. The colored regions show the exclusions obtained under different assumptions for the flavor structure of the ALP couplings, while the horizontal dotted line denotes the corresponding constraint derived from the SMEFT analysis.}
  \label{fig:global_D_ALP_sec5_sum}
\end{figure}

While the QCD axion provides a particularly simple realization of the framework, much of the phenomenological richness emerges once the relation between the axion mass and decay constant is relaxed. We therefore now turn to the more general ALP case, where the mass and decay constant can be treated as independent parameters and a wider range of experimental signatures becomes accessible.

In order to connect these considerations to phenomenology in a more quantitative manner, it is not sufficient to specify only the production mechanisms of the ALP. The experimental signature also depends crucially on how the ALP decays, and in particular on whether it decays promptly, with a displaced vertex, or escapes the detector before decaying. In the framework developed here, the relevant production channels are induced by the effective ALP couplings to fermions, while the observable final states are determined by the ALP decay pattern.

The dominant decay modes depend sensitively on the ALP mass and on the structure of its effective couplings~\cite{MartinCamalich:2020dfe,Bauer:2021mvw,Alda:2025nsz,Alda:2025uwo}. For sufficiently light ALPs, the only kinematically accessible final states are typically photons and light leptons, while hadronic modes become increasingly important once the ALP mass exceeds the corresponding hadronic thresholds. This feature is particularly relevant in the present framework, as all scenarios considered here correspond to hadronic ALP models possessing an anomalous coupling to QCD. As a result, hadronic decays play a central role in determining both the ALP lifetime and its branching-ratio pattern. In recent years, significant effort has been devoted to modeling these decays, including the challenging intermediate region between the domains of validity of Chiral Perturbation Theory (ChPT) and perturbative QCD~\cite{Aloni:2018vki,Bai:2024lpq,Ovchynnikov:2025gpx,Bai:2025fvl,Balkin:2025enj}. At higher masses, decays into heavier fermions and multi-hadron final states may become dominant, depending on the relative size of the fermionic and anomaly-induced gauge-boson couplings.

For the KSVZ-like models considered in this work, ALP decays are typically dominated by anomaly-induced couplings to gluons. This follows from the structure of the fermionic interactions: in the ALP basis, the leading ALP-fermion couplings arise from dimension-seven operators and are therefore suppressed by an additional factor of order $v^2/m_\Psi^2$. By contrast, the production of KSVZ ALPs in meson decays is especially sensitive to direct ALP couplings to fermions, in particular when these couplings are flavor violating. As a result, gluon-induced interactions control the dominant decay patterns, while flavor-changing fermionic couplings provide the most powerful probes in rare meson production channels. Notably, universal couplings can also be tested in radiative quarkonia decays, see Refs.~\cite{DiLuzio:2024jip,Merlo:2019anv}. The resulting phenomenology is therefore governed by the interplay between production through flavor-changing fermionic couplings and decay through the anomaly-induced gluonic interactions. Consequently, the same underlying ALP production process can lead to qualitatively different experimental signatures across the ALP mass range. For the phenomenological study in this section, we have used the \texttt{ALP-aca} package~\cite{Alda:2025nsz}, which consistently incorporates running effects, as well as computes the relevant bounds considering all the aforementioned details.

Before analyzing representative benchmark scenarios associated with specific flavor transitions, it is useful to first obtain a broader overview of the ALP parameter space predicted by the framework. In Figure~\ref{fig:global_D_ALP_sec5_sum} we present the constraints in the $(m_a,f_a)$ plane for the representative $D$-type VLF scenario regarding the flavor structure of the underlying ALP couplings. In particular, we consider fully anarchic couplings, flavor-aligned scenarios associated with individual down-type quarks, as well as a $\U(2)$-inspired structure in which the third generation carries the dominant coupling, while the first two generations enter only through CKM-suppressed effects~\cite{Allwicher:2023shc,Greljo:2022cah,Faroughy:2020ina,Greljo:2025mwj}. The resulting exclusion patterns demonstrate that the phenomenology is strongly shaped by the flavor structure of the underlying couplings. More importantly, the comparison with the SMEFT constraint reveals regions where indirect flavor information competes with or surpasses direct ALP searches, motivating more detailed benchmark studies of the $b\to s$ and $s\to d$ sectors presented below.
Similar conclusions are found in FN ALPs~\cite{Greljo:2024evt}.

Furthermore, the low-mass ALP region is predominantly constrained by bounds from the $s\to d$ sector, while for masses above the kaon threshold the strongest limits typically arise from $b\to s$ transitions. Radiative quarkonium decays also provide relevant probes in scenarios with sizable couplings to the $b$ quark~\cite{DiLuzio:2024jip}. Since the $s\to d$ and $b\to s$ sectors give the most sensitive probes in these two mass regions, we investigate their phenomenology in more detail in the following subsections.

\begin{figure}[t]
  \centering
  \includegraphics[width=\textwidth]{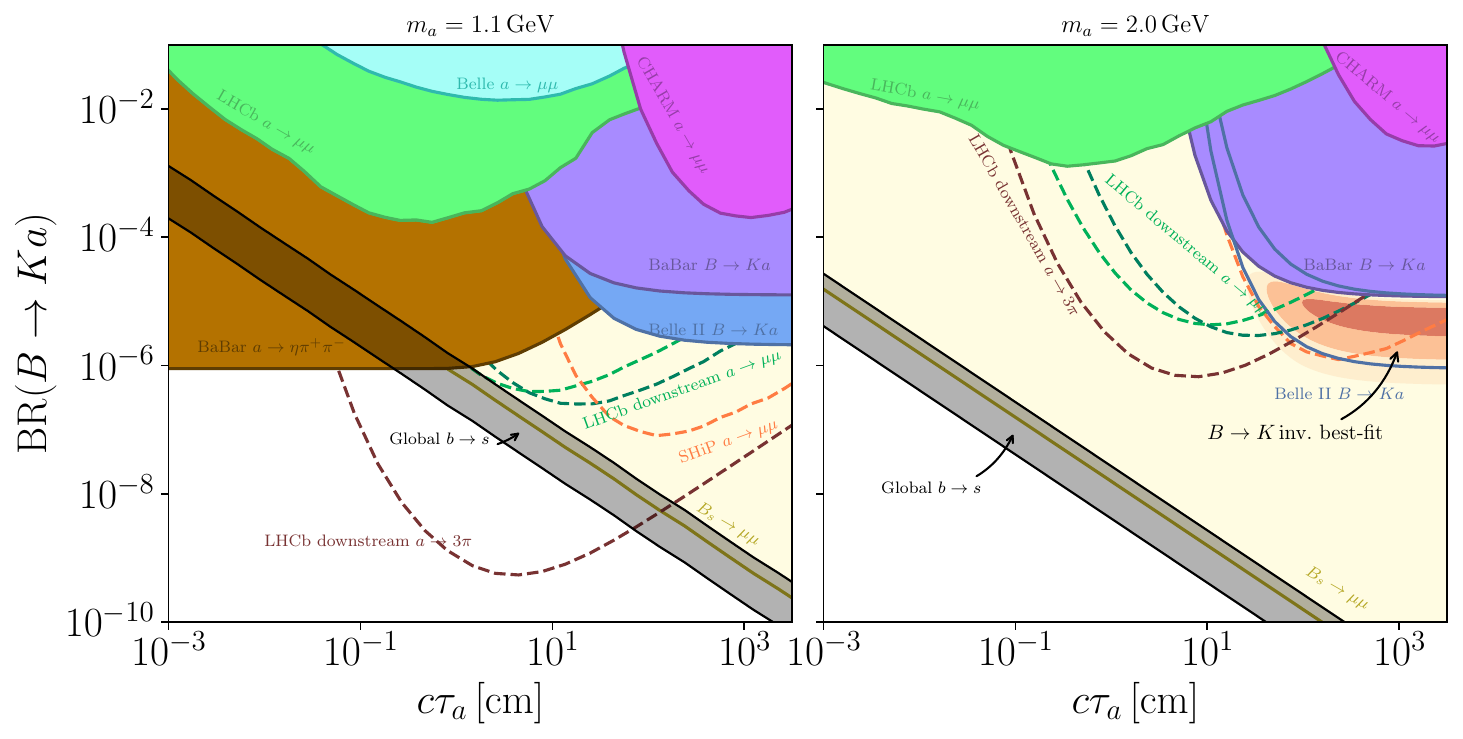}
  \caption{Constraints in the $(c\tau_a,\mathrm{BR}(B\to Ka))$ plane for the $D$-type VLF scenario with $(ij)=(23)$, shown for the benchmark masses $m_a=1.1\,\gev$ (left) and $m_a=2.0\,\gev$ (right). The colored regions indicate current and projected exclusions from direct ALP searches, while the gray diagonal band corresponds to the constraint obtained from the global $b\to s$ SMEFT analysis translated into the ALP parameter space through the unified SMEFT–ALP framework.}
  \label{fig:lifetime_plot}
\end{figure}

\subsubsection{Benchmark Analysis of $b\to s$ Sector}
Let us first illustrate this interplay between production and decay in the $b\to s$ sector, which provides one of the most sensitive probes of flavor-violating ALP interactions. We once again focus on the representative $D$-type VLF scenario and consider the $(ij)=(23)$ flavor structure, for which the same UV parameters induce both the SMEFT operators constrained by rare $B$-meson observables and the ALP coupling responsible for $B\to K a$.

In order to simultaneously capture the production and decay properties of the ALP, we first present the results in the $(c\tau_a,\, \mathrm{BR}(B\to Ka))$ plane, shown in Figure~\ref{fig:lifetime_plot}. This representation proves to be particularly useful because it directly connects the ALP lifetime, which determines the nature of the decay (prompt, displaced, or effectively invisible), with the production rate governing the sensitivity of flavor experiments. The two benchmark masses, $m_a=1.1\,\gev$ and $m_a=2.0\,\gev$, illustrate different decay regimes and therefore different experimental reach as well as potential search strategies. Their phenomenological implications are discussed separately in the following.

\paragraph{Benchmark I.} For the lighter benchmark, $m_a=1.1\,\gev$, the translated global $b\to s$ constraint occupies a diagonal band, stemming from the scaling $\textrm{Br}(B\to K a) \sim v^4 (\mathscr{C}_{32}^D)^2 /f_a^2$, spanning several orders of magnitude in lifetime. A notable feature is that a sizable fraction of this region lies below the current direct ALP limits, indicating that the indirect SMEFT information already probes parameter space beyond the reach of existing searches.

The origin of the present exclusion pattern can be traced to the ALP decay channels available in this mass range. In particular, for $m_a \gtrsim 3m_\pi$, hadronic decays become kinematically accessible, with the dominant modes initially involving three-pion final states, such as $a \to \pi^0(\gamma)\pi^+\pi^-$, followed at higher masses by channels such as $a \to \eta \pi^+\pi^-$. In the scenario under consideration, these hadronic final states currently provide the strongest direct constraints.

\begin{figure}[t]
  \centering
  \includegraphics[width=\textwidth]{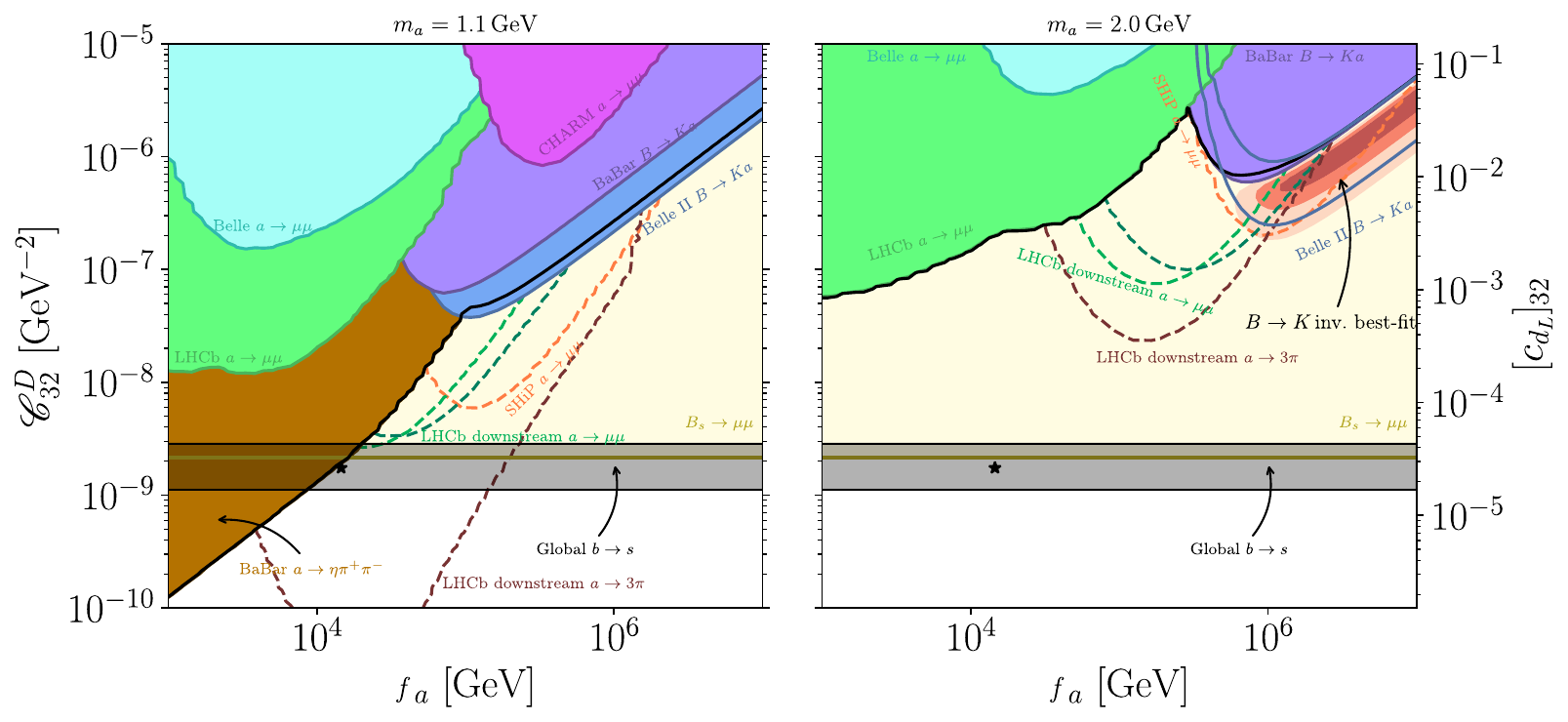}
  \caption{Constraints in the $(f_a,\mathscr{C}^D_{32})$ plane for the $D$-type VLF scenario with flavor structure $(ij)=(23)$, shown for the benchmark masses $m_a=1.1\,\gev$ (left) and $m_a=2.0\,\gev$ (right). The colored regions correspond to the direct ALP-search constraints displayed in Figure~\ref{fig:lifetime_plot}, translated into the underlying EFT parameter space using the relations between the ALP lifetime, branching ratios, and effective couplings. The gray horizontal band denotes the constraint obtained from the global $b\to s$ SMEFT fit.}
  \label{fig:global_comb_alp_plot}
\end{figure}

By contrast, decays into muons, which often constitute one of the most sensitive probes of long-lived ALPs~\cite{Bauer:2021mvw,Alda:2025uwo}, do not yet reach the region selected by the SMEFT analysis. This can be understood from the structure of the effective interactions: the ALP coupling to muons is generated only radiatively, originating from the ALP coupling to top quarks, which itself arises through RGE effects~\cite{Bauer:2020jbp,DasBakshi:2023lca,Chala:2020wvs}. As a result, the corresponding branching ratio is suppressed by several orders of magnitude despite the superior experimental sensitivity associated with muonic final states. At the same time, the figure highlights a strong complementarity between prompt, displaced, and invisible search strategies, which collectively probe different portions of the parameter space. Future facilities are expected to significantly improve the sensitivity in the long-lifetime regime~\cite{AldaImplications25}, approaching the region selected by the global flavor fit.

\paragraph{Benchmark II.} For the case of the heavier benchmark, $m_a=2.0\,\gev$, the modified decay pattern alters both the ALP lifetime and the relative sensitivity of the different search channels, leading to a visibly different exclusion landscape. While the global $b\to s$ constraint remains largely unchanged, as it originates from the underlying flavor structure rather than the ALP decay properties, the direct limits become concentrated in different regions of the $(c\tau_a,\mathrm{BR}(B\to Ka))$ plane.

This behavior can be understood from the opening of additional hadronic decay channels at larger ALP masses. These channels increase the total decay width, resulting in a shorter ALP lifetime, while simultaneously reducing the branching fractions into lighter hadronic final states such as $a\to3\pi$, which become subdominant as heavier hadronic modes start to dominate. As a consequence, the experimental sensitivity is redistributed across the parameter space, leaving portions of this mass region comparatively less explored. Nevertheless, future dedicated searches are expected to substantially improve the coverage and probe a significant fraction of the currently unconstrained parameter space~\cite{Gorkavenko:2023nbk}.

This benchmark is also particularly interesting in light of the mild excess recently reported by Belle II in $B^+\to K^+ + \mathrm{inv.}$, which can be interpreted in terms of a light and sufficiently long-lived ALP~\cite{Craik:2022riw,Altmannshofer:2023hkn,Alda:2025uwo}. Although the significance of this deviation remains limited and its interpretation requires further scrutiny, it provides a useful case study for illustrating the framework developed in this work. In particular, the SMEFT--ALP map allows us to directly assess whether the ALP interpretation of the excess is compatible with the indirect constraints implied by the same underlying flavor structure. We find that accommodating the excess requires an ALP with a sufficiently long lifetime, controlled by the same flavor-changing coupling that generates the SMEFT contribution to $B_s\to\mu^+\mu^-$. The corresponding bound from $B_s\to\mu^+\mu^-$ therefore severely restricts, and for the benchmark considered here excludes, the parameter region favored by the Belle II excess. Lastly, this example neatly illustrates the complementarity between direct ALP searches and indirect SMEFT probes, and highlights the predictive power of the unified SMEFT–ALP framework developed in this work.

\vspace{+0.3cm}
\noindent
Additionally, while Figure~\ref{fig:lifetime_plot} provides a representation of the bounds in terms of the ALP lifetime and production rate, clearer for experimental targets, it is also instructive to recast the same information directly in terms of the parameters appearing in the effective Lagrangian, which is more representative in terms of the theoretical landscape. Figure~\ref{fig:global_comb_alp_plot} presents the corresponding constraints in the $(f_a,\, \mathscr{C}_{32}^D)$ plane for the same benchmark masses. The various exclusion regions are obtained by translating the bounds shown in Figure~\ref{fig:lifetime_plot} through the relations between the ALP lifetime, branching ratios, and the underlying effective couplings, thereby providing an equivalent EFT-oriented interpretation of the results.

When comparing Figures~\ref{fig:lifetime_plot} and \ref{fig:global_comb_alp_plot}, it is important to note that the mapping between the two parameter spaces is governed by a non-trivial rescaling. Prompt decay searches, such as those targeting $a\to \eta \pi^- \pi^+$, are primarily sensitive to relatively small values of $f_a$, whereas invisible ALP searches require larger decay constants in order for the ALP to be sufficiently long lived. Despite probing different lifetime regimes, both classes of observables rely on the same underlying production mechanism, governed by the combination $|\mathscr{C}^D_{32}|/f_a$. Consequently, the exclusion regions in the $(c\tau_a,\mathrm{BR}(B\to Ka))$ plane do not translate into simple rescalings in the $(f_a,\, \mathscr{C}_{32}^D)$ plane, leading to the qualitatively different structure observed in Figure~\ref{fig:global_comb_alp_plot}.

This observation also makes it possible to understand, directly from Figure~\ref{fig:global_comb_alp_plot}, why the ALP interpretation of the Belle II excess is disfavored in the present framework. In order for the ALP to remain sufficiently long lived and contribute to the $B^+\to K^+ + \mathrm{inv.}$ signature, hadronic decay searches impose a lower bound on the decay constant of roughly $f_a \gtrsim 10^6\,\gev$. At the same time, the production rate is controlled by the same flavor-changing coupling that contributes to $B_s\to\mu^+\mu^-$, leading to the constraint $\mathscr C^D_{32}\lesssim 2\times10^{-3}\,\tev^{-2}$. Taken together, these requirements exclude the region of parameter space needed to account for the Belle II excess. This example provides a particularly transparent illustration of the predictive power of the unified SMEFT--ALP framework: direct ALP searches and indirect flavor constraints probe complementary manifestations of the same underlying UV dynamics, and their combination can exclude scenarios that would otherwise appear viable when considered in isolation.

Regarding this anomaly, one can scrutinize more models using our framework, for instance the $D$-KSVZ with $(\chi_L,\, \chi_R)=(0, \, -1)$, where the relation between the SMEFT coefficient and the ALP is given by
  \begin{equation}
    [c_{d_R}]_{32}=[C_{D,R}^{(2)}]_{32} = - m_D^2 y_b^{-1} \mathscr{C}_{32}^D y_s^{-1} \, .
  \end{equation}
  If we assume that the $B^+\to K^+ a$ deviation would be generated by an ALP of $m_a=2\,$GeV we can fit the coefficient as $|C_{D,R}^{(2)}|/f_a \approx 4\times 10^{\eminus8}\,$GeV for $f_a \gtrsim 10^6\,$GeV in order for the ALP to be sufficiently long-lived. Using this value and substituting in the previous expression we get
  \begin{equation}
    \left(\frac{f_a }{4\times 10^{8} \, \textrm{GeV}}\right)^{1/2} = \frac{m_D}{115 \, \textrm{GeV}} \left(\frac{\mathscr{C}^D_{32}}{10^{-3}\, \textrm{TeV}^{-2}}\right)^{1/2} \, .
  \end{equation}
This expression makes the tension transparent. For values of $f_a$ close to the minimum required for invisibility, the matching would require a very light vector-like fermion, outside the regime in which the EFT description is appropriate and in conflict with direct searches. Conversely, imposing a TeV-scale VLF mass pushes the required decay constant to much larger values, while the production rate remains controlled by the same flavor-changing coefficient $\mathscr C^D_{32}$ constrained by $B_s\to\mu^+\mu^-$. Thus, even in this mass-mixing benchmark, where the direct ALP coupling is enhanced by inverse Yukawa factors, the Belle II interpretation is highly restricted once the correlated SMEFT constraint is imposed.

\subsubsection{Benchmark Analysis of $s\to d$ Sector}
\begin{figure}[t]
  \centering
  \includegraphics[width=\textwidth]{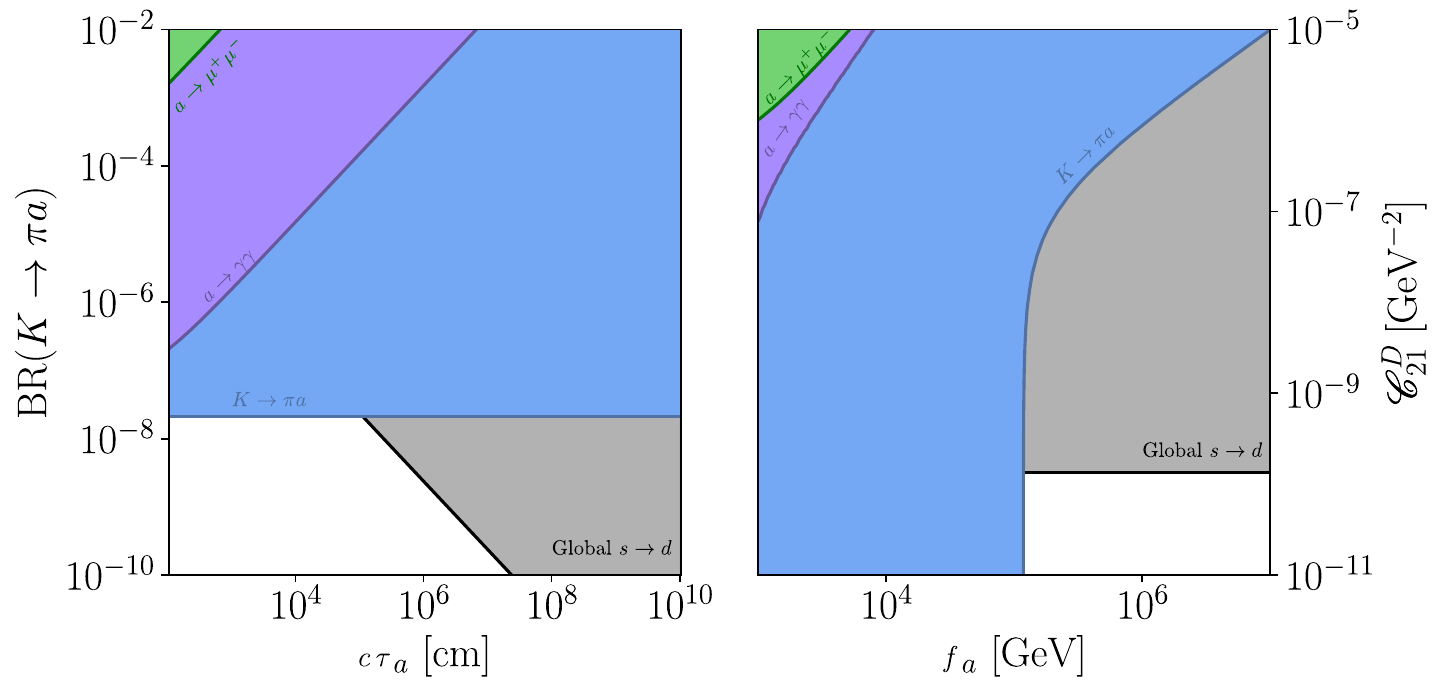}
  \caption{Constraints in the $(c\tau_a,\mathrm{BR}(K\to\pi a))$ plane (left) and the corresponding EFT parameter space $(f_a,\mathscr C_{21}^D)$ (right) for the $s\to d$ benchmark with $m_a=230\,\mev$. The colored regions indicate the current limits from direct ALP searches in different decay channels, while the gray regions correspond to the translated constraint obtained from the global $s\to d$ SMEFT analysis.}
  \label{fig:ALP_s_d_sector}
\end{figure}
While the $b\to s$ sector discussed above highlights the role of $B$-physics observables in constraining flavor-violating ALP interactions, it does not exhaust the phenomenological possibilities offered by the framework. Another particularly interesting benchmark is provided by the $s\to d$ transition, where the combination of sensitive kaon measurements and dedicated searches for rare kaon decays allows one to probe flavor-violating ALP interactions in a qualitatively different manner.

To illustrate these features, we consider a benchmark ALP mass $m_a=230\,\mev$, which lies in a region where rare kaon decays provide strong sensitivity. The resulting constraints are shown in Figure~\ref{fig:ALP_s_d_sector}, both in the $(c\tau_a,\mathrm{BR}(K\to\pi a))$ plane and in the corresponding EFT representation in terms of $(f_a,\mathscr C_{21}^D)$. As in the previous example, the framework allows the indirect SMEFT constraint obtained from the global $s\to d$ analysis to be translated directly into the ALP parameter space and compared with dedicated ALP searches.

In contrast to the $b\to s$ benchmark, the phenomenology is now dominated by kaon probes. The exceptional sensitivity of rare kaon decay searches shifts the phenomenological emphasis toward direct probes, which become comparable to or stronger than the translated SMEFT constraint over large regions of parameter space. Moreover, at this mass point hadronic decay channels are kinematically inaccessible. As a result, the ALP lifetime is significantly enhanced, (see left plot in Fig.~\ref{fig:ALP_s_d_sector}), causing it to escape the detector and appear invisible to low-energy probes.
Consequently, the $s\to d$ sector provides an important example of how flavor-violating ALP production can be probed directly through rare meson decays, while, at the same time, the comparison with the global $s\to d$ constraint once again demonstrates that both approaches can be consistently interpreted within the same SMEFT–ALP framework. 

\section{Conclusions}
\label{sec:Conclusions}
In this work, we developed a unified SMEFT–ALP framework for a broad class of KSVZ-like models containing vector-like fermions and a spontaneously broken Peccei–Quinn symmetry. The framework is applicable to arbitrary VLF representations and PQ-charge assignments that admit linear couplings to SM fermions, and systematically captures the correlated effective interactions generated in the SMEFT and ALP sectors. This provides a direct connection between low-energy ALP phenomenology and the precision, Higgs, and flavor observables encoded in the SMEFT description.

Building on this framework, we first performed a comprehensive phenomenological analysis of the SMEFT operators generated by the different VLF representations. Using global fits to electroweak precision, Higgs, and flavor observables, we derived robust constraints on the corresponding Wilson coefficients and identified the dominant observables across the various flavor structures. The resulting bounds can subsequently be translated into the ALP parameter space, allowing indirect constraints from precision measurements to be directly compared with dedicated ALP searches.

A central result of our analysis is that the common UV origin of the SMEFT and ALP interactions leads to non-trivial phenomenological correlations that would be missed in separate treatments of the two sectors. We showed that, over large regions of parameter space, indirect constraints derived from the SMEFT analysis provide sensitivity that is competitive with, and often stronger than, current direct QCD-axion and ALP searches. An important exception arises in scenarios where the PQ-charge assignments permit mass mixing with SM fermions, for which flavor-changing meson decays such as $K\to\pi a$ and $B\to Ka$ become particularly powerful probes of the QCD-axion sector.

To illustrate the predictive power of our framework, we investigated several representative examples of the correlated SMEFT–ALP phenomenology. In particular, we studied flavor-violating ALP production in rare $B$-meson decays and demonstrated how direct ALP searches and indirect flavor constraints probe complementary aspects of the same underlying dynamics. We further showed that the unified SMEFT–ALP map provides a powerful tool for assessing proposed ALP interpretations of experimental anomalies. As an example, we considered the recent Belle II excess in $B^+\to K^+ + \mathrm{inv.}$ and found that the parameter region required to explain the excess is excluded by the corresponding flavor constraints arising from the same UV structure.

Overall, our results demonstrate that KSVZ-like models give rise to a rich phenomenology extending well beyond the ALP sector itself. The framework developed in this work provides a systematic bridge between UV completions, SMEFT analyses, and ALP searches, enabling precision, flavor, collider, and ALP observables to be interpreted within a common theoretical setting. Importantly, its utility extends beyond the derivation of bounds: should evidence for an ALP emerge in future experiments, the SMEFT–ALP map developed here would provide a direct route for interpreting such a signal in terms of the underlying UV dynamics and for identifying correlated signatures across complementary observables. As the sensitivity of both precision measurements and dedicated ALP experiments continues to improve, the correlated SMEFT–ALP approach presented here offers an increasingly compelling framework for probing the underlying UV dynamics.

Several natural extensions of this work deserve further investigation. First, although we have focused on KSVZ-like models with colored vector-like fermions and renormalizable interactions with SM quarks, the matching strategy developed here is more general and can be applied to other classes of axion and ALP completions, including leptonic or electroweak vector-like fermions, models with several mediator multiplets, and scenarios in which the SM fields themselves carry non-trivial charges under the spontaneously broken symmetry. Such extensions would allow one to study more general flavor textures, complex CP-violating structures, and correlations involving charged-lepton, Higgs, and collider observables. Second, it would be important to go beyond the leading tree-level matching considered in this work by performing a systematic one-loop matching of the full UV theory onto the SMEFT and ALP EFT. This would provide a complete treatment of loop-induced four-fermion operators, dipole interactions, anomalous gauge couplings, and SMEFT--ALP operators, and would further sharpen the connection between precision observables and direct ALP searches. Together with future improvements in global SMEFT fits and dedicated ALP searches, these developments would turn the SMEFT--ALP map into an increasingly powerful tool for diagnosing the UV origin of possible deviations from the Standard Model.

\section*{Acknowledgements}
We thank Admir Greljo and Aleks Smolkovi\v c for useful discussions. The work of AP is supported by MCIU/AEI/10.13039/501100011033 (grants CEX2023-001292-S and PID2023-146220NB-I00).

\clearpage
\appendix

\section{Generalized Matching Framework}
\label{app:gen_match_framework}
In this section, we derive a fully general matching of a VLF theory containing a pNGB onto the SMEFT and the ALP EFT. The most general Lagrangian including all renormalizable interactions is given by
\begin{equation}\label{eq:general_VLF_PQ}
  {\scalebox{0.98}{
      $
      \begin{alignedat}{2}
        \mathcal L_{\sscript{int}}^{\sscript{gen}}&=
        \overline{\Psi}_L\,\left[M_\Psi \delta_{\chi_{\Psi_R},\chi_{\Psi_L}}+\delta_{\chi_{\Psi_R},\chi_{\Psi_L}-1}Y_\Psi\phi\right]\,\Psi_R\,
        +\overline{f}_L\,\mathcal M_{f\Psi}(\phi,\phi^\ast)\,\Psi_R
        +\overline{\Psi}_L\,\mathcal M_{\Psi f}(\phi,\phi^\ast)\,f_R
        \\&
        ~~+\overline{f}_L\,\lambda^\varphi_{f\Psi}\,\Delta^{\varphi,I}_{f\Psi}\,\varphi\,\Psi_R^I
        +\overline{\Psi}_L\,\lambda^\varphi_{\Psi f}\,\Delta^\varphi_{\Psi f}\,\varphi\,f_R
        +\text{h.c.}\,.
      \end{alignedat}$
  }}
\end{equation}
The first term in Eq.~\eqref{eq:general_VLF_PQ} corresponds to the two possible mass terms for the VLFs: a purely vector-like mass term and a chiral contribution arising from the PQ symmetry after spontaneous symmetry breaking (SSB).\footnote{Terms involving $\phi^\ast$, $\phi^2$, and higher powers can be reabsorbed into the operators shown above through a redefinition of the couplings, resulting only in an overall rescaling of the corresponding coefficients.} The Lagrangian also allows for multiple generations of VLFs. The second term contains mixing interactions between the SM fermions and the VLFs. Depending on the PQ charge assignment, these couplings may involve the scalar field $\phi$, and after SSB provide an additional source of mixing. In general, these interactions can be written as
\begin{align}
  \mathcal M_{f\Psi}(\phi,\phi^\ast)
  &\equiv
  M_{f\Psi}\,\Delta^M_{f\Psi}
  +Y_{f\Psi}\,\phi\,\Delta^\phi_{f\Psi}
  +\widetilde Y_{f\Psi}\,\phi^\ast\,\widetilde\Delta^\phi_{f\Psi}\,,
  \\[2pt]
  \mathcal M_{\Psi f}(\phi,\phi^\ast)
  &\equiv
  M_{\Psi f}\,\Delta^M_{\Psi f}
  +Y_{\Psi f}\,\phi\,\Delta^\phi_{\Psi f}
  +\widetilde Y_{\Psi f}\,\phi^\ast\,\widetilde\Delta^\phi_{\Psi f}\,.
\end{align}
The $\Delta^X_{YZ}$ factors denote generalized Kronecker delta functions that enforce the compatibility of the VLF with the quantum numbers of the fermions under $\textrm{G}_\sscript{SM}\times \U(1)_{\sscript{PQ}}$. Explicitly, they can be written as
\begin{align}
  \Delta^{M}_{f\Psi}&=\delta_{R(f_L),R(\Psi)}\,\delta_{\chi_{f_L},\,\chi_{\Psi_R}}\,,
  &\quad
  \Delta^{M}_{\Psi f}&=\delta_{R(\Psi),R(f_R)}\,\delta_{\chi_{\Psi_L},\,\chi_{f_R}}\,,
  \\
  \Delta^{\phi}_{f\Psi}&=\delta_{R(f_L),R(\Psi)}\,\delta_{\chi_{f_L},\,\chi_{\Psi_R}+1}\,,
  &\quad
  \Delta^{\phi}_{\Psi f}&=\delta_{R(\Psi),R(f_R)}\,\delta_{\chi_{f_R},\,\chi_{\Psi_L}-1}\,,
  \\
  \widetilde{\Delta}^{\phi}_{f\Psi}&=\delta_{R(f_L),R(\Psi)}\,\delta_{\chi_{f_L},\,\chi_{\Psi_R}-1}\,,
  &\quad
  \widetilde{\Delta}^{\phi}_{\Psi f}&=\delta_{R(\Psi),R(f_R)}\,\delta_{\chi_{f_R},\,\chi_{\Psi_L}+1}\,.
\end{align}
In these expressions, the subscript $R$ in the first Kronecker delta labels the SM representation of the corresponding field, and we normalize the PQ charge of the singlet field $\phi$ to $\chi_\phi = 1$. A given mixing term is allowed only if the VLF matches the gauge representation of the corresponding SM fermion, a condition enforced by the first Kronecker delta. The second Kronecker delta ensures compatibility of the PQ charges between SM fermions and VLFs. The three terms are mutually exclusive, so that for a given model only one of them is present. Finally, Eq.~\eqref{eq:general_VLF_PQ} contains an additional term describing mixing with the Higgs field, where $\varphi = H,\, \tilde{H}$. The corresponding Kronecker delta functions are defined analogously as
\begin{align}
  \Delta^{\varphi,I}_{f\Psi}
  &=
  \delta_{\mathbf c_\Psi,\mathbf c_f}\,
  \delta_{\chi_{\Psi_R},\chi_{f_L}}\,
  \delta_{y_\Psi,\,y_{f_L}-y_\varphi}\,
  \Big(\delta_{\boldsymbol\ell_\Psi,\mathbf 1}+\sigma^I\delta_{\boldsymbol\ell_\Psi,\mathbf 3}\Big)\,,\\
  \Delta^{\varphi}_{\Psi f}
  &=
  \delta_{\mathbf c_\Psi,\mathbf c_f}\,
  \delta_{\chi_{\Psi_L},\chi_{f_R}}\,
  \delta_{y_{f_R},\,y_\Psi-y_\varphi}\,
  \delta_{\boldsymbol\ell_\Psi,\mathbf 2}\,,
\end{align}
where the representations under $\SU(3)_{\sscript{C}}$ and $\SU(2)_{\sscript{L}}$ are denoted in boldface as $\bm{\mathrm{c}}$ and $\bm{\ell}$, respectively. In this case, since all left-handed SM fermions $f_L$ transform as doublets under $\SU(2)_{\sscript{L}}$, the first type of coupling allows the VLF to be either a singlet or a triplet. Conversely, the second type of coupling requires the VLF to transform as a doublet under $\SU(2)_{\sscript{L}}$.

The general framework outlined above can be applied to two distinct regimes, depending on the hierarchy between the PQ-breaking scale and the VLF mass: the KSVZ limit ($v_\phi > M_\Psi$), and the Froggatt–Nielsen limit ($v_\phi < M_\Psi$). In the former case, the VLFs and the PQ singlet can be integrated out simultaneously, yielding a direct matching onto the SMEFT and the ALP EFT. In the latter case, the matching proceeds in two steps: first, the VLFs are integrated out, resulting in an intermediate EFT in which the scalar field remains dynamical (which we refer to as the FN basis). Subsequently, the PQ singlet acquires a vacuum expectation value and is integrated out, completing the matching onto the SMEFT. Below, we apply this framework to KSVZ-like models, while the extension to Froggatt–Nielsen scenarios proceeds analogously.

\subsubsection*{Matching for KSVZ Models}
\label{app:Match_KSVZ}
Consistency of the matching requires the VLF mass to be set by the PQ-breaking scale. This determines the structure of the first term in Eq.~\eqref{eq:general_VLF_PQ}, and after spontaneous breaking of the PQ symmetry, the VLF mass term reads
\begin{equation}
  \bm{M}_\Psi = \frac{v_\Phi}{\sqrt2}Y_\Psi\,.
\end{equation}
The Lagrangian with a global $\U(1)_{\sscript{PQ}}$ symmetry can then be written, after making a chiral field redefinition to remove axions from the mass terms as
\begin{equation}
  {\scalebox{0.95}{$
      \begin{alignedat}{2}
        \mathcal L_{\Psi}&=
        \bar\Psi\, i \slashed D\, \Psi
        -\, \bar\Psi \left(\bm{M}_\Psi P_R+ \bm{M}_\Psi^\dagger P_L\right)\Psi
        +\bar f \Big(\bm{M}_{f\Psi} P_R+\bm{M}_{\Psi f}^\dagger P_L\Big)\Psi
        +\bar\Psi \Big(\bm{M}_{f\Psi}^\dagger P_L+\bm{M}_{\Psi f} P_R\Big)f
        \\[2pt]
        &~~+\bar f \Big(\bm{\lambda}^\varphi_{f\Psi}\,\varphi\, P_R+\bm{\lambda}_{\Psi f}^{\varphi\,\dagger}\,\varphi^\dagger P_L\Big)\Psi
        +\bar\Psi \Big(\bm{\lambda}_{f\Psi}^{\varphi\,\dagger}\,\varphi^\dagger P_L+\bm{\lambda}_{\Psi f}^{\varphi}\,\varphi\,P_R\Big)f
        \\[2pt]
        &~~+\frac{\partial_\mu a}{v_\Phi}
        \left[
          \bar f \gamma^\mu \Big(\chi_{f_L}P_L+\chi_{f_R}P_R\Big)f
          + \bar\Psi\gamma^\mu
        \left(\chi_{\Psi_L}P_L+ \chi_{\Psi_R}P_R\right)\Psi\right]\,.
      \end{alignedat}
  $}}
\end{equation}
For convenience, we introduce the boldface couplings as
\begin{equation}
  \bm{M}_{f\Psi}\equiv \mathcal{M}_{f\Psi}(\phi,\phi^\ast)\Big|_{\phi=\phi^\ast=v_\phi/\sqrt{2}}\,,\qquad
  \bm{\lambda}_{f\Psi}^\varphi \equiv \lambda^\varphi_{f\Psi}\,\Delta^{\varphi}_{f\Psi}\,,
\end{equation}
and analogously for the $\bm{M}_{\Psi f},\, \bm{\lambda}_{\Psi f}^\varphi$ couplings. From now on, we drop possible $I$ indices in the couplings. The mixing terms can be eliminated through the following field redefinitions
\begin{align}
  f_R &\to \cR_{f_R} f_R+  \bm{M}_{\Psi f}^\dagger\bm{M}_{\Psi}^{-1}\cR_{\Psi_R}  \Psi_R\,,
  &\quad
  f_L&\to \cR_{f_L} f_L+  \bm{M}_{f \Psi}\bm{M}_{\Psi}^{-1}\cR_{\Psi_L}  \Psi_L\,,
  \\
  \Psi_R &\to \cR_{\Psi_R} \Psi_R-\bm{M}_\Psi^{-1}\bm{M}_{\Psi f} \cR_{f_R} f_R \,,
  &\quad
  \Psi_L&\to\cR_{\Psi_L} \Psi_L-\bm{M}_\Psi^{-1}\bm{M}_{f \Psi}^\dagger \cR_{f_L} f_L\,,
\end{align}
with the kinetic terms canonically normalized using
\begin{align}
  \cR_{f_{R}} &\equiv \left(\mathbf{1}+\bm{M}_{ \Psi f}^\dagger \bm{M}_{\psi}^{\dagger\,-1}\bm{M}_{\Psi}^{-1}\bm{M}_{\Psi f}\right)^{\eminus\frac{1}{2}}\,,
  ~~&
  \cR_{f_{L}} &\equiv \left(\mathbf{1}+\bm{M}_{f \Psi } \bm{M}_{\psi}^{\dagger\,-1}\bm{M}_{\Psi}^{-1}\bm{M}_{f \Psi }^\dagger\right)^{\eminus\frac{1}{2}}\,,
  \nonumber
  \\
  \cR_{\Psi_{R}} &\equiv \left(\mathbf{1}+ \bm{M}_{\psi}^{\dagger\,-1}\bm{M}_{ \Psi f}\bm{M}_{\Psi f}^\dagger \bm{M}_{\Psi}^{-1}\right)^{\eminus\frac{1}{2}}\,,
  &~~
  \cR_{\Psi_{L}} &\equiv \left(\mathbf{1}+\bm{M}_{\Psi}^{\dagger\,-1}\bm{M}_{f \Psi}^\dagger \bm{M}_{f\Psi }\bm{M}_{\Psi}^{-1}\right)^{\eminus\frac{1}{2}}\,.
\end{align}
As a result of these transformations, the Lagrangian is correspondingly redefined, with induced shifts in the SM Yukawa couplings, and takes the form
\begin{equation}
  {\scalebox{0.92}{$
      \begin{alignedat}{2}
        \mathcal L_{\Psi}&=
        \bar\Psi\, i \slashed D\, \Psi
        -\, \bar\Psi \left(\widetilde{\bm{M}}_\Psi P_R+ \widetilde{\bm{M}}_\Psi^\dagger P_L\right)\Psi
        +\bar f\left( \tilde{Y}_f P_R\varphi +\tilde{Y}_f^\dagger P_L\varphi^\dagger \right) f
        +\bar f \Big(\tilde{\bm{\lambda}}^\varphi_{f\Psi}\,\varphi\, P_R+\tilde{\bm{\lambda}}_{\Psi f}^{\varphi\,\dagger}\,\varphi^\dagger P_L\Big)\Psi
        \\[3pt]
        &~~+\bar\Psi \Big(\tilde{\bm{\lambda}}_{f\Psi}^{\varphi\,\dagger}\,\varphi^\dagger P_L+\tilde{\bm{\lambda}}_{\Psi f}^{\varphi}\,\varphi\,P_R\Big)f+\frac{\partial_\mu a}{v_\Phi}
        \Big[\bar\Psi\gamma^\mu\left(C_{\Psi,L}^{(1)}P_L+C_{\Psi,R}^{(1)}P_R\right)\Psi
          \\[3pt]
        &~~+\bar f \gamma^\mu \Big(C_{\Psi,L}^{(2)}P_L+C_{\Psi,L}^{(2)}P_R\Big)f+\overline{\Psi} \gamma^\mu \Big(C_{\Psi,L}^{(3)}P_L+C_{\Psi,R}^{(3)}P_R\Big)f+\textrm{h.c.}\Big]\,,
  \end{alignedat}$}}
\end{equation}
with the mass and Yukawa couplings given as
\begin{align}
  \widetilde{M}_\Psi &\equiv \cR_{\Psi_L}^\dagger \bm{M}_\Psi \cR_{\Psi_R}\,,
  &\quad
  \tilde{\bm{\lambda}}^\varphi_{f \Psi}&\equiv \cR_{f_L}\bm{\lambda}^\varphi_{f \Psi} \cR_{\Psi_R}+\cR_{f_L}Y_f\bm{M}_{\Psi f}^\dagger \bm{M}_\Psi^{-1} \cR_{\Psi_R}\,,
  \\[3pt]
  \tilde{Y}_f&\equiv\cR_{f_L} Y_f \cR_{f_R}\,,
  &\quad
  \tilde{\bm{\lambda}}^\varphi_{ \Psi f}&\equiv \cR_{\Psi_L}\bm{\lambda}^\varphi_{\Psi f}\cR_{f_R}+\cR_{\Psi_L}\bm{M}_\Psi^{\dagger \,-1}\bm{M}_{f\Psi }^\dagger Y_f \cR_{f_R}\,.
\end{align}
The crossed terms, such as $\lambda_{\Psi f}^\varphi\, \bm{M}_{\Psi f}$, are absent, as the corresponding structures are mutually exclusive. A mass-mixing term requires the VLF to transform in the same representation as the corresponding SM fermion (either left- or right-handed), whereas a Yukawa-type interaction imposes a different representation requirement. In particular, for a term of the form $\overline{\Psi}_L \varphi f_R$ to be allowed, $\Psi$ must transform as an $\SU(2)_\sscript{L}$ doublet. This forbids right-handed mixing, and consequently terms of the form $\bm{\lambda}_{\Psi f}\,\bm{M}_{\Psi f}$ do not arise.

The ALP couplings generated in this way are given by
\begin{align}
  C^{(1)}_{\Psi_{L}}&= \cR_{\Psi_{L}} \chi_{\Psi_{L}} \cR_{\Psi_{L}}+\cR_{\Psi_L}\bm{M}_{\Psi}^{\dagger \, -1}\bm{M}_{f\Psi }^\dagger\chi_{f_{L}} \bm{M}_{f\Psi } \bm{M}_{\Psi}^{-1}\cR_{\Psi_L}\,,
  \\[2.5pt]
  C^{(1)}_{\Psi_{R}}&= \cR_{\Psi_{R}} \chi_{\Psi_{R}} \cR_{\Psi_{R}}+\cR_{\Psi_R}\bm{M}_{\Psi}^{\dagger \, -1}\bm{M}_{\Psi f} \chi_{f_{R}}\bm{M}_{\Psi f}^\dagger \bm{M}_{\Psi}^{-1}\cR_{\Psi_R}\,,
  \\[2.5pt]
  C^{(2)}_{\Psi{L}}&= \cR_{f_{L}} \chi_{f_{L}} \cR_{f_{L}}+\cR_{f_L}\bm{M}_{f\Psi }\bm{M}_{\Psi}^{\dagger \, -1}  \chi_{\Psi_L}\bm{M}_{\Psi}^{-1}\bm{M}_{f\Psi }^\dagger\cR_{f_L}\, ,
  \\[2.5pt]
  C^{(2)}_{\Psi_{R}}&= \cR_{f_{R}} \chi_{f_{R}} \cR_{f_{R}}+\cR_{f_R}\bm{M}_{\Psi f}^\dagger  \bm{M}_{\Psi}^{-1}\chi_{\Psi_R}\bm{M}_{\Psi}^{\dagger \, -1}\bm{M}_{\Psi f} \cR_{f_R}\,,
  \\[2.5pt]
  C^{(3)}_{\Psi,L} &= \cR_{\Psi_L} \bm{M}_\Psi^{\dagger\, -1} \bm{M}_{\Psi f}\chi_{f_{L}} R_{f_L}-\cR_{\Psi_L} \chi_{\Psi_{L}}\bm{M}_\Psi^{ -1}\bm{M}_{f\Psi}^\dagger R_{f_L}\, ,
  \\[2.5pt]
  C^{(3)}_{\Psi,R} &= \cR_{\Psi_R} \bm{M}_\Psi^{-1} \bm{M}_{f\Psi}^\dagger  \chi_{f_{R}}R_{f_R}-\cR_{\Psi_R} \chi_{\Psi_{R}}\bm{M}_\Psi^{ -1}\bm{M}_{\Psi f} R_{f_R} \, .
\end{align}
We now turn to integrating out the heavy VLFs. The corresponding classical equations of motion are
\begin{equation}
  \Big[
    i\slashed D
    -\widetilde{\bm{M}}_\Psi P_R-\widetilde{\bm{M}}_\Psi^\dagger P_L
    +\frac{\slashed\partial a}{v_\Phi}
    \big(C_{\Psi,L}^{(1)}P_L+C_{\Psi,R}^{(1)}P_R\big)
  \Big]\Psi=-\mathcal J_\Psi\,,
\end{equation}
with
\begin{equation}\label{eq:source_JJ}
  \mathcal J_\Psi
  \equiv
  \left[
    \tilde{\bm{\lambda}}_{\Psi f}^{\varphi}\,\varphi\,P_R +\tilde{\bm{\lambda}}_{f\Psi}^{\varphi\,\dagger}\,\varphi^\dagger P_L
    +\frac{\slashed{\partial} a}{v_\Phi}
  \Big(C_{\Psi f,L}^{(3)}P_L+C_{\Psi f,R}^{(3)}P_R\Big)\right]f\,.
\end{equation}
The solution to the equations of motion for $\Psi$ can be written explicitly as
\begin{equation}
  \label{eq:Psi_solution_simplified}
  \begin{aligned}
    \Psi
    &=
    \Big[\widetilde{\bm M}_\Psi^{-1}P_R+(\widetilde{\bm M}_\Psi^\dagger)^{-1}P_L\Big]\mathcal J_\Psi+
    \widetilde{\bm M}_\Psi^{-1}
    \left[
      i\slashed D+\frac{\slashed{\partial}a}{v_\Phi}C_{\Psi,L}^{(1)}
    \right]
    (\widetilde{\bm M}_\Psi^\dagger)^{-1}P_L\,\mathcal J_\Psi
    \\[0.2cm]
    &~~+(\widetilde{\bm M}_\Psi^\dagger)^{-1}
    \left[
      i\slashed D+\frac{\slashed{\partial}a}{v_\Phi}C_{\Psi,R}^{(1)}
    \right]
    \widetilde{\bm M}_\Psi^{-1}P_R\,\mathcal J_\Psi
    +\mathcal O(\widetilde{\bm M}_\Psi^{-3}) \,.
  \end{aligned}
\end{equation}
Substituting this solution back into the Lagrangian and truncating the expansion at $\mathcal O(\widetilde{\bm M}_\Psi^{-3})$, we obtain
\begin{equation}\label{eq:Leff_generic_simplified}
  \begin{alignedat}{2}
    \mathcal L_{\sscript{eff}}
    =&-\overline{\mathcal J}_\Psi\,\widetilde{\bm M}_\Psi^{-1}P_R\,\mathcal J_\Psi-\overline{\mathcal J}_\Psi\,(\widetilde{\bm M}_\Psi^\dagger)^{-1}P_L\,\mathcal J_\Psi
    -\overline{\mathcal J}_\Psi\,\widetilde{\bm M}_\Psi^{-1}\left[i\slashed D+\frac{\slashed{\partial}a}{v_\Phi}C_{\Psi,L}^{(1)}\right](\widetilde{\bm M}_\Psi^\dagger)^{-1}P_L\,\mathcal J_\Psi
    \\&
    -\overline{\mathcal J}_\Psi\,(\widetilde{\bm M}_\Psi^\dagger)^{-1}\left[i\slashed D+\frac{\slashed{\partial}a}{v_\Phi}C_{\Psi,R}^{(1)}\right]\widetilde{\bm M}_\Psi^{-1}P_R\,\mathcal J_\Psi
    +\mathcal O(\widetilde{\bm M}_\Psi^{-3})\,.
  \end{alignedat}
\end{equation}
Finally, matching onto the SMEFT and ALP EFT is achieved by substituting $\mathcal{J}_\Psi$ from Eq.~\eqref{eq:source_JJ}. Simplifying the resulting expressions using Fierz identities and equations of motion, the SMEFT Lagrangian becomes
\begin{align}
  \mathcal{L}_{\sscript{SMEFT}} &=  \bar{f}_L \cC_{H f_L}^{(1)}\gamma^\mu f_L H^\dagger \overleftrightarrow{D}_\mu H+ \bar{f}_L \cC_{H f_L}^{(3)}\gamma^\mu \sigma^I f_L H^\dagger \overleftrightarrow{D}_\mu^I H+ \bar{f}_R \cC_{H f_R}\gamma^\mu f_R H^\dagger \overleftrightarrow{D}_\mu H \nonumber
  \\[3pt]
  &+\bar{f}_L \cC_{ f_RH}\,H f_R H^\dagger H +\textrm{h.c.} \,,
\end{align}
where terms involving products of mutually exclusive couplings vanish identically and are therefore omitted. The resulting SMEFT Wilson coefficients are given by
\begin{align}
  \cC_{H f_L}^{(1)} &=\frac{ (\delta_{\varphi,\widetilde{H}}-\delta_{\varphi,H})\textrm{dim}(\bm{\ell}_\Psi)}{4}\tilde{\bm{\lambda}}_{f\Psi}^\varphi \left(\widetilde{\bm{M}}_\Psi^{\dagger}\right)^{-1} \widetilde{\bm{M}}_\Psi^{-1}\tilde{\bm{\lambda}}_{ f\Psi}^{\varphi\, \dagger}\,,
  \\[2pt]
  \cC_{H f_R}&= \frac{(\delta_{\varphi,H}-\delta_{\varphi,\widetilde{H}})}{2}\tilde{\bm{\lambda}}_{\Psi f}^{\varphi\,\dagger}\widetilde{\bm{M}}_\Psi^{-1} \left(\widetilde{\bm{M}}_\Psi^{\dagger}\right)^{-1} \tilde{\bm{\lambda}}_{\Psi f}^\varphi \,,
  \\[2pt]
  \cC_{H f_R f_R'} &= (\delta_{\varphi,H}-\delta_{\varphi,\widetilde{H}})\tilde{\bm{\lambda}}_{\Psi f}^{\varphi\,\dagger}\widetilde{\bm{M}}_\Psi^{-1} \left(\widetilde{\bm{M}}_\Psi^{\dagger}\right)^{-1} \tilde{\bm{\lambda}}_{\Psi f'}^\varphi\,,
  \\[2pt]
  \cC_{f_R H} &= \frac{1}{2}\tilde{\bm{\lambda}}_{\Psi f}^{\varphi\,\dagger}\left(\widetilde{\bm{M}}_\Psi^{\dagger}\right)^{-1} \widetilde{\bm{M}}_\Psi^{-1} \tilde{\bm{\lambda}}_{\Psi f}^\varphi\tilde{Y}_f \,,
\end{align}
where $\cC_{H f_L}^{(1)}$ and $\cC_{H f_L}^{(3)}$ are related through the relation
\begin{equation}
  \cC_{H f_L}^{(1)}
  = (\delta_{\varphi,H}-\delta_{\varphi,\widetilde{H}})(2-\textrm{dim}(\bm{\ell}_\Psi))\textrm{dim}(\bm{\ell}_\Psi)\cC_{H f_L}^{(3)}  \,.
\end{equation}
To match onto the ALP EFT, we retain operators up to dimension seven and neglect terms involving more than one ALP field, i.e. $\mathcal{O}(f_a^{\eminus2})$. Under these assumptions, the general Lagrangian is given by
\begin{align}
  \mathcal{L}_{\sscript{ALP}}
  = \frac{\partial^\mu a}{f_a} &\left[  \bar{f}_R \cC_{\Psi,R}^{(2)} \gamma_\mu f_R+\bar{f}_L \cC_{\Psi,L}^{(2)}\gamma_\mu f_L + \bar{f}_R \cC_{a H f_R}\gamma_\mu f_R H^\dagger H +\bar{f}_L \cC^{(1)}_{a H f_L}\gamma_\mu f_L H^\dagger H \right. \nonumber
  \\ &\left.+\bar{f}_L \cC^{(3)}_{a H f_L}\gamma_\mu \sigma^I f_L H^\dagger \sigma^I H +\cC_{aDf_R} \bar{f}_{L} D^\nu\varphi \gamma_\mu \gamma_\nu f_R +\textrm{h.c.}\right] \, .
\end{align}
With the exception of $\cC^{(2)}_{\Psi, L,R}$, all Wilson coefficients are generated through the matching and are given by
\begin{align}
  \cC_{aHf_L}^{(1)} &=\frac{\textrm{dim}(\bm{\ell}_\Psi)}{2}\left[ \tilde{\bm{\lambda}}^\varphi_{f\Psi }\widetilde{\bm{M}}_\Psi^{-1}  \cC_{\Psi,L}^{(1)}\widetilde{\bm{M}}_\Psi^{\dagger\,-1} \tilde{\bm{\lambda}}^{\varphi\, \dagger}_{f\Psi } +\left( \tilde{Y}_f^\dagger\tilde{\bm{\lambda}}_{\Psi f}^{\varphi\,\dagger}\widetilde{\bm{M}}_\Psi^{\dagger\, -1}\widetilde{\bm{M}}_\Psi^{-1}C_{\Psi,L}^{(3)} +\textrm{h.c.}\right)\right] \,,
  \\[2pt]
  \cC_{aHf_R} &= \tilde{\bm{\lambda}}^{\varphi\,\dagger}_{\Psi f}\widetilde{\bm{M}}_\Psi^{\dagger\,-1}  \cC_{\Psi,R}^{(1)}\widetilde{\bm{M}}_\Psi^{-1} \tilde{\bm{\lambda}}^\varphi_{\Psi f}+\left(\tilde{Y}_f^\dagger\tilde{\bm{\lambda}}_{f\Psi }^{\varphi} \widetilde{\bm{M}}_\Psi^{-1}\widetilde{\bm{M}}_\Psi^{\dagger\, -1} C_{\Psi,R}^{(3)}  +\textrm{h.c.}\right)\, ,
  \\[2pt]
  \cC_{aD f_R}&=\tilde{\bm{\lambda}}_{f\Psi}^\varphi \widetilde{\bm{M}}_\Psi^{-1} \widetilde{\bm{M}}_\Psi^{\dagger\,-1}  C_{\Psi,R}^{(3) }\,,
\end{align}
where $\cC_{aHf_L}^{(1)}$ and $\cC_{aHf_L}^{(3)}$ are related by
\begin{equation}
  \cC_{aHf_L}^{(1)} = (\delta_{\varphi,H}-\delta_{\varphi,\widetilde{H}})(2-\textrm{dim}(\bm{\ell}_\Psi))\cC_{aHf_L}^{(3)}\,.
\end{equation}
We note that these results follow the fact that there exist terms which are mutually exclusive and cancel several cross-terms. Additionally, these results apply to models with renormalizable interactions to the fermions of the SM model. Other scenarios including higher-dimensional operators have not been considered, but could be implemented in a similar manner. These results can be used to obtain the low-energy interactions from Tables~\ref{tab:VLFs_SMEFT_tree} and \ref{tab:ALP_EFT_Lagr}, and can be readily extended to scenarios in which the SM fermions also carry charges under the underlying symmetry, such as Froggatt–Nielsen models~\cite{Greljo:2024evt}.

\section{Details of the Matching Procedure}
\label{app:det_mat_proc_app}
\addtocontents{toc}{\setcounter{tocdepth}{1}}
\subsection{Field Redefinitions}
\label{app:field_redefs}
In Table~\ref{tab:field_redefs_apps} we collect the relevant field redefinitions required to remove mass-mixing terms between vector-like fermions and SM fields following PQ symmetry breaking. For each VLF representation and corresponding PQ charge assignment, the table shows the redefinitions applied to both the VLFs and SM fields. As indicated in Section~\ref{subsec:spont_U1br_FRs}, $Q_{5,7}$ and $T_{1,2}$ VLF representations do not admit any field redefinitions, as no SM-VLF mass mixing is generated in these cases.

\begin{table}[t]
  \centering
  \scalebox{0.7}{
\begin{tabular}{ccc@{\hspace{1.5cm}}c@{\hspace{1.5cm}}c}
\toprule
\textbf{\textbf{VLF}}&$\bm{\cX_{L}}$ &$\bm{\cX_R}$& \textbf{VLF Field Redefinition}&\textbf{SM Field Redefinition}
\\
\midrule
\multirow{3}{*}{\vspace{-1.0cm}$D\sim(\bm 3,\bm 1)_{-1/3}$}
&1
&0
&$D_R\to \lzs1+\frac{y_{2,d}y_{2,d}^\dag}{y_D^2}\dzs^{\eminus\frac{1}{2}} D_R-\frac{y_{2,d}}{y_D}\cR_d d_R$
&$d_R\to \cR_d d_R+\frac{y_{2,d}^\dagger}{y_D} \lzs1+\frac{y_{2,d}y_{2,d}^\dag}{y_D^2}\dzs^{\eminus\frac{1}{2}} D_R$
\\
&0
&-1
&$D_R\to \lzs1+\frac{2M_dM_d^\dag}{y_D^2v_\Phi^2}\dzs^{\eminus\frac{1}{2}} D_R-\frac{\sqrt2 M_d}{y_Dv_\Phi}\cR_d d_R$
&$d_R\to \cR_d d_R+\frac{\sqrt2 M_d^\dag}{y_D v_\Phi}  \lzs1+\frac{2M_dM_d^\dag}{y_D^2v_\Phi^2}\dzs^{\eminus\frac{1}{2}} D_R$
\\
&-1
&-2
&$D_R\to \lzs1+\frac{y_{1,d}y_{1,d}^\dag}{y_D^2}\dzs^{\eminus\frac{1}{2}} D_R-\frac{y_{1,d}}{y_D}\cR_d d_R$
&$d_R\to \cR_d d_R +\frac{y^\dag_{1,d}}{y_D}\lzs1+\frac{y_{1,d}y_{1,d}^\dag}{y_D^2}\dzs^{\eminus\frac{1}{2}}D_R$
\\
\midrule
\multirow{3}{*}{\vspace{-1.0cm}$U\sim(\bm 3,\bm 1)_{2/3}$}
&1
&0
&$U_R\to \lzs1+\frac{y_{2,u}y_{2,u}^\dag}{y_U^2}\dzs^{\eminus\frac{1}{2}}U_R-\frac{y_{2,u}}{y_U}\cR_u u_R$
&$u_R\to \cR_u u_R+\frac{y_{2,u}^\dagger}{y_U} \lzs1+\frac{y_{2,u}y_{2,u}^\dag}{y_U^2}\dzs^{\eminus\frac{1}{2}} U_R$
\\
&0
&-1
&$U_R\to \lzs1+\frac{2M_uM_u^\dag}{y_U^2v_\Phi^2}\dzs^{\eminus\frac{1}{2}} U_R-\frac{\sqrt2 M_u}{y_Uv_\Phi}\cR_u u_R$
&$u_R\to \cR_u u_R+\frac{\sqrt2 M_u^\dag}{y_U v_\Phi}  \lzs1+\frac{2M_uM_u^\dag}{y_U^2v_\Phi^2}\dzs^{\eminus\frac{1}{2}} U_R$
\\
&-1
&-2
&$U_R\to \lzs1+\frac{y_{1,u}y_{1,u}^\dag}{y_U^2}\dzs^{\eminus\frac{1}{2}} U_R-\frac{y_{1,u}}{y_U}\cR_u u_R$
&$u_R\to \cR_u u_R +\frac{y^\dag_{1,u}}{y_U} \lzs1+\frac{y_{1,u}y_{1,u}^\dag}{y_U^2}\dzs^{\eminus\frac{1}{2}} U_R$
\\
\midrule
\multirow{3}{*}{\vspace{-1.0cm}$Q\sim(\bm 3,\bm 2)_{1/6}$}
&1
&0
&$Q_L\to \lzs1+\frac{2M_qM_q^\dag}{y_Q^2v_\Phi^2}\dzs^{\eminus\frac{1}{2}} Q_L-\frac{\sqrt2 M_q}{y_Qv_\Phi}\cR_q q_L$
&$q_L\to \cR_q q_L+\frac{\sqrt2 M_q^\dag}{y_Qv_\Phi} \lzs1+\frac{2M_qM_q^\dag}{y_Q^2v_\Phi^2}\dzs^{\eminus\frac{1}{2}} Q_L$
\\
&0
&-1
&$Q_L\to\lzs1+\frac{y_q y_q^\dag}{y_Q}\dzs^{\eminus\frac{1}{2}} Q_L-\frac{y_q}{y_Q}\cR_q q_L$
&$q_L\to \cR_q q_L+\frac{y_q^\dag}{y_Q} \lzs1+\frac{y_q y_q^\dag}{y_Q}\dzs^{\eminus\frac{1}{2}} Q_L$
\\
&2
&1
&$Q_L\to \lzs1+\frac{y_q y_q^\dag}{y_Q}\dzs^{\eminus\frac{1}{2}} Q_L-\frac{y_q}{y_Q}\cR_q q_L$
&$q_L\to \cR_q q_L+\frac{y_q^\dag}{y_Q}\lzs1+\frac{y_q y_q^\dag}{y_Q}\dzs^{\eminus\frac{1}{2}} Q_L$
\\
\bottomrule
\end{tabular}
}
  \caption{Field redefinitions of vector-like and SM fermions following the spontaneous breaking of the $\U(1)_{\sscript{PQ}}$ symmetry. The table displays the PQ charge assignments $(\cX_L, \cX_R)$ for each VLF representation, together with the corresponding field redefinitions required to eliminate the mass-mixing terms induced by the VEV of $\Phi$.}
  \label{tab:field_redefs_apps}
\end{table}

\subsection{Mapping to UV parameters}
\label{app:map_UV_params}
Here we provide explicit expressions that relate the parameters appearing in the unified interaction Lagrangians of Section~\ref{sec:unified_effdctive_descriptions} (see Table~\ref{tab:interactions_after_redef}) to the underlying UV couplings of the theory. These mappings are obtained after performing the field redefinitions required by spontaneous $\U(1)_{\sscript{PQ}}$ breaking and express all effective parameters in terms of the original Yukawa couplings, VLF masses, and mixing matrices. We separate the discussion into two sectors: Table~\ref{tab:SMEFTmap} covers the parameters relevant for SMEFT matching, while Table~\ref{tab:ALPmap} addresses the corresponding ALP interactions.

\begin{table}[h]
  \centering
  \centering
\scalebox{0.83}{
\begin{tabular}{ccccccc}
\toprule
\textbf{\textbf{VLF}}
&$\bm{\cX_{L}}$ 
&$\bm{\cX_R}$
&$\bm{\widetilde m_\Psi}$
&$\bm{\widehat Y_{u,d}}$
&$\bm{\lambda_\Psi}$
\\
\midrule
\multirow{3}{*}{\vspace{-1.6cm}$D\sim(\bm 3,\bm 1)_{-1/3}$}
&1
&0
&$\frac{y_Dv_\Phi}{\sqrt2}\lzs 1+\frac{y_{2,d}y_{2,d}^\dag}{y_D^2} \dzs^{\frac{1}{2}}$
&$\lzs Y_{d}-\frac{y_{1,d}y_{2,d}}{y_D} \dzs \cR_d$
&$\frac{v_\Phi}{\sqrt{2}\tilde{m}_D}\lzs y_{1,d}y_D+Y_d y_{2,d}^\dag \dzs$
\vspace{+0.3cm}\\
&0
&-1
&$\frac{y_Dv_\Phi}{\sqrt2}\lzs 1+\frac{2M_dM_d^\dag}{y_D^2v_\Phi^2} \dzs^{\frac{1}{2}}$
&$Y_d\cR_d$
&$\frac{1}{\tilde{m}_D}Y_d M_d^\dag $
\vspace{+0.3cm}\\
&-1
&-2
&$\frac{y_Dv_\Phi}{\sqrt2}\lzs 1+\frac{y_{1,d}y_{1,d}^\dag}{y_D^2} \dzs^{\frac{1}{2}}$
&$Y_d\cR_d$
&$\frac{v_\Phi}{\sqrt{2}\tilde{m}_D} Y_d y_{1,d}^\dag $
\\
\midrule
\multirow{3}{*}{\vspace{-1.6cm}$U\sim(\bm 3,\bm 1)_{2/3}$}
&1
&0
&$\frac{y_U v_\Phi}{\sqrt2}\lzs 1+\frac{y_{2,u}y_{2,u}^\dag}{y_U^2} \dzs^{\frac{1}{2}}$
&$\lzs Y_{u}-\frac{y_{1,u}y_{2,u}}{y_U} \dzs \cR_u$
&$\frac{ v_\Phi}{\sqrt{2}\tilde{m}_U}\lzs y_{1,u} y_U+Y_u y_{2,u}^\dag \dzs $
\vspace{+0.3cm}\\
&0
&-1
&$\frac{y_Uv_\Phi}{\sqrt2}\lzs 1+\frac{2M_uM_u^\dag}{y_U^2v_\Phi^2} \dzs^{\frac{1}{2}}$
&$Y_u\cR_u$
&$\frac{1}{\tilde{m}_U}Y_u M_u^\dag $
\vspace{+0.3cm}\\
&-1
&-2
&$\frac{y_Uv_\Phi}{\sqrt2}\lzs 1+\frac{y_{1,u}y_{1,u}^\dag}{y_U^2} \dzs^{\frac{1}{2}}$
&$Y_u\cR_u$
&$\frac{v_\Phi}{\sqrt{2}\tilde{m}_U} Y_u y_{1,u}^\dag $
\\
\midrule
\multirow{3}{*}{\vspace{-1.2cm}$Q\sim(\bm 3,\bm 2)_{1/6}$}
&1
&0
&$\frac{y_Qv_\Phi}{\sqrt2}\lzs 1+\frac{2M_qM_q^\dag}{y_Q^2v_\Phi^2} \dzs^{\frac{1}{2}}$
&$\cR_q^\dag Y_{u,d}$
&$\frac{1}{\tilde{m}_Q}M_q Y_{u,d}$
\\[10pt]
&\multirow{1}{*}{\vspace{-0.0cm}0}
&\multirow{1}{*}{\vspace{-0.0cm}-1}
&$\frac{y_Qv_\Phi}{\sqrt2}\lzs 1+\frac{y_{q}y_{q}^\dag}{y_Q^2} \dzs^{\frac{1}{2}}$
&$\cR_q^\dag \lzm Y_{u,d}-\frac{y_q^\dag y_Q^{u,d}}{y_Q} \dzm$
&$\frac{v_\Phi}{\sqrt{2}\tilde{m}_Q}\lzs y_{Q}^{u,d}y_Q+y_q Y_{u,d} \dzs $
\\[10pt]
&2
&1
&$\frac{y_Qv_\Phi}{\sqrt2}\lzs 1+\frac{y_{q}y_{q}^\dag}{y_Q^2} \dzs^{\frac{1}{2}}$
&$\cR_q^\dag Y_{u,d}$
&$\frac{y_Q v_\Phi}{\sqrt{2}\tilde{m}_Q}y_q Y_{u,d}$
\\
\midrule
$Q_5\sim(\bm 3,\bm 2)_{-5/6}$
&0&-1
&$\frac{y_{Q_5}v_\Phi}{\sqrt2}$
&$Y_d$
&$\lambda_{Q_5}$
\\
\midrule
$Q_7\sim(\bm 3,\bm 2)_{7/6}$
&0&-1
&$\frac{y_{Q_7}v_\Phi}{\sqrt2}$
&$Y_u$
&$\lambda_{Q_7}$
\\
\midrule
$T_1\sim(\bm 3,\bm 3)_{-1/3}$
&1&0
&$\frac{y_{T_1}v_\Phi}{\sqrt2}$
&$Y_{u,d}$
&$\lambda_{T_1}$
\\
\midrule
$T_2\sim(\bm 3,\bm 3)_{2/3}$
&1&0
&$\frac{y_{T_2}v_\Phi}{\sqrt2}$
&$Y_{u,d}$
&$\lambda_{T_2}$
\\
\bottomrule
\end{tabular}
}
  \caption{Relations between the effective parameters in the unified Lagrangians and the UV inputs for each VLF representation. For each charge assignment $(\cX_L, \cX_R)$, we provide expressions for the rescaled mass $\widetilde{m}_\Psi$, the induced Yukawa structures $\widehat{Y}_{u,d}$, and the effective couplings $\lambda_\Psi$. These quantities form the input for the SMEFT matching analysis (see Table~\ref{tab:interactions_after_redef} for more details).}
  \label{tab:SMEFTmap}
\end{table}
\begin{table}[h]
  \centering
  \centering
\scalebox{0.74}{
\begin{tabular}{ccc@{\hspace{0.8cm}}cccccc}
\toprule
\textbf{\textbf{VLF}}
&$\bm{\cX_{L}}$ 
&$\bm{\cX_R}$
&$\bm{C_{\Psi,L}^{(1)}}$
&$\bm{C_{\Psi,L}^{(2)}}$
&$\bm{C_{\Psi,L}^{(3)}}$
&$\bm{C_{\Psi,R}^{(1)}}$
&$\bm{C_{\Psi,R}^{(2)}}$
&$\bm{C_{\Psi,R}^{(3)}}$
\\
\midrule
\multirow{3}{*}{\vspace{-1.8cm}$D\sim(\bm 3,\bm 1)_{-1/3}$}
&1
&0
&1
&0&0&0&0&0
\vspace{+0.5cm}\\
&0
&-1
&0&0&0
&$-\frac{y_D^2 v_\Phi^2}{2 \tilde{m}_D^2}$
&$-\frac{2}{y_D^2v_\Phi^2}\cR_d^\dag M_d^\dag M_d \cR_d$
&$\frac{1}{\tilde{m}_D}  M_d \cR_d$
\vspace{+0.5cm}\\
&-1
&-2
&$-1$&0&0
&$-\frac{y_D^2 v_\Phi^2}{\tilde{m}_D^2} $
&$-\frac{2}{y_D^2}\cR_d^\dag y_{1,d}^\dag y_{1,d}\cR_d$
&$\frac{\sqrt{2} v_\Phi y_{1,d}}{\tilde{m}_D}\cR_d$
\\[10pt]
\midrule
\multirow{3}{*}{\vspace{-1.6cm}$U\sim(\bm 3,\bm 1)_{2/3}$}
&1
&0
&1
&0&0&0&0&0
\vspace{+0.5cm}\\
&0
&-1
&0&0&0
&$-\frac{y_U^2 v_\Phi^2}{2 \tilde{m}_U^2}$
&$-\frac{2}{y_U^2v_\Phi^2}\cR_u^\dag M_u^\dag M_u \cR_u$
&$\frac{1}{\tilde{m}_U} M_u\cR_u  $
\vspace{+0.5cm}\\
&-1
&-2
&$-1$&0&0
&$-\frac{y_U^2 v_\Phi^2}{ \tilde{m}_U^2}$
&$-\frac{2}{y_U^2}\cR_u^\dag y_{1,u}^\dag y_{1,u}\cR_u$
&$\frac{\sqrt{2} v_\Phi y_{1,u}}{\tilde{m}_U}\cR_u$
\\[10pt]
\midrule
\multirow{3}{*}{\vspace{-1.6cm}$Q\sim(\bm 3,\bm 2)_{1/6}$}
&1
&0
&$\frac{y_Q^2 v_\Phi^2}{2 \tilde{m}_Q^2}$
&$\frac{2}{y_Q^2v_\Phi^2}\cR_q^\dag M_q^\dag M_q \cR_q$
&$-\frac{1}{\tilde{m}_Q}M_q\cR_q$
&0&0&0
\vspace{+0.5cm}\\
&\multirow{1}{*}{\vspace{-0.0cm}0}
&\multirow{1}{*}{\vspace{-0.0cm}-1}
&0&0&0
&$-1$&0&0
\vspace{+0.5cm}\\
&2
&1
&$\frac{y_Q^2 v_\Phi^2}{\tilde{m}_Q^2}$
&$\frac{2}{y_Q^2}\cR_q^\dag y_q^\dag y_q\cR_q$
&$-\frac{\sqrt{2}v_\Phi}{\tilde{m}_Q}y_q \cR_q$
&1
&0&0
\\[10pt]
\midrule
$Q_5\sim(\bm 3,\bm 2)_{-5/6}$
&0&-1
&0&0&0&$-1$
&0&0
\\
\midrule
$Q_7\sim(\bm 3,\bm 2)_{7/6}$
&0&-1
&0&0&0&$-1$
&0&0
\\
\midrule
$T_1\sim(\bm 3,\bm 3)_{-1/3}$
&1&0
&1&0&0&0&0&0
\\
\midrule
$T_2\sim(\bm 3,\bm 3)_{2/3}$
&1&0
&1&0&0&0&0&0
\\
\bottomrule
\end{tabular}
}
  \caption{UV expressions for the coefficients $C^{(i)}_{\Psi,(L,R)}$ governing ALP–fermion interactions, after field redefinitions and in terms of the original parameters of the UV theory. Each entry corresponds to a specific VLF irrep and PQ charge assignment, following the conventions used in the main text (see Table~\ref{tab:interactions_after_redef} for more details). }
  \label{tab:ALPmap}
\end{table}

\section{Overview of SMEFT Observables}
\label{app:SMEFT_obs_overview}

Here we briefly review the observables used in the analysis of Section~\ref{sec:SMEFT_pheno}.

\subsection{Electroweak Precision Observables}
Electroweak precision observables used in the analysis are primarily extracted from high-precision measurements performed at the $Z$ pole~\cite{ALEPH:2005ab,UA2:1983mlz,UA1:1983mne,Abrams:1989aw,TwoFermionWorkingGroup:2000nks,ALEPH:1994ljn,ALEPH:1993jml,ALEPH:1991btx,ALEPH:1990gko,ALEPH:1999smx,DELPHI:1991hau,DELPHI:1994bya,DELPHI:1994tlg,DELPHI:2000wje,L3:1991gfs}, as well as from precision determinations of electroweak parameters such as the $W$ boson mass. In the following, we review two EWPT observables entering the analysis: the forward-backward asymmetry parameter $A_e$ and the $W$ boson mass $m_W$.

\vspace{0.2cm}
\noindent
$\bm{A_{e,\mu,\tau}.}$ The polarization asymmetry parameter associated with the coupling of the $Z$ boson to charged leptons is extracted from asymmetry measurements at the $Z$ pole~\cite{ATLAS:2015ihy,CMS:2022uul,ALEPH:2005ab}. It is defined as
\begin{equation}
  A_\ell = \frac{\Gamma(Z \to \ell_L^+ \ell_L^-) - \Gamma(Z \to \ell_R^+ \ell_R^-)}{\Gamma(Z \to \ell^+ \ell^-)}=\frac{[Z_{e_L}]_{ww}^2 - [Z_{e_R}]_{ww}^2}{[Z_{e_L}]_{ww}^2 + [Z_{e_R}]_{ww}^2}
  \,,
  \qquad
  \ell=e,\mu,\tau\,,
\end{equation}
where
\begin{equation}
  [Z_{e_L}]_{pr}=\lzm s_\sscript{W}^2-\frac{1}{2} \dzm\delta_{pr}-\frac{v^2}{2}[\cC_{H\ell}^{(1)}+\cC_{H\ell}^{(3)}]_{pr}\,,
  \qquad
  [Z_{e_R}]_{pr}=s_\sscript{W}^2\delta_{pr}-\frac{v^2}{2}[\cC_{H e}]_{pr}\,.
\end{equation}
In the models considered here, the VLF completions generate SMEFT operators that modify $Z$ couplings to quarks at tree level. Consequently, direct contributions to the leptonic couplings $[Z_{e_L}]{pr}$ and $[Z{e_R}]_{pr}$ are absent at this order. However, quark-sector operators mix into leptonic ones through RGE~\cite{Alonso:2013hga,Jenkins:2013wua}, which we account for by evolving the Wilson coefficients with the \texttt{wilson} package~\cite{Aebischer:2018bkb}.

\vspace{0.2cm}
\noindent
$\bm{A_{b}.}$ The bottom-quark polarization asymmetry parameter $A_b$ enters the forward-backward asymmetry $A_{\mathrm{FB}}^{0,b}$ measured at the $Z$ pole and is defined in terms of the left- and right-handed $Z$ couplings to the bottom quark. Within the SMEFT, $A_b$ can be written as
\begin{equation}
  A_b = \frac{\Gamma(Z \to b_L^+ b_L^-) - \Gamma(Z \to b_R^+ b_R^-)}{\Gamma(Z \to b^+ b^-)}=\frac{[Z_{d_L}]_{22}^2 - [Z_{d_R}]_{22}^2}{[Z_{d_L}]_{22}^2 + [Z_{d_R}]_{22}^2}
  \,,
\end{equation}
where
\begin{equation}
  [Z_{d_L}]_{pr}=\lzm \frac{1}{3}s^2_{\sscript{W}}-\frac{1}{2} \dzm\delta_{pr}-\frac{v^2}{2}[\cC_{Hq}^{(1)}+\cC_{Hq}^{(3)}]_{pr}\,,
  \qquad
  [Z_{d_R}]_{pr}=\frac{1}{3}s_\sscript{W}^2\delta_{pr}-\frac{v^2}{2}[\cC_{H d}]_{pr}\,.
\end{equation}
In the VLF scenarios considered, these operators are generated at tree level for representations coupling to the down-type quark sector, leading to direct modifications of the $Zb\bar b$ vertex. For representations coupling only to the up-type sector, such effects arise only radiatively.

\vspace{0.2cm}
\noindent
$\bm{m_W.}$ In the SMEFT, shifts in $m_W$ arise from dimension-six operators modifying the electroweak gauge sector together with fermionic operators such as $\cC_{H\ell}^{(3)}$ and $\cC_{\ell\ell}$, and can be written as~\cite{Greljo:2023bdy,Bagnaschi:2022whn}
\begin{equation}\label{eq:mW_SMEFT}
  \frac{\delta m_W^2}{m_W^2} =
  -\frac{s_{2\sscript{W}}}{c_{2\sscript{W}}} \frac{v^2}{4}
  \Bigg[\frac{c_\sscript{W}}{s_\sscript{W}} \cC_{HD}
    + \frac{s_\sscript{W}}{c_\sscript{W}} (4\,\cC_{H\ell}^{(3)} - 2\,\cC_{\ell\ell})
  + 4\,\cC_{HWB}\Bigg]\,,
\end{equation}
where fermionic flavor indices are omitted assuming minimal flavor violation (MFV). The operators in Eq.~\eqref{eq:mW_SMEFT} are not generated at tree level in our setup but arise through RG mixing of quark-sector operators, which is included using the \texttt{wilson} package.

\begin{table}[t]
  \centering
\scalebox{0.9}{
\begin{tabular}{c@{\hspace{.5cm}}c@{\hspace{.5cm}}c@{\hspace{.5cm}}c@{\hspace{.5cm}}c}
\toprule
\textbf{VLF}
&$\boldsymbol{\alpha_F}$
&$\boldsymbol{\beta_F}$
&$\boldsymbol{\gamma_F}$
&$\boldsymbol{\delta_F}$
\\
\midrule
\addlinespace[0.2cm]
$U$&$-8/45$&$+7/6$&$-3/2$&$-1/3$
\\[0.2cm]
$D$&$-2/45$&$+1/6$&$-3/2$&$+1/3$
\\[0.2cm]
$Q_u$&$-1/45$&$-4/3$&$-2$&$+4/3$
\\[0.2cm]
$Q_d$&$-1/45$&$-1/3$&$-2$&$+2/3$
\\[0.2cm]
\bottomrule
\end{tabular}
\quad
\begin{tabular}{c@{\hspace{.5cm}}c@{\hspace{.5cm}}c@{\hspace{.5cm}}c@{\hspace{.5cm}}c}
\toprule
\textbf{VLF}
&$\boldsymbol{\alpha_F}$
&$\boldsymbol{\beta_F}$
&$\boldsymbol{\gamma_F}$
&$\boldsymbol{\delta_F}$
\\
\midrule
\addlinespace[0.2cm]
$Q_5$&$-5/9$&$+5/3$&$-2$&$-2/3$
\\[0.2cm]
$Q_7$&$-49/45$&$+8/3$&$-2$&$-4/3$
\\[0.2cm]
$T_1$&$-2/15$&$+1/8$&$-19/32$&$+1/4$
\\[0.2cm]
$T_2$&$-8/15$&$+7/8$&$-19/32$&$-1/4$
\\[0.2cm]
\bottomrule
\end{tabular}
}
  \caption{Numerical coefficients for the VLF representations entering Eq.~\eqref{eq:CHD_general_one_loop}.}
  \label{tab:CHD_coeffs}
\end{table}

\vspace{0.2cm}
\noindent
$\bm{T}$ \textbf{parameter.} In our analysis, the Wilson coefficient $\cC_{HD}$, which encodes contributions to the electroweak $T$-parameter, enters indirectly through its effect on the $W$-boson mass and is therefore already included in the EWPT fits discussed above. In the class of models considered here, however, this coefficient is not generated at tree level, but arises only radiatively through renormalization group evolution.

For completeness, we also report the one-loop matching contributions obtained by integrating out the VLF representations. Neglecting terms proportional to Yukawa couplings, the coefficient $\cC_{HD}$ can be written as~\cite{Gargalionis:2024jaw}
\begin{equation}\label{eq:CHD_general_one_loop}
  \cC^F_{HD}(\mu)\approx\frac{1}{16\pi^2}\frac{1}{\widetilde m_F^2}\lzs \alpha_F\,g_1^4+\beta_F g_1^2|\lambda_F|^2+\gamma_F |\lambda_F|^4+\delta_F g_1^2|\lambda_F|^2\log\frac{\widetilde m_F^2}{\mu^2} \dzs\,,
\end{equation}
where $\mu$ denotes the matching scale, and $\alpha_F,\beta_F,\gamma_F,\delta_F$ are numerical coefficients that depend on the specific VLF representation, which we collect in Table~\ref{tab:CHD_coeffs}. Evaluated at the matching scale $\mu \simeq \widetilde m_F$, this expression can be translated into a bound on the VLF scale~\cite{Ellis:2020unq}:
\begin{equation}
  \widetilde{m}_F\sim 1\,\tev\times\lzu \lzs \alpha_F\,g_1^4+\beta_F g_1^2|\lambda_F|^2+\gamma_F |\lambda_F|^4 \dzs^{1/2} \dzu\,.
\end{equation}
Although numerically subleading compared to the dominant bounds discussed in the main text, the one-loop contribution to $\cC_{HD}$ yields a generic electroweak constraint within the minimal VLF setup considered here. The corresponding scale is typically of order $\mathcal{O}(1\,\tev)$, and is therefore comparable to the current collider reach (see Section~\ref{sec:coll_constraints}).

\subsection{Rare $B$-Meson Decays}
Flavor-changing neutral current processes considered in this analysis are generated by a subset of SMEFT operators that modify the couplings of down-type quarks to electroweak gauge bosons. In particular, we focus on the operators $\cO_{Hq}^{(1)}$, $\cO_{Hq}^{(3)}$, and $\cO_{Hd}$, which arise at tree level in the VLF completions discussed in Section~\ref{sec:UV_IR_Framework}. After EWSB, these operators induce semileptonic $b\to s\ell^+\ell^-$ transitions. The corresponding low-energy phenomenology is described by the effective Hamiltonian
\begin{equation}\label{eq:B_phys_eff_H}
  \cH_{\sscript{eff}}\supset -\frac{4G_F}{\sqrt{2}}\lzs \cC_9\cO_9+\cC_9'\cO_9'+\cC_{10}\cO_{10}+\cC_{10}'\cO_{10}' \dzs+\hermc\,,
\end{equation}
where
\begin{equation}\label{eq:WET_SL_ops}
  \cO_{9}^{(\prime)}=\frac{\alpha}{4\pi}(\bar s\gamma^\mu P_{L(R)}b)(\bar\ell\gamma_\mu \ell)\,,
  \qquad
  \cO_{10}^{(\prime)}=\frac{\alpha}{4\pi}(\bar s\gamma^\mu P_{L(R)}b)(\bar\ell\gamma_\mu\gamma^5 \ell)\,.
\end{equation}
Matching the SMEFT onto the weak effective theory at the electroweak scale generates contributions to the Wilson coefficients $\cC_{9,10}^{(\prime)}$~\cite{Jenkins:2017jig,Ali:2025xkw}:
\begin{equation}\label{eq:C9_C10_WET}
  \begin{alignedat}{2}
    \cC_9&=\frac{\sqrt2}{G_F}\frac{\pi}{2\alpha}(4s_\sscript{W}^2-1)[\cC_{H q}^{(1)}+\cC_{H q}^{(3)}]_{23}\,,
    &\qquad
    \cC_{10}&=\frac{\sqrt2}{G_F}\frac{\pi}{2\alpha}[\cC_{H q}^{(1)}+\cC_{H q}^{(3)}]_{23}\,,
    \\
    \cC'_9&=\frac{\sqrt2}{G_F}\frac{\pi}{2\alpha}(4s_\sscript{W}^2-1)[\cC_{H d}]_{23}\,,
    &\qquad
    \cC'_{10}&=\frac{\sqrt2}{G_F}\frac{\pi}{2\alpha}[\cC_{H d}]_{23}\,.
  \end{alignedat}
\end{equation}
Using the tree-level matching relations summarized in Table~\ref{tab:VLFs_SMEFT_tree}, these coefficients can be expressed in terms of the UV parameters of the VLF models. Below we review the three observables discussed in Section~\ref{ref:SMEFT_res_dis} that provide the dominant constraints in this sector.

\vspace{0.2cm}
\noindent
$\bm{\langle P_5^\prime\rangle(B^0\to K^{\ast 0}\mu^+\mu^-).}$ Among the many angular observables accessible in the $B^0 \to K^{\ast 0}\mu^+\mu^-$ decay and included in the global fit, the optimized quantity $P_5^\prime$ provides one of the most sensitive probes of short-distance contributions to the $b\to s\ell^+\ell^-$ transition~\cite{LHCb:2020lmf,ATLAS:2018gqc,Belle:2016xuo,CMS:2017rzx}. It is constructed from the angular distribution of $B^0\to K^{\ast}(\to K\pi)\mu^+\mu^-$ and designed to reduce hadronic uncertainties through ratios of decay amplitudes. The observable integrated over a given $q^2$ bin is defined as
\begin{equation}
  \langle P^{\prime}_5 \rangle = \frac{\int_{\sscript{bin}}\dd q^2\, \cI_{5}}{2 \sqrt{-\int_{\sscript{bin}}\dd q^2\, \cI_{2}^c\, \int_{\sscript{bin}}\dd q^2 \,\cI_{2}^s}}\,,
\end{equation}
where $\mathcal I_5$, $\mathcal I_2^c$, and $\mathcal I_2^s$ are angular coefficients appearing in the differential decay distribution. These coefficients can be expressed in terms of the transversity amplitudes as~\cite{Rajeev:2020aut,Ali:2025xkw,Altmannshofer:2026cwk}
\begin{equation}\label{eq:I2sc5_defs}
  \begin{alignedat}{2}
    \cI_2^s&=\frac{1}{4} \beta_{\mu}^2 \Big[|A_{L\perp}|^2 + |A_{L\parallel}|^2 + |A_{R\perp}|^2 + |A_{R\parallel}|^2\Big]\,,
    \qquad
    \cI_2^c=-\beta_{\mu}^2 \Big[|A_{L0}|^2 + |A_{R0}|^2\Big]\,,
    \\
    \cI_5&=\sqrt{2} \beta_\mu \Big[ \re(A_{L0}A_{L\perp}^{*}) - \re(A_{R0}A_{R\perp}^{*}) \Big]\,,
  \end{alignedat}
\end{equation}
where the transversity amplitudes encode the dependence on the Wilson coefficients as
\begin{equation}\label{eq:transversity_amp}
  \begin{alignedat}{2}
    A_{(L,R)0}&=\frac{\mathcal N}{2m_{{{K^*}}}\sqrt{q^2}} \bigg[(m_{B}^2 - m_{{K^*}}^2 - q^2)(m_{B} + m_{{K^*}})A_1 - \frac{\lambda}{m_{B} + m_{{K^*}}}A_2\bigg](\cC_{9} \mp \cC_{10})\,,
    \\
    A_{(L,R)\perp}&=-
    \frac{\mathcal N\sqrt{2\lambda}\,V}{m_{B} + m_{{K^*}}} (\cC_{9} \mp  \cC_{10})\,,
    \quad
    A_{(L,R)\parallel} = \mathcal N\sqrt{2}
    (m_{B} + m_{{K^*}}) A_1 (\cC_{9} \mp \cC_{10})\,,
    \\
    A_{(L,R)t}&=  \frac{\mathcal N\sqrt{\lambda}A_0}{\sqrt{q^2}} (\cC_{9} \mp \cC_{10})\,,
  \end{alignedat}
\end{equation}
with $\lambda=m_{B}^4+m_{K^*}^4+q^4-2(m_{B}^2m_{K^*}^2+m_{K^*}^2 q^2+q^2 m_{B}^2)$ and $\mathcal N$ defined as
\begin{equation}
  \mathcal N= \bigg[\frac{G_{F}^2\,\alpha^2}{512\,\pi^5\,m_{B}^3}\frac{q^2 \sqrt{\lambda}}{6}|V_{tb}V_{ts}^{*}|^2 \beta_\mu\bigg]^{1/2}\,.
\end{equation}
Lastly, the $V$ and $A_{0,1,2}$ form factors appearing in Eq.~\eqref{eq:transversity_amp} are defined as~\cite{Dutta:2019wxo,Bouchard:2013eph}
\begin{equation}
  \begin{alignedat}{2}
    \langle K^{\ast}|\bar{s}\gamma_{\mu}b|B \rangle &= \frac{2\,i\,V(q^2)}{m_{B}+m_{K^{\ast}}}\,\epsilon_{\mu\nu\rho\sigma}\epsilon^{\ast\nu}\,
    p_{B}^\rho\,p_{{K^{\ast}}}^\sigma\,,
    \\
    \langle K^{\ast}|\bar{s}\gamma^{\mu}\gamma_5\,b|B \rangle &= 2\,m_{K^{\ast}}\,A_0(q^2)\frac{\epsilon^{\ast}\cdot q}{q^2}\,q^{\mu} +
    (m_{B} + m_{K^{\ast}})\,A_1(q^2)\Big(\epsilon^{{\ast\mu}}-\frac{\epsilon^{\ast}\cdot q}{q^2}\,q^{\mu}\Big)
    \\&-
    A_2(q^2)\frac{\epsilon^{\ast}\cdot q}{m_{B} + m_{K^{\ast}}}\,\Big[p_{B}^{\mu}+p_{K^{\ast}}^{\mu}-\frac{m_{B}^2-m_{K^{\ast}}^2}
    {q^2}\,q^{\mu}\Big]\,,
  \end{alignedat}
\end{equation}
where $q^{\mu}=(p_B-p_{K^*})^\mu$ denotes four-momentum transfer and $\epsilon_{\mu}$ is the polarization vector of the $K^{\ast}$ meson.

\vspace{0.2cm}
\noindent
$\bm{B \to K^{\ast0} \mu^+ \mu^-.}$ Another class of observables relevant for the analysis involves exclusive decays of the form $B\to V\mu^+\mu^-$, where $V$ denotes a vector meson. In particular, we consider the mode with $V=K^{\ast0}$, which probes the same underlying $b\to s\mu^+\mu^-$ transition discussed above. The differential branching ratio for these processes can be written in terms of the angular coefficients appearing in the full decay distribution as~\cite{Yilmaz:2008pa,Dubnicka:2016nyy,Jin:2020qfp,Xu:2013lms,Ali:1999mm}
\begin{equation}
  \dv{\mathrm{BR}}{q^2}\,(B \to V \mu^+ \mu^-) = \frac{\tau_B}{\hbar}  \frac{1}{4} \Big[ 3 \,\cI_1^c + 6 \,\cI_1^s - \cI_2^c - 2\, \cI_2^s \Big]\,,
\end{equation}
with $\cI_2^s$ and $\cI_2^c$ defined in Eq.~\eqref{eq:I2sc5_defs}, while $\cI_1^s$ and $\cI_1^c$ take the form
\begin{equation}\small
  \begin{alignedat}{2}
    \cI_1^s &= \frac{3}{4} \Big[ |A_{L\perp}|^2 + |A_{L\parallel}|^2 + |A_{R\perp}|^2 + |A_{R\parallel}|^2 \Big]
    \left( 1 - \frac{4 m_\mu^2}{3 q^2} \right)
    + \frac{4 m_\mu^2}{q^2} \, \re \left[ A_{L\perp} A_{R\perp}^* + A_{L\parallel} A_{R\parallel}^* \right]\,,
    \\
    \cI_1^c &= \Big[ |A_{L0}|^2 + |A_{R0}|^2 \Big]
    + 8 \frac{m_\mu^2}{q^2} \, \re \left[ A_{L0} A_{R0}^* \right]
    + 4 \frac{m_\mu^2}{q^2} \Big[|A_{Lt}|^2+|A_{Rt}|^2\Big]\,,
  \end{alignedat}
\end{equation}
where the transversity amplitudes encode the dependence on the weak effective theory Wilson coefficients $\cC_9^{(\prime)}$ and $\cC_{10}^{(\prime)}$.

\vspace{0.2cm}
\noindent
$\bm{B_s\to \mu^+\mu^-.}$ In the SM, this decay mode proceeds through electroweak penguin and box diagrams and is strongly suppressed by the GIM mechanism and helicity effects, making it particularly sensitive to short-distance contributions to the $b\to s\ell^+\ell^-$ transition, which in the weak effective theory lead to the branching ratio~\cite{Huang:2002ni,DeBruyn:2012wk}
\begin{equation}
  \mathrm{BR}(B_s \to \mu^+ \mu^-)=\frac{G_F^2\, \alpha^2}{16 \pi^3} f_{B_s}^2 m_{B_s}^3 \tau_{B_s} |V_{tb} V_{ts}^*|^2 \sqrt{1 - \frac{4 m_\mu^2}{m_{B_s}^2}} \left| \cC_{10} \right|^2\,,
\end{equation}
where $f_{B_s}$ and $\tau_{B_s}$ denote the decay constant and lifetime of the $B_s$ meson.

\subsection{Rare Kaon Decays}
Rare kaon decays provide complementary constraints on flavor-changing $s\to d$ transitions. In the VLF scenarios considered here, the relevant SMEFT contributions arise from the operators $\cO_{Hq}^{(1)}$, $\cO_{Hq}^{(3)}$, and $\cO_{Hd}$ generated at tree level. After matching to the weak effective theory, these operators induce contributions to the semileptonic Wilson coefficients $\cC_{9}^{(\prime)}$ and $\cC_{10}^{(\prime)}$
\begin{equation}\label{eq:C9_C10_WET_Kaon}
  \begin{alignedat}{2}
    \cC_9&=\frac{\sqrt2}{G_F}\frac{\pi}{2\alpha}(4s_\sscript{W}^2-1)[\cC_{H q}^{(1)}+\cC_{H q}^{(3)}]_{12}\,,
    &\qquad
    \cC_{10}&=\frac{\sqrt2}{G_F}\frac{\pi}{2\alpha}[\cC_{H q}^{(1)}+\cC_{H q}^{(3)}]_{12}\,,
    \\
    \cC'_9&=\frac{\sqrt2}{G_F}\frac{\pi}{2\alpha}(4s_\sscript{W}^2-1)[\cC_{H d}]_{12}\,,
    &\qquad
    \cC'_{10}&=\frac{\sqrt2}{G_F}\frac{\pi}{2\alpha}[\cC_{H d}]_{12}\,.
  \end{alignedat}
\end{equation}
In the following, we briefly summarize the kaon observables used in the analysis.

\vspace{0.2cm}
\noindent
$\bm{K_L^0\to\mu^+\mu^-.}$ This decay mode probes the flavor-changing transition $s\to d\mu^+\mu^-$. While the total rate receives sizable long-distance contributions from the two-photon process $K_L\to\gamma^\ast\gamma^\ast\to\mu^+\mu^-$, the short-distance component provides sensitivity to the Wilson coefficients $\cC_{10}^{(\prime)}$. The branching ratio can be written as~\cite{Chobanova:2017rkj}
\begin{equation}
  \mathrm{BR}(K_L^0\to\mu^+\mu^-)=\tau_L\frac{f_K^2 m_K^3}{16\pi}\sqrt{1-\frac{4m_\mu^2}{m_K^2}}\lvert \mathcal A_L \rvert^2\,,
\end{equation}
where $f_K$ and $\tau_L$ denote the kaon decay constant and lifetime. The decay amplitude takes the form
\begin{equation}
  \mathcal A_L=\frac{2G_F^2 m_W^2 m_\mu}{\pi^2 m_K}A_{L\gamma\gamma}^\mu-\frac{\sqrt2 m_\mu}{m_K}\frac{\alpha G_F}{\pi}\re(\cC_{10}-\cC_{10}')\,,
\end{equation}
with the first term describing the long-distance contribution from two-photon intermediate states~\cite{Ecker:1991ru,Isidori:2003ts,DAmbrosio:2017klp}:
\begin{equation}
  \frac{2G_F^2 m_W^2 m_\mu}{\pi^2 m_K}A_{L\gamma\gamma}^\mu=\pm(0.54-3.96\,i)\times10^{\eminus11}\,\gev^{\eminus2}\,.
\end{equation}

\vspace{0.2cm}
\noindent
$\bm{K\to\pi\nu\bar\nu.}$ The rare decays $K^+ \to \pi^+ \nu \bar{\nu}$ and $K_L \to \pi^0 \nu \bar{\nu}$ probe the flavor-changing transition $s\to d\nu\bar\nu$. In the weak effective theory, these processes are described by the weak Hamiltonian
\begin{equation}
  \cH_{\sscript{eff}}\supset -\frac{4G_F}{\sqrt2}\sum_{i=e,\mu,\tau}\lzs \cC_9^\nu \cO_9^{\nu_i}+\cC_9^{\prime\nu}\cO_{9}^{\prime\nu_i} \dzs+\hermc\,,
\end{equation}
where
\begin{equation}
  \cC_9^\nu=\frac{\sqrt2}{G_F}\frac{\pi}{\alpha}[\cC_{Hq}^{(1)}+\cC_{Hq}^{(3)}]_{12}\,,
  \qquad
  \cC_9^{\prime\nu}=\frac{\sqrt2}{G_F}\frac{\pi}{\alpha}[\cC_{Hd}]_{12}\,
  \qquad
  \cO^{(\prime)\nu_j}_9=\frac{\alpha}{4\pi}(\bar d\gamma^\mu P_{L(R)}s)(\bar\nu_j \gamma_\mu\nu_j)\,.
\end{equation}
The branching ratio for $K^+ \to \pi^+ \nu \bar{\nu}$ can be expressed as~\cite{Hou:2024vyw}
\begin{equation}
  \begin{alignedat}{2}
    \mathrm{BR}(K^+\to \pi^+\nu\bar\nu)&
    =\kappa_+(1+\Delta_{\sscript{EM}})\lzs \frac{\re(V_{cs}^*V_{cd})}{|V_{us}|}P_c-\frac{8\pi^2}{\alpha^2}\frac{s_\sscript{W}^2}{|V_{us}|^2}\re(\cC_9^\nu) \dzs^2
    \\&
    +\frac{4\pi^2}{\alpha^2}\frac{\kappa_+ s_\sscript{W}^4}{|V_{us}|^{10}}(1+\Delta_{\sscript{EM}})\lzs\im\bigg( \frac{V_{ts}^* V_{td}}{V_{ts}V_{td}^*}\cC_9^\nu \bigg)\dzs^2\,,
  \end{alignedat}
\end{equation}
where $\kappa_+\approx (5.173\pm 0.025)\times10^{\eminus 11}$ captures the hadronic matrix element extracted from the $K_{\ell3}$ data~\cite{Buras:2015qea}, $\Delta_{\sscript{EM}}$ parametrizes the isospin-breaking EM corrections~\cite{Mescia:2007kn,Cirigliano:2011ny}, while $P_c$ accounts for the short- and long-distance charm contributions~\cite{Buras:2015qea,Buras:2006gb,Brod:2008ss}.

For the CP-violating mode $K_L \to \pi^0 \nu \bar{\nu}$, the branching ratio reads
\begin{equation}
  \mathrm{BR}(K_L^0\to \pi^0\nu\bar\nu)=\frac{s_\sscript{W}^4 \kappa_L}{4|V_{us}|^{10}}(1-\delta_\varepsilon)\,\im\bigg(\frac{V_{ts}^* V_{td}}{V_{ts}V_{td}^*}\cC_9^\nu\bigg)\,,
\end{equation}
where $\kappa_L\approx(2.231 \pm0.013)\times10^{\eminus10}$ encodes the hadronic matrix element, whereas $\delta_\varepsilon$ denotes the indirect CP-violating contribution~\cite{Buchalla:1998ux}.

\subsection{Collider Constraints}
\label{sec:coll_constraints}
Collider searches probe VLFs through both pair and single production mechanisms at high-energy colliders. While pair production is largely insensitive to electroweak couplings, single production directly probes their mixing with SM fermions. Experimental results are typically presented in simplified scenarios with free masses and branching ratios. In the UV constructions considered in this work, these quantities are correlated by the underlying flavor and PQ structure; nevertheless, existing collider limits provide a robust and useful benchmark for the parameter space. The corresponding constraints can be organized into several distinct categories discussed below.

\vspace{0.2cm}
\noindent
\textbf{Single production.} Single production searches probe directly the mixing of VLFs with SM quarks and are therefore particularly sensitive in scenarios with sizable electroweak couplings. ATLAS analyses targeting the production of a vector-like top partner $T$ decaying via $T \to Zt$ and $T \to Ht$, using the full Run 2 dataset of $139\,\mathrm{fb}^{\eminus1}$ at $\sqrt{s}=13\,\tev$, set limits in the $(m_T,\kappa)$ plane~\cite{ATLAS:2023pja,ATLAS:2023bfh}. For singlet scenarios $T\equiv U \sim (\rep3,\rep1)_{2/3}$, values $\kappa \gtrsim 0.5$ are excluded for $m_T \lesssim 2.3\,\mathrm{TeV}$, while for doublet embeddings $T \subset Q \sim (\rep3,\rep2)_{1/6}$, values $\kappa \gtrsim 0.72$ are excluded for $m_T \lesssim 1.7\,\mathrm{TeV}$, with sensitivity extending to $\kappa \gtrsim 0.55$ near $m_T \sim 1\,\mathrm{TeV}$. Complementary fully hadronic searches exploiting boosted topologies exclude $\kappa_T \gtrsim 0.3$ for $m_T \gtrsim 1.1\,\mathrm{TeV}$, with sensitivity reaching $\kappa_T \gtrsim 1.6$ at $m_T = 2.3\,\mathrm{TeV}$, and masses up to $1.7\,\mathrm{TeV}$ for broad-width scenarios~\cite{ATLAS:2022ozf,CMS:2024qdd}. Analogous searches for a vector-like bottom partner $B$ decaying via $B \to bH$ constrain the couplings $c_W \gtrsim 0.4$ at $m_B \approx 1.1\,\mathrm{TeV}$, strengthening to $c_W \gtrsim 0.5{-}0.6$ in the range $1.3\text{--}2.0\,\mathrm{TeV}$~\cite{ATLAS:2023ixh}.

\vspace{0.2cm}
\noindent
\textbf{Pair production.} Pair production of VLFs proceeds dominantly through QCD interactions and therefore provides robust bounds largely independent of the mixing parameters. Searches targeting final states with leptonic $Z$ decays exclude masses up to $m_T = 1.27\,\mathrm{TeV}$ and $m_B = 1.46\,\mathrm{TeV}$ for singlet representations, and $m_T = 1.20\,\mathrm{TeV}$ and $m_B = 1.32\,\mathrm{TeV}$ for doublets~\cite{ATLAS:2022hnn,CMS:2022fck,CMS:2024xbc}. Analyses focusing on mixing with light generations exclude VLF masses below $1.53\,\mathrm{TeV}$ under the assumption $\mathrm{BR}(Q \to Wq)=1$, with more general branching scenarios yielding limits up to $1.15\,\mathrm{TeV}$~\cite{ATLAS:2024zlo,ATLAS:2024gyc}. Additional searches employing missing transverse momentum and multivariate techniques further constrain the parameter space, with exclusions reaching $m_T = 1.26\,\mathrm{TeV}$ and $m_B = 1.33\,\mathrm{TeV}$ for singlets, as well as $m_T = 1.41\,\mathrm{TeV}$ for doublets, and up to $m_{T,B} > 1.59\,\mathrm{TeV}$ in degenerate doublet scenarios~\cite{ATLAS:2022tla}.

\vspace{0.2cm}
\noindent
\textbf{Exotic representations.} Several VLF representations considered in this work involve exotic electric charges or $\SU(2)_{\sscript{L}}$ triplet structures, which are not fully covered by standard searches. Dedicated CMS searches for vector-like quarks with electric charge $+5/3$, denoted $X_{5/3}$, exclude masses up to $1.33\,\mathrm{TeV}$ and $1.30\,\mathrm{TeV}$ for right- and left-handed couplings, respectively, assuming $\mathrm{BR}(X_{5/3}\to tW)=1$~\cite{CMS:2018ubm}. Similarly, searches for vector-like quarks with electric charge $-4/3$, denoted $Y_{\eminus4/3}$, set bounds up to $1.20\,\mathrm{TeV}$ for single production with coupling $0.5$~\cite{CMS:2017fpk}. Constraints on mixing angles have also been derived, reaching $|\sin\theta_L| \sim 0.18$ and $|\sin\theta_R| \sim 0.16$ for masses around $800\,\mathrm{GeV}$~\cite{ATLAS:2018dyh}. In our analysis, these results are particularly relevant for the $Q_5$, $Q_7$, $T_1$, and $T_2$ representations, where different components of the multiplets are constrained individually.

\section{Phenomenology of $Q_7$}
\label{app:Q7_SMEFT_ALP}
In Section~\ref{sec:SMEFT_pheno}, we focused on the $D$-type VLF as a representative example, which couples to the down-quark sector and leads to the characteristic pattern of constraints discussed. In this appendix, we perform an analogous analysis for the $Q_7$ VLF, which couples to right-handed up-type quarks.

Starting with the SMEFT analysis, the resulting constraints on the relevant Wilson coefficients, together with the dominant observables driving the fit, are shown in Figure~\ref{fig:Q7_VLF_SMEFT_observables_pheno}. As in the previous sections, we distinguish between flavor-diagonal and off-diagonal configurations:

\begin{figure*}[t]
  \centering
  \begin{tabular}{cc}
    \includegraphics[width=0.475\linewidth]{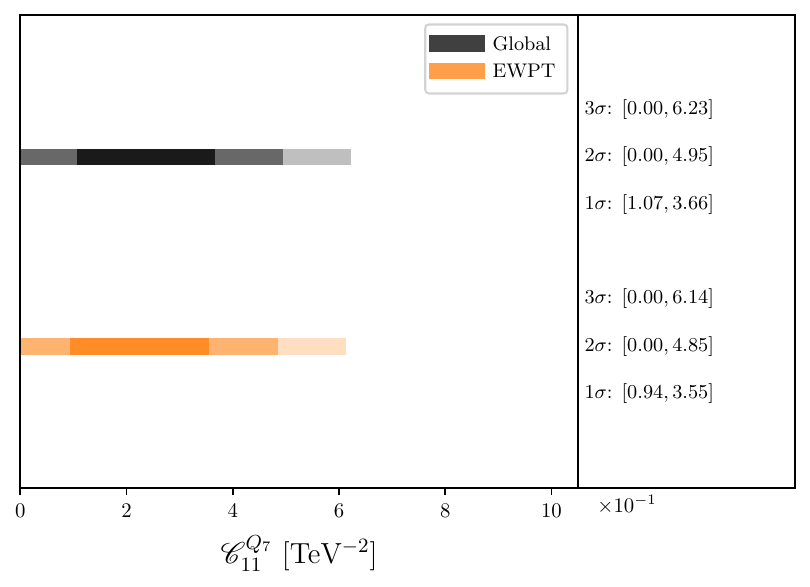}
    &
    \includegraphics[width=0.475\linewidth]{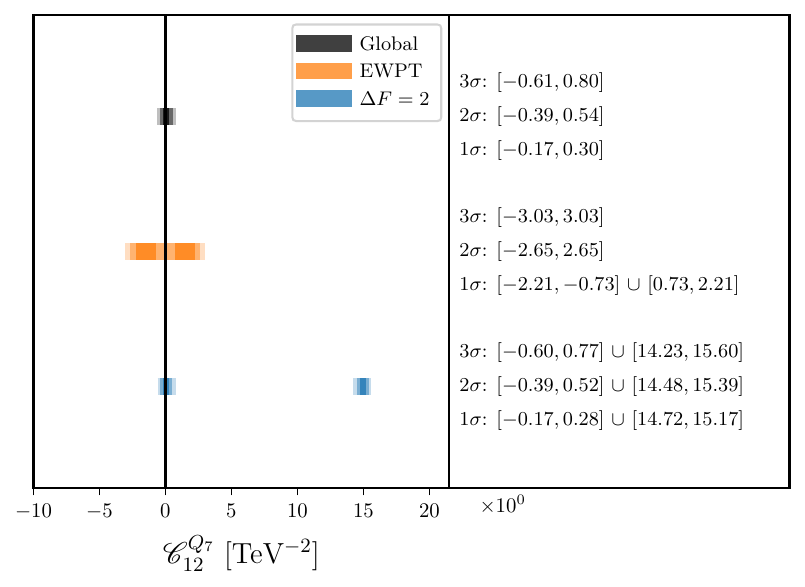}
    \\
    \includegraphics[width=0.475\linewidth]{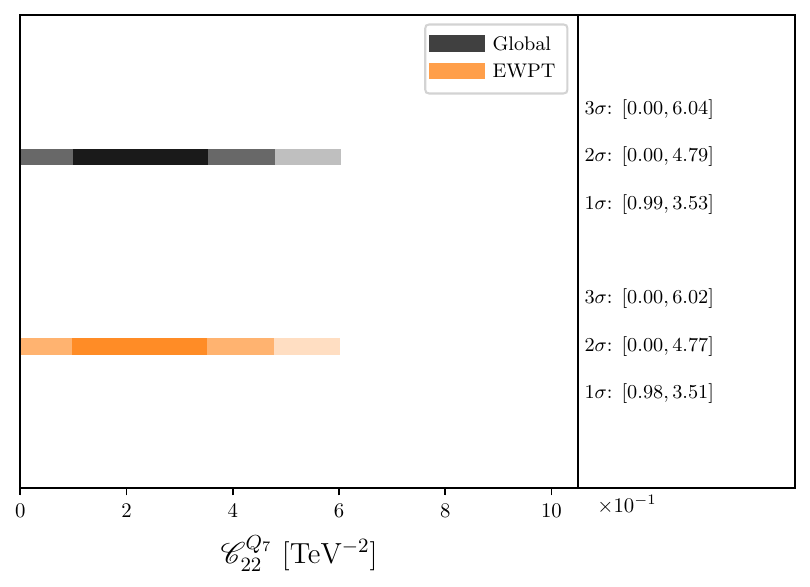}
    &
    \includegraphics[width=0.475\linewidth]{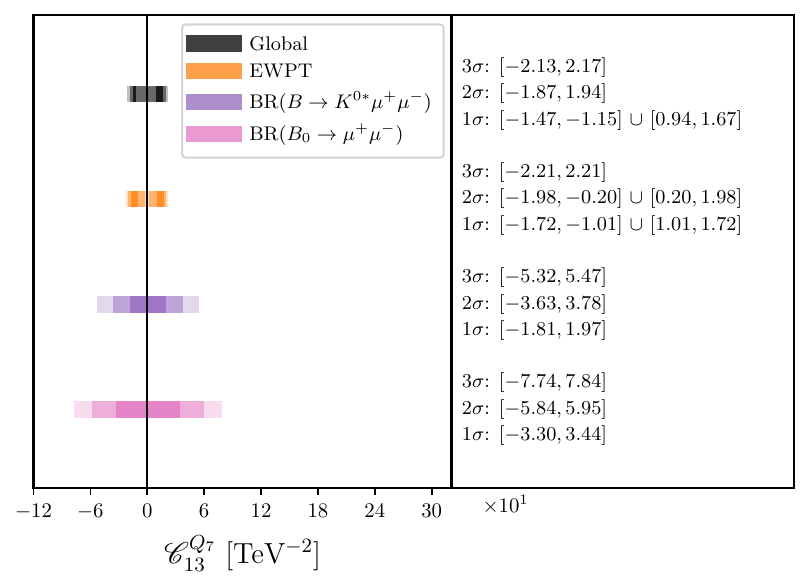}
    \\
    \includegraphics[width=0.475\linewidth]{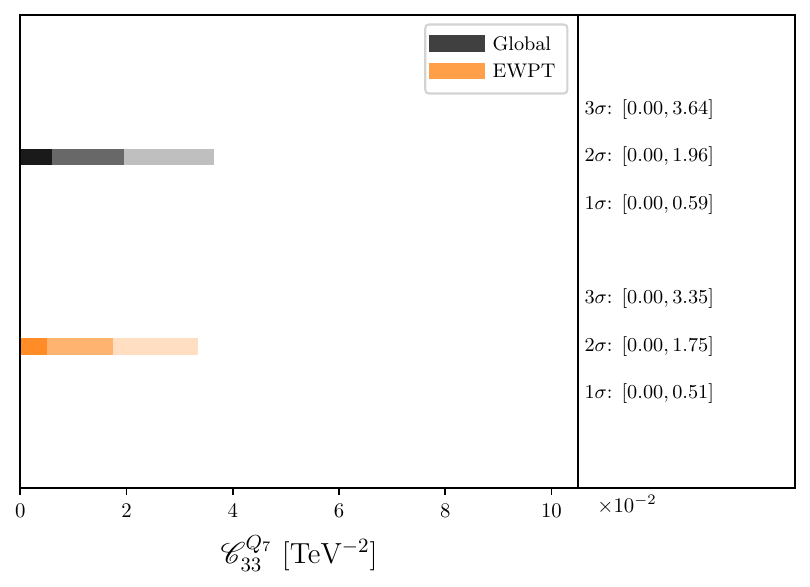}
    &
    \includegraphics[width=0.475\linewidth]{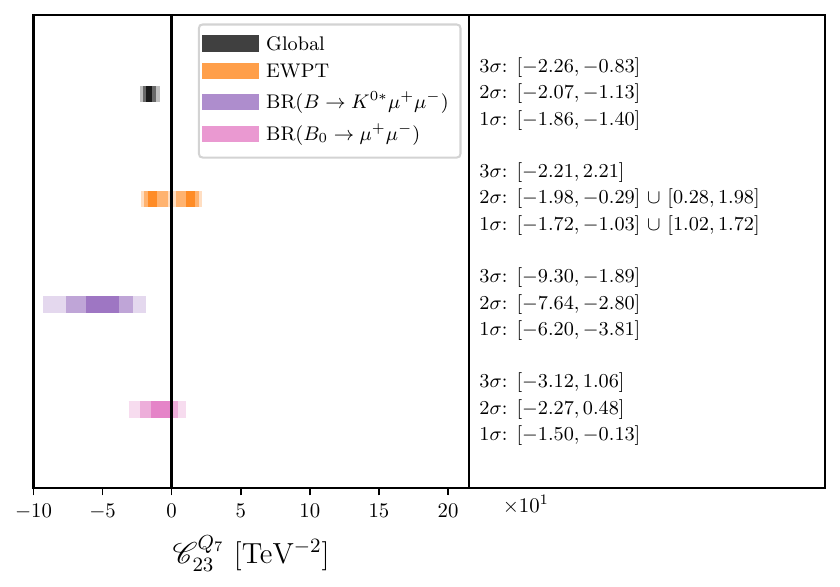}
  \end{tabular}
  \caption{Decomposition of the constraints on the $\mathscr{C}^{Q_7}_{ij}$ parameter for the $Q_7$-type VLF obtained from different subsets of observables. Each panel corresponds to a specific flavor configuration $(ij)$. The horizontal bars show the $1\sigma$, $2\sigma$, and $3\sigma$ intervals derived from fits restricted to selected classes of observables, compared with the result of the full global fit.}
  \label{fig:Q7_VLF_SMEFT_observables_pheno}
\end{figure*}

\begin{itemize}
  \item For the flavor-diagonal entries $(ij)=(11),(22),(33)$, the constraints are uniformly dominated by electroweak precision observables and are of order $\cO(10^{\eminus1}\,\tev^{\eminus2})$.
  \item The off-diagonal flavor configurations exhibit a richer phenomenological structure. For $(12)$, in addition to the EWPT contributions, $\Delta F=2$ observables become relevant, with the dominant sensitivity arising from the imaginary part of the charm-mixing amplitude. Although these effects strengthen the constraints, the resulting bounds remain weaker than in the flavor-diagonal case. For the $(13)$ and $(23)$ configurations, EWPT continues to provide an important contribution, while additional sensitivity arises from B-physics observables, most notably the branching ratios $\mathrm{BR}(B\to K^{\ast0}\mu^+\mu^-)$ and $\mathrm{BR}(B_s\to\mu^+\mu^-)$. These effects are induced through renormalization group evolution driven by the Yukawa sector, which propagates the $(13)$ flavor structure into the $(23)$ sector and leads to comparable constraints in both cases. Nevertheless, the resulting bounds remain weaker than those obtained for the flavor-diagonal configurations, in agreement with the general trends identified in the discussion of Tables~\ref{tab:SMEFT_intervals_sbs_diag} and \ref{tab:SMEFT_intervals_sbs_offdiag} in Section~\ref{sec:SMEFT_pheno}.
\end{itemize}
\begin{figure}[t]
  \centering
  \includegraphics[width=1\textwidth]{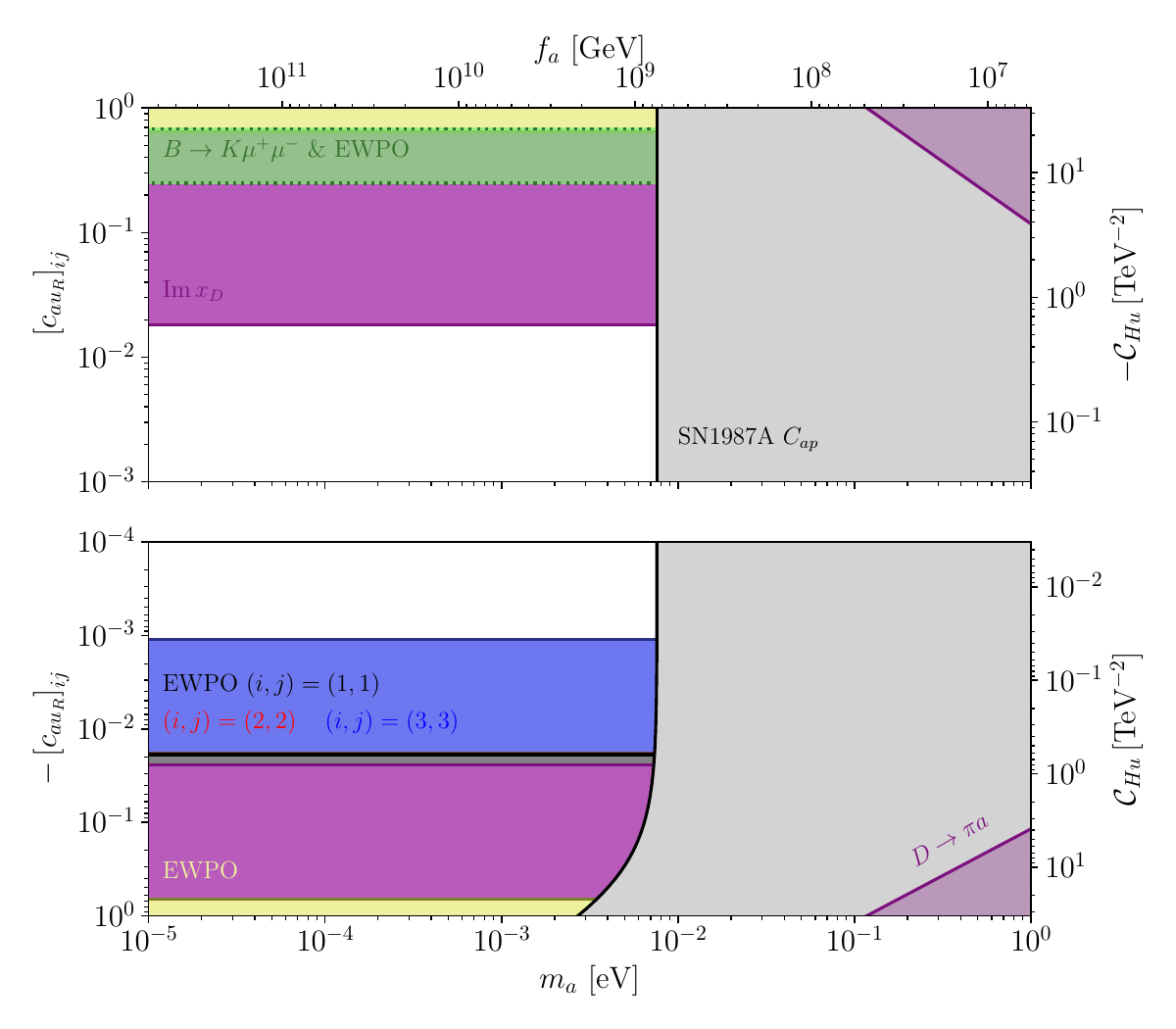}
  \caption{Bounds on the QCD-axion parameter space in the $(m_a,\,c_{au_R})$ plane for the $Q_7$ model. The black, red, and blue regions denote the constraints on the first-, second-, and third-generation diagonal coefficients, respectively, as derived from EWPO. The purple, yellow, and green regions correspond instead to the flavor-violating transitions.}
  \label{fig:axion_Q7_model}
\end{figure}
Lastly, we consider the QCD axion constraints in the $Q_7$ model. The resulting bounds are shown in Figure~\ref{fig:axion_Q7_model}, where the translated SMEFT constraints can be directly compared with conventional axion limits. In this case, axion couplings involving the second generation, as well as the flavor-changing $t\to c$ and $t\to u$ transitions, are probed most efficiently through indirect SMEFT bounds. These constraints arise from RGE effects that mix the up-sector interactions into down-quark flavor operators, which provide the strongest sensitivity in the off-diagonal cases, except for the $u-c$ sector. For the flavor-diagonal interactions, instead, the dominant constraints are set by EWPO. Once again, for scenarios in which the axion couplings do not arise from mass mixing, $D\to\pi$ a provides only limited sensitivity. Consequently, the parameter space selected by the SMEFT analysis remains well beyond the reach of both current and projected future searches.

\bibliographystyle{JHEP}
\bibliography{References}

\end{document}